\documentclass[preprint2]{aastex}

\usepackage{graphicx}
\usepackage{natbib}
\usepackage{color}
\newcommand{\Spitzer}{\textit{Spitzer}}
\newcommand{\Herschel}{\textit{Herschel}}

\newcommand{\Av}{$\rm{A}_V$}
\newcommand{\Rv}{$\rm{R}_V$}
\newcommand{\Lsun}{L$_\odot$}

\newcommand{\lkha}{LkH$\alpha$~}
\newcommand{\alphaexcess}{$\alpha_{\rm{excess}}$}
\newcommand{\lambdaturnoff}{$\lambda_{\rm{turnoff}}$}
\newcommand{\Lstar}{$L_{\rm{star}}$}
\newcommand{\Ldisk}{$L_{\rm{disk}}$}
\newcommand{\Lcrit}{$\rm{L}_{\rm{crit}}$}
\newcommand{\Reff}{$R_{\rm{eff}}$}
\newcommand{\Rcirc}{$R_{\rm{circ}}$}
\newcommand{\NYSO}{$N_{\rm{YSO}}$}
\newcommand{\NII}{$N_{\rm{II}}$}
\newcommand{\NF}{$N_{\rm{F}}$}
\newcommand{\NI}{$N_{\rm{I}}$}
\newcommand{\NIF}{$N_{\rm{I+F}}$}

\newcommand{\ch}{\colhead}

\newcommand{\tnm}{\tablenotemark}
\newcommand{\tnt}{\tablenotetext}

\shorttitle{GBS VI. Star Formation in the AMC}
\shortauthors{Broekhoven-Fiene et al.}

\begin{document}

\title{The \Spitzer\ Survey of Interstellar Clouds in the Gould Belt. VI. The Auriga-California Molecular Cloud observed with IRAC and MIPS }

\author{Hannah Broekhoven-Fiene\altaffilmark{1}, %\\ 
Brenda C.~Matthews\altaffilmark{2,1}, %\\ %\email{brenda.matthews@nrc-cnrc.gc.ca}
Paul M. Harvey\altaffilmark{3}, %\\
Robert A. Gutermuth\altaffilmark{4}, %\\
Tracy L. Huard\altaffilmark{5,6}, %\\ 
Nicholas F. H. Tothill\altaffilmark{7}, %\\ 
David Nutter\altaffilmark{8}, %\\ 
Tyler L. Bourke\altaffilmark{9}, %\\ 
James DiFrancesco\altaffilmark{2}, %\\
Jes K. J{\o}rgensen\altaffilmark{10,11}, %\\
Lori E. Allen\altaffilmark{12}, %\\ 
Nicholas L. Chapman\altaffilmark{13}, %\\
Michael M. Dunham\altaffilmark{14}, %\\ 
Bruno Mer\'{i}n\altaffilmark{15}, %\\ 
Jennifer F. Miller\altaffilmark{5,9}, %\\
Susan Terebey\altaffilmark{16}, %\\ 
Dawn E. Peterson\altaffilmark{17}, %\\ 
Karl R. Stapelfeldt\altaffilmark{18}}

\altaffiltext{1}{Department of Physics and Astronomy, University of Victoria, Victoria, BC, V8W 3P6, Canada} % Hannah and Brenda
\altaffiltext{2}{National Research Council Herzberg Astronomy \& Astrophysics, Victoria, BC, V9E 2E7, Canada} % Brenda
\altaffiltext{3}{Astronomy Department, University of Texas at Austin, 1 University Station C1400, Austin, TX 78712-0259, USA} % Paul
\altaffiltext{4}{Department of Astronomy, University of Massachusetts, Amherst, MA, USA} % Rob
\altaffiltext{5}{Department of Astronomy, University of Maryland, College Park, MD 20742, USA} % Tracy 1
\altaffiltext{6}{Columbia Astrophysics Laboratory, Columbia University, New York, NY 10027, USA} % Tracy 2
\altaffiltext{7}{School of Computing, Engineering \& Mathematics, University of Western Sydney, Locked Bag 1797, Penrith, NSW 2751, Australia} % Nick
\altaffiltext{8}{School of Physics and Astronomy, Cardiff University, Queen's Buildings, The Parade, Cardiff CF24 3AA, UK} % David
\altaffiltext{9}{Harvard-Smithsonian Center for Astrophysics, 60 Garden Street, Cambridge, MA 02138, USA}%; tbourke@cfa.harvard.edu}
\altaffiltext{10}{Niels Bohr Institute, University of Copenhagen, Juliane Maries Vej 30, DK-DK-2100 Copenhagen {\O.}, Denmark} % Jes 1
\altaffiltext{11}{Centre for Star and Planet Formation, Natural History Museum of Denmark, {\O}ster Voldgade 5-7, DK-1350 Copenhagen K., Denmark} % Jes 2
\altaffiltext{12}{National Optical Astronomy Observatories, Tucson, AZ, USA} % Lori
\altaffiltext{13}{Center for Interdisciplinary Exploration and Research in Astrophysics (CIERA) \& Department of Physics \& Astronomy, Northwestern University, 2145 Sheridan Road, Evanston, IL 60208} % Nicholas Chapman
\altaffiltext{14}{Department of Astronomy, Yale University, P.O. Box 208101, New Haven, CT 06520, USA} % Mike
\altaffiltext{15}{Herschel Science Centre, ESAC-ESA, P.O. Box 78, 28691 Villanueva de la Ca\~{n}ada, Madrid, Spain} % Bruno
\altaffiltext{16}{Department of Physics and Astronomy PS315, 5151 State University Drive, California State University at Los Angeles, Los Angeles, CA 90032, USA} % Susan
\altaffiltext{17}{Space Science Institute, 4750 Walnut Street, Suite 205, Boulder, CO 80301, USA}%; \email{dpeterson@spacescience.org}} 
\altaffiltext{18}{Code 667, NASA Goddard Space Flight Center, Greenbelt, MD 20771, USA} % Karl

\begin{abstract}
We present observations of the Auriga-California Molecular Cloud (AMC) at 3.6, 4.5, 5.8, 8.0, 24, 70 and 160 \micron\ observed with the IRAC and MIPS detectors as part of the {\it Spitzer} Gould Belt Legacy Survey. The total mapped areas are 2.5 deg$^2$ with IRAC and 10.47 deg$^2$ with MIPS. This giant molecular cloud is one of two in the nearby Gould Belt of star-forming regions, the other being the Orion A Molecular Cloud (OMC). We compare source counts, colors and magnitudes in our observed region to a subset of the SWIRE data that was processed through our pipeline. Using color-magnitude and color-color diagrams, we find evidence for a substantial population of 166 young stellar objects (YSOs) in the cloud, many of which were previously unknown. Most of this population is concentrated around the \lkha 101 cluster and the filament extending from it. We present a quantitative description of the degree of clustering and discuss the fraction of YSOs in the region with disks relative to an estimate of the diskless YSO population.  Although the AMC is similar in mass, size and distance to the OMC, it is forming about 15 -- 20 times fewer stars. 
\end{abstract}


%%%%%%%%%%%%%%%%%%%%%%%%%%%%%%%%%%%%%%%%%%%%%%%%
%%%%%%%%%%%%%%%%%%%%%%%%%%%%%%%%%%%%%%%%%%%%%%%%

\section{Introduction}\label{sec:intro}

The cycle 4 {\it Spitzer Space Telescope} Legacy project ``The Gould Belt: Star Formation in the Solar Neighborhood'' (PID: 30574; PI: L.E. Allen) completed the {\it Spitzer} survey of the large, nearby star-forming regions begun by the c2d Legacy Project \citep{Evans2003, Evans2009}.  The cloud with the least prior study included in the survey is the cloud we have designated as ``Auriga'' which lies on the Perseus-Auriga border. This cloud has also been designated the California Molecular Cloud by \cite{Ladaetal2009} since it extends from the California Nebula in the west to the \lkha 101 region and associated NGC 1529 cloud in the east. We adopt the name Auriga-California Molecular Cloud (AMC) to encompass both nomenclatures.  

Despite the AMC's proximity to two of the most well-examined star-forming clouds, Taurus-Auriga and Perseus, it is a relatively unstudied region. Several dark nebulae were noted along its length by \cite{Lynds1962}, and CO associated with many Lynds objects was measured by \cite{Ungerechts1987}, who note the presence of a CO ``cloud extending from the California nebula (NGC 1499) in Perseus along NGC 1579 and \lkha 101 well into Auriga'' (their cloud 12). Only very recently has a giant molecular cloud been unambiguously associated with the series of Lynds nebulae through high resolution extinction maps by \cite{Ladaetal2009} who placed its distance firmly within the Gould Belt (GB) at $450 \pm 23$ pc.  At this distance, the cloud's extent of 80 pc and mass of $\sim10^5 \ M_{\odot}$ rivals that of the Orion Molecular Cloud (L1641) for the most massive in the Gould Belt.  For the remainder of this paper, we adopt this distance of 450 pc for the entire AMC. This is consistent with the distance of $510^{+100}_{-40}$ pc found by \citep{Wolketal2010} on their study of \lkha 101 with Chandra. We note that this distance differs from that adopted by \cite{Gutermuthetal2009} for \lkha 101  of 700 pc. 

We have mapped a significant fraction of the AMC with the Infrared Array Camera (IRAC; \citealt{Fazio2004}) and the Mid-Infrared Photometer for \Spitzer\ (MIPS; \citealt{Rieke2004}) on board the \textit{Spitzer Space Telescope} \citep{Werner2004}, with a total overlapping coverage of 2.5 deg$^2$ in the four IRAC bands (3.6, 4.5, 5.8 and 8.0 \micron) and 10.47 deg$^2$ in the three MIPS bands (24, 70, and 160 \micron). The mapped areas are not all contiguous and were chosen to include the areas with \Av~$> 3$, as given by the \cite{Dobashi2005} extinction maps. The goal of these observations is to identify and characterize the young stellar object (YSO) and substellar object populations. The data presented here are the first mid-IR census of the YSO population in this region. The area around \lkha 101 and its associated cluster was observed as part of a survey of 36 clusters within 1 kpc of the Sun with \Spitzer\ by \cite{Gutermuthetal2009} and those data have been incorporated into our dataset through the c2d pipeline.  

More recently, the AMC has been observed by the {\it Herschel Space Observatory} at 70 -- 500 \micron, and by the Caltech Submillimeter Observatory with the Bolocam 1.1 mm camera \citep{Harveyetal2013}. These observations characterize the diffuse dust emission and the cooler Class 0 and Class I objects which can be bright in the far-IR. We do not analyze the large scale structure of the cloud in this paper as \cite{Harveyetal2013} present such an analysis with the \Herschel~observations, which are more contiguous and have a higher resolution than our MIPS observations. \cite{Harveyetal2013} also include a comparison to these MIPS data and so further analysis is not required here.

We describe the observations and data reduction (briefly as it is well-documented elsewhere) in $\S$ \ref{sec:obs}. In $\S$ \ref{sec:yso}, we describe the source statistics, the criteria for identifying and classifying YSO candidates and we compare the YSO population to other clouds. The SEDs and disk properties of YSOs are modeled in $\S$ \ref{sec:sed}. We characterize the spatial distribution of YSOs in $\S$ \ref{sec:spatial} and summarize our findings in $\S$ \ref{sec:summary}.

%%%%%%%%%%%%%%%%%%%%%%%%%%%%%%%%%%%%%%%%%%%%%%%%
%%%%%%%%%%%%%%%%%%%%%%%%%%%%%%%%%%%%%%%%%%%%%%%%

\section{Observations and Data Reduction}\label{sec:obs}

\begin{figure*}
\includegraphics[width=6 in,clip=True,trim=2.45cm 7cm 3.85cm 8cm]{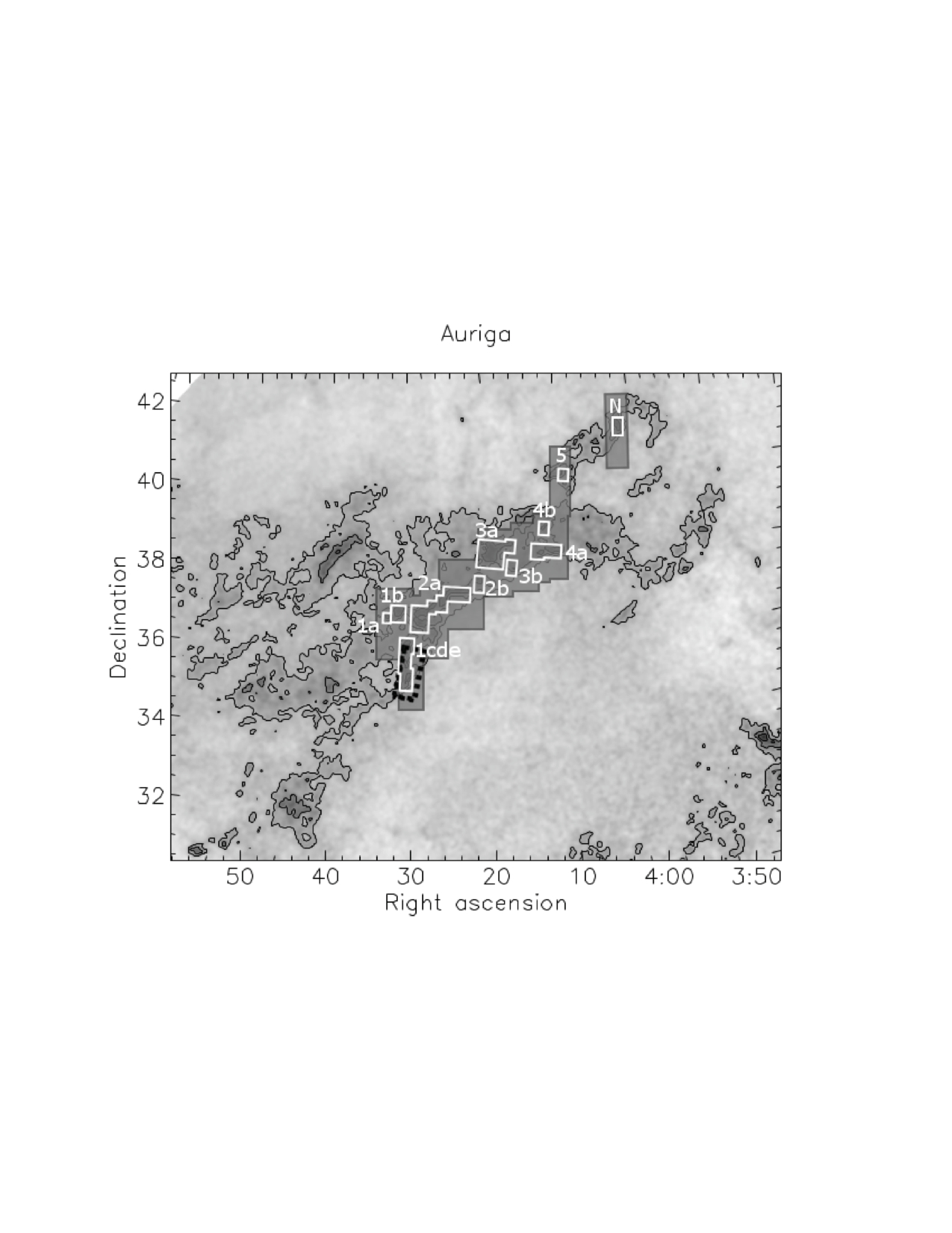}
\caption{Integrated \Spitzer\ mapped areas from the Gould Belt Survey and other projects. The grey boxed area shows the MIPS coverage; the white boxes show the IRAC coverage (with the sub-regions labelled); and the hatched black box shows the non-GBS survey data in the field from \cite{Gutermuthetal2009}. These regions are schematic to give a general picture of the layout of the coverage and to identify the subregions. The greyscale is the extinction map of \cite{Dobashi2005}. Contours show the $A_V$ levels of 1, 3 and 5 mag.}\label{fig:areacoverage}
\end{figure*}

The areas mapped are shown in Figure \ref{fig:areacoverage}.  The MIPS coverage is more contiguous than the IRAC coverage due to the mapping modes of the two instruments. Observations were designed to cover regions with \Av~$> 3$ within the extinction maps of \cite{Dobashi2005}. All areas were observed twice with IRAC and MIPS cameras with the AORs and dates of the observations compiled in Tables \ref{iracobssum} and \ref{mipsobssum}.  The two epochs were compared to remove transient asteroids that are numerous at the low ecliptic latitude of these observations.

The GBS survey data and the \lkha 101 data from \cite{Gutermuthetal2009} were processed through the c2d pipeline. Details of the data processing are available in \cite{Evans2007}. Briefly, the data processing starts with a check of the images whereupon image corrections are made for obvious problems. Mask files are created to remove problematic pixels.  The individual frames are then mosaicked together, with one mosaic created for each epoch and one joint mosaic as well.  Sources are detected in each mosaic and then re-extracted from the stack of individual images which include the source position. Finally, the source lists for each wavelength are band-merged, and sources not detected at some wavelengths are ``band-filled'' to find appropriate fluxes or upper limits at the positions which had clear detections at other wavelengths.

As noted by \cite{Harvey2008}, the details of this data reduction are essentially the same as that of the original c2d datasets except that the input to the c2d pipeline are products of later versions of the \Spitzer\ BCD pipeline.  The c2d processing of IRAC data was described by \cite{Harvey2006}, and the MIPS data processing was described by \cite{Young2005} and \cite{Rebull2007}.  \cite{Harvey2007} describe additional reduction processes which we have used for the AMC data.

%%%%%%%%%%%%%%%%%%%%%%%%%%%%%%%%%%%%%%%%%%%%%%%%
%%%%%%%%%%%%%%%%%%%%%%%%%%%%%%%%%%%%%%%%%%%%%%%%

\section{Star-forming Objects in the AMC}
\label{sec:yso}

\begin{figure}%[b]
\includegraphics[width=3.5 in, clip=True,trim=2.5cm 0cm 3cm 0cm]{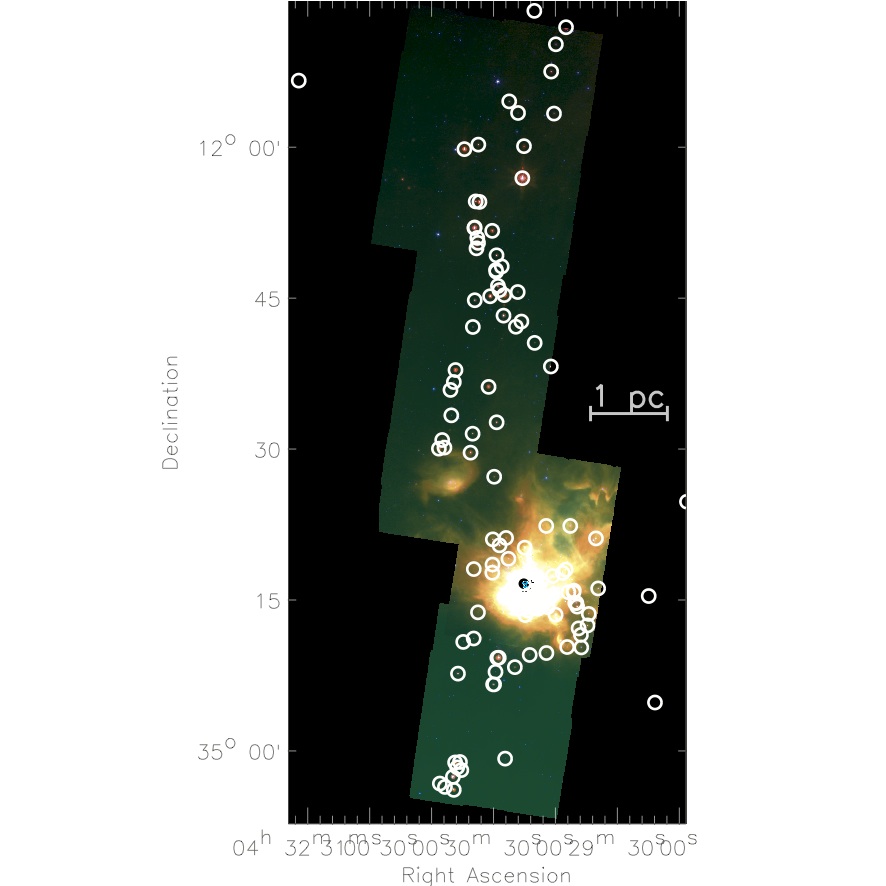}
\caption[AUR_reg1cde_IRAC24_MIPS1.pdf]{
False colour image with 4.5 \micron\ (blue), 8 \micron\ (green), and 24 \micron\ (red) of the IRAC 1cde fields with YSO positions are overlaid.
(Similar figures for other IRAC regions are shown in Figures \ref{fig:rgb2} -- \ref{fig:rgb4}.)}
\label{fig:rgb1}
\end{figure}

\begin{figure*}%[h]
\includegraphics[width=4.5 in, clip=True,trim=0cm 0.5cm 0cm 0.5cm]{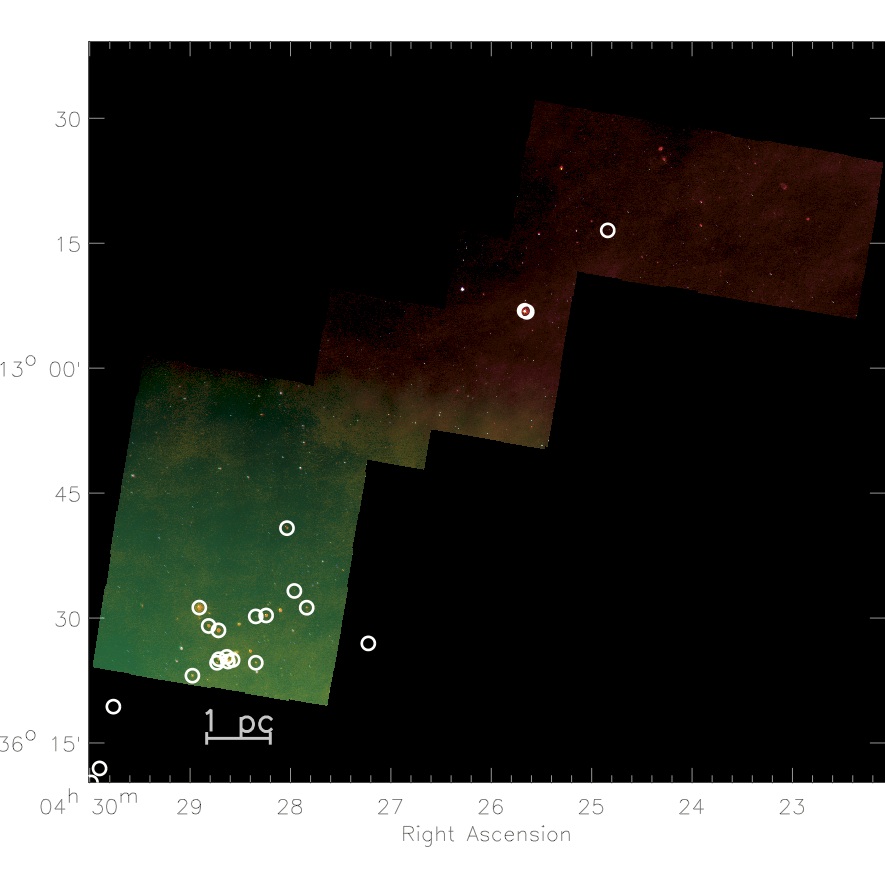}\centering
\caption[AUR_reg2a_IRAC24_MIPS1.pdf]{
False colour image with 4.5 \micron\ (blue), 8 \micron\ (green), and 24 \micron\ (red) of the IRAC 2a field with YSO positions are overlaid.
(Similar figures for other IRAC regions are shown in Figures \ref{fig:rgb1}, \ref{fig:rgb3}, and \ref{fig:rgb4}.)}
\label{fig:rgb2}
\end{figure*}

\begin{figure*}%[h]
\begin{centering}
\includegraphics[width=3in, clip=True, trim=0cm 0.25cm 0cm 0.25cm]{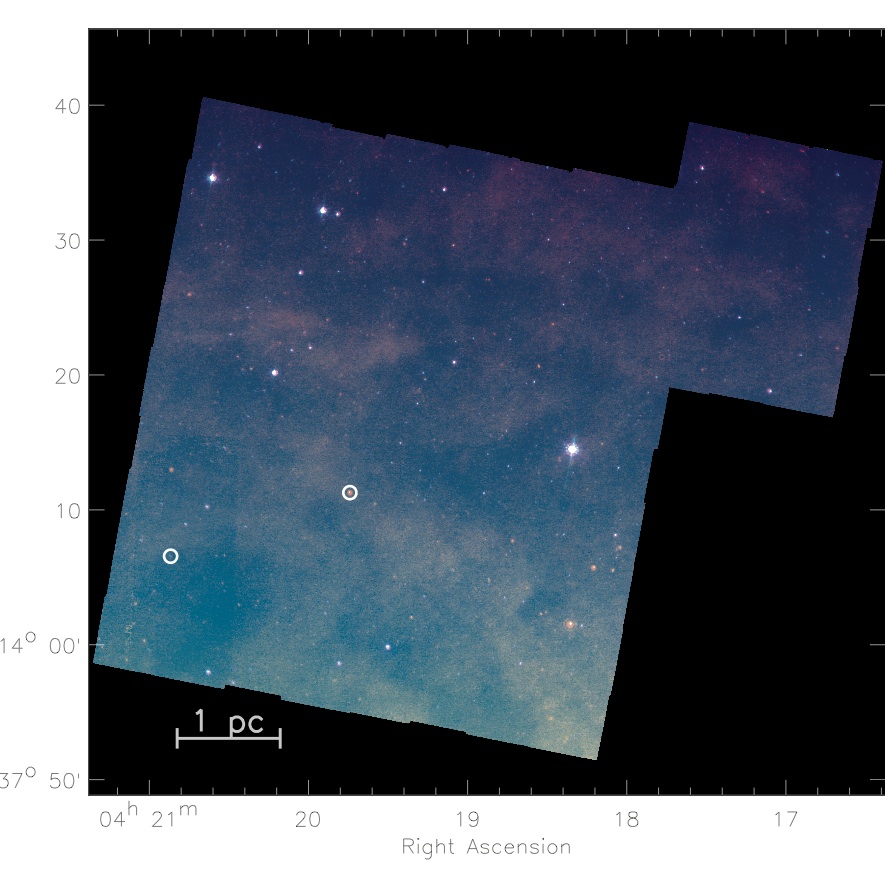}
\includegraphics[width=3.5in, clip=True, trim=0cm 2cm 0cm 2cm]{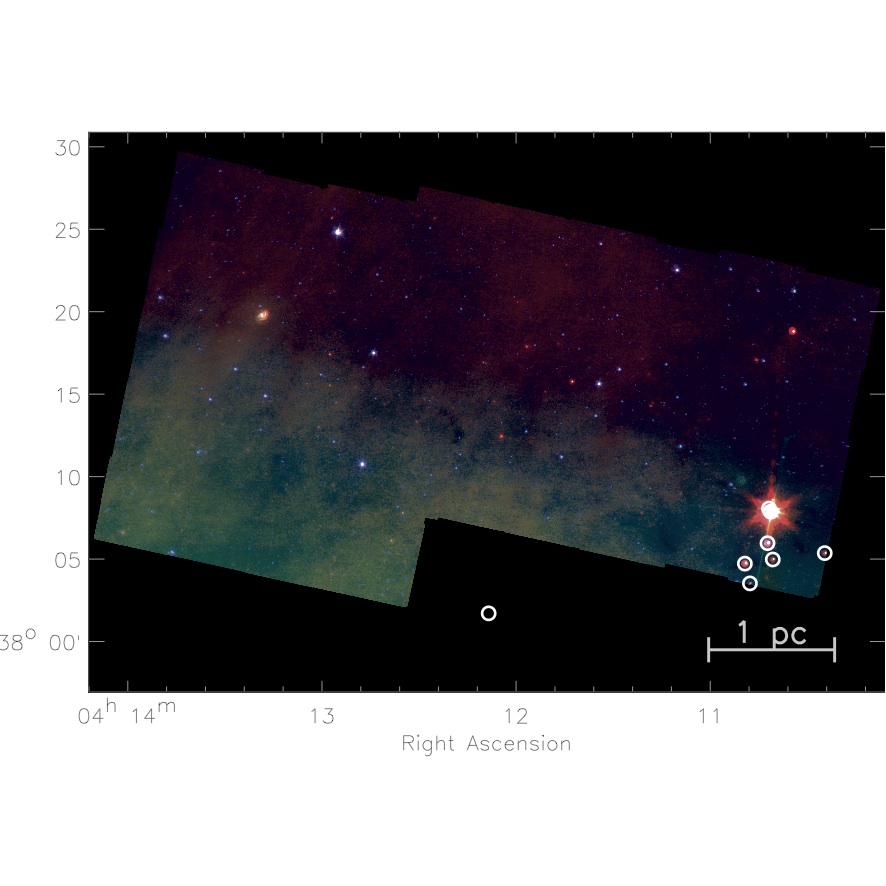}
\includegraphics[width=2.2in, clip=True, trim=1cm 0cm 1.5cm 0cm]{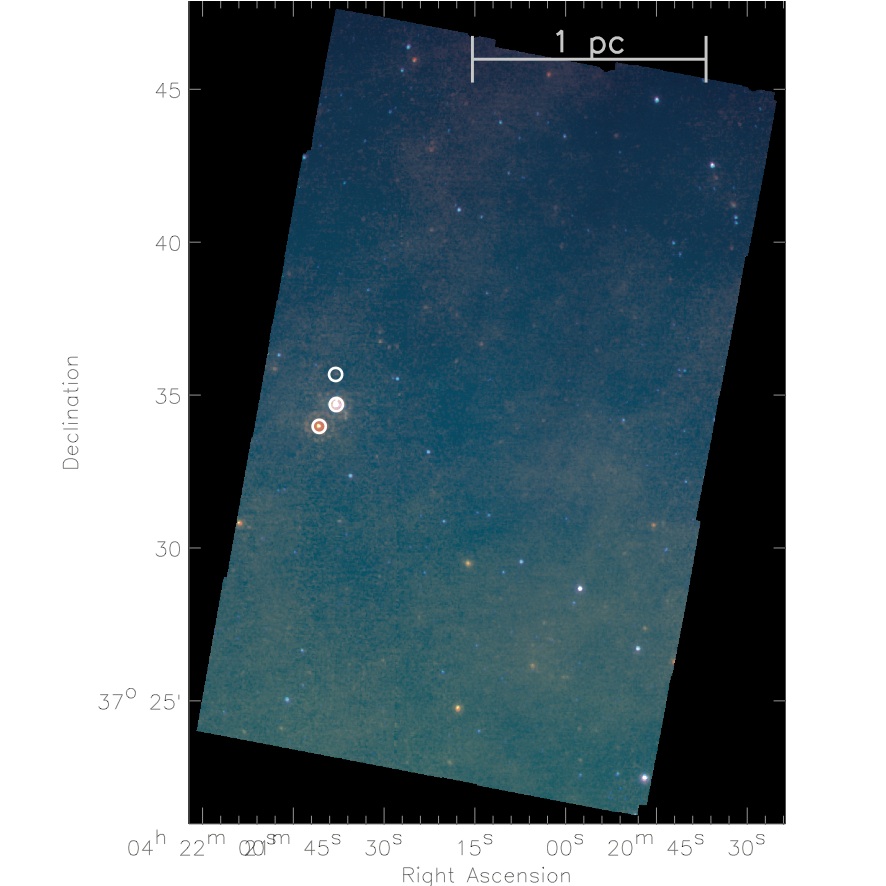}
\includegraphics[width=2.2in, clip=True, trim=0.5cm 0cm 1.25cm 0cm]{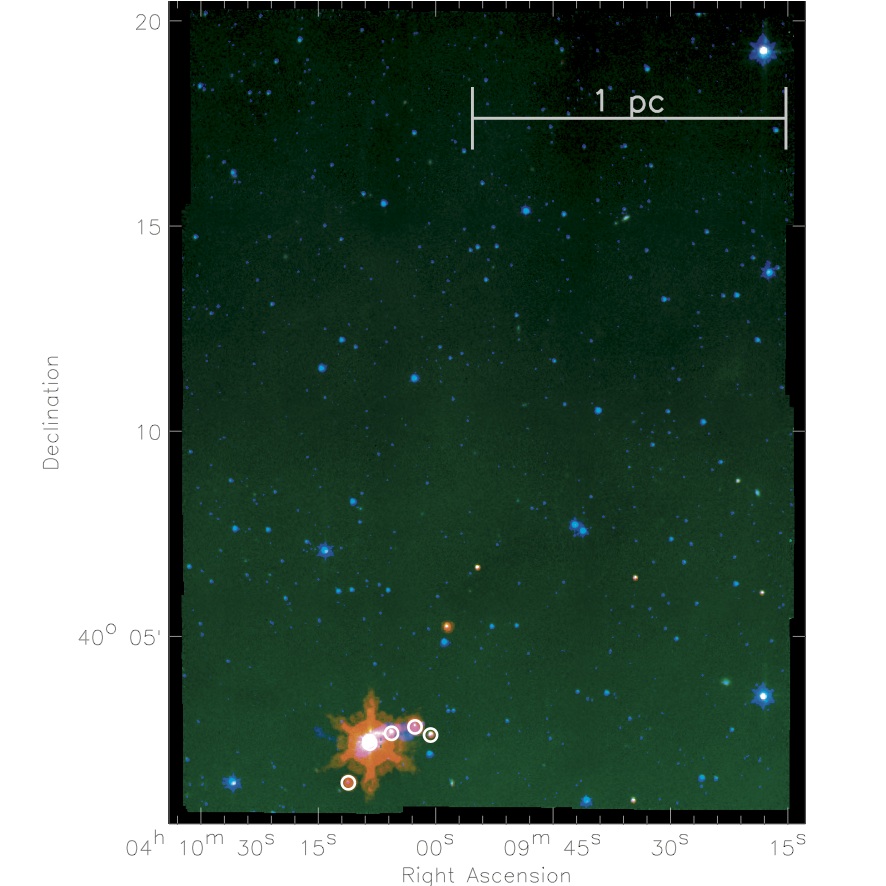}
\includegraphics[width=1.8in, clip=True, trim=2.25cm 0cm 3cm 0cm]{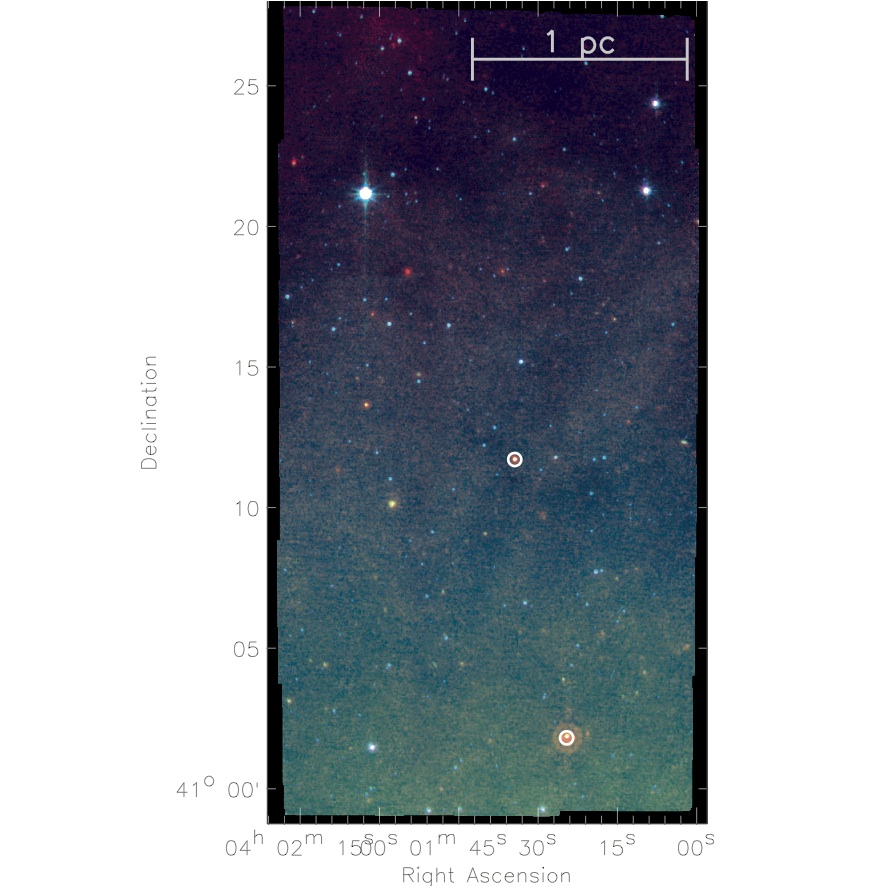}
\end{centering}
\caption{ 
False colour image with 4.5 \micron\ (blue), 8 \micron\ (green), and 24 \micron\ (red) of the IRAC fields 3a, 4a, 2b, 5, and North (left to right, top to bottom) with YSO positions are overlaid. These regions contain only a few YSOs each.
(Similar figures for other IRAC regions are shown in Figures \ref{fig:rgb1}, \ref{fig:rgb2}, and \ref{fig:rgb4}.)}
\label{fig:rgb3}
\end{figure*}

\begin{figure*}%[h]
\begin{centering}
\includegraphics[scale=0.6, clip=True, trim=0.5cm 0cm 1.25cm 0cm]{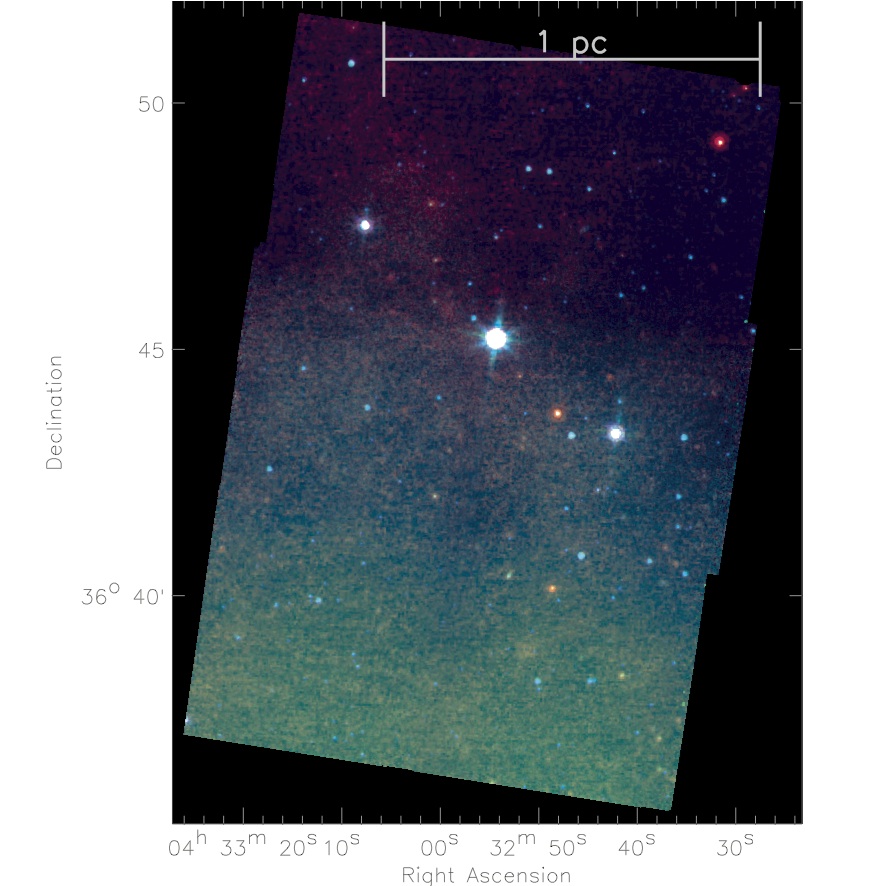}
\includegraphics[scale=0.6, clip=True, trim=0cm 0cm 0.5cm 0cm]{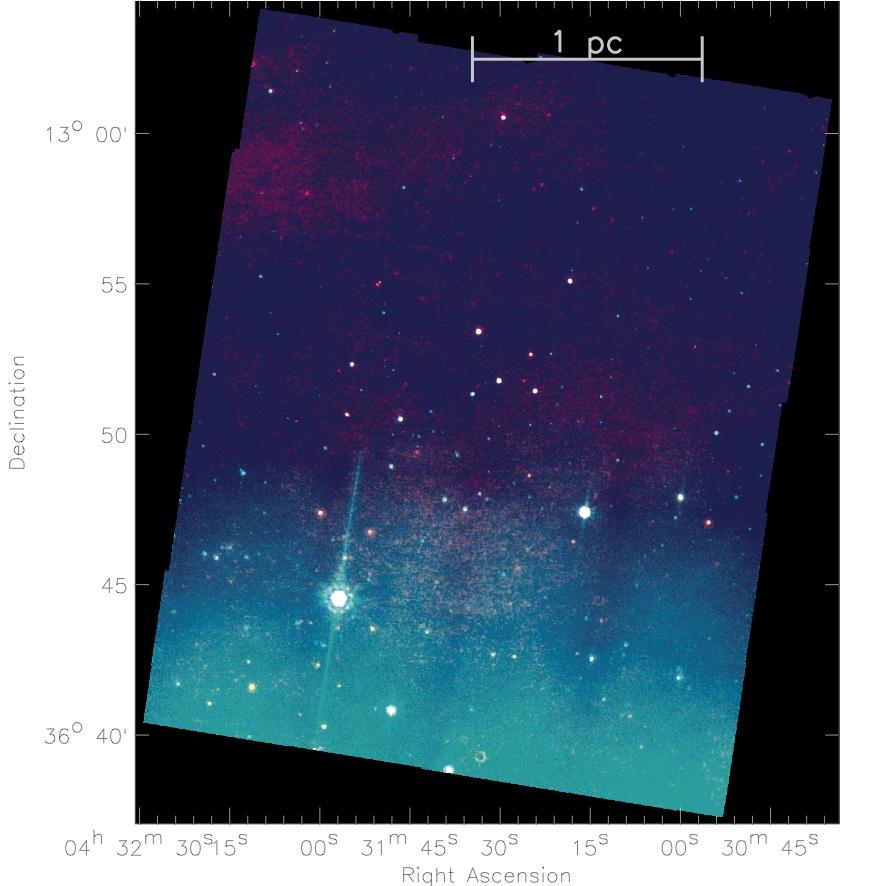}
\includegraphics[scale=0.6, clip=True, trim=1cm 0cm 1.5cm 0cm]{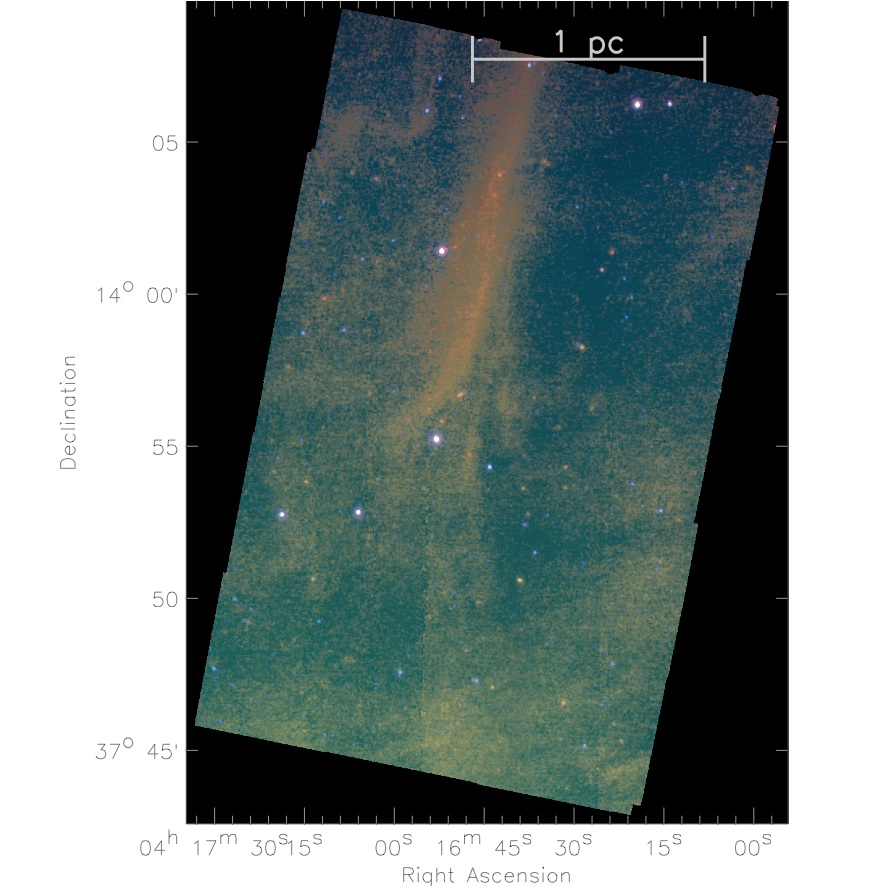}
\includegraphics[scale=0.6, clip=True, trim=0.25cm 0cm 0.75cm 0cm]{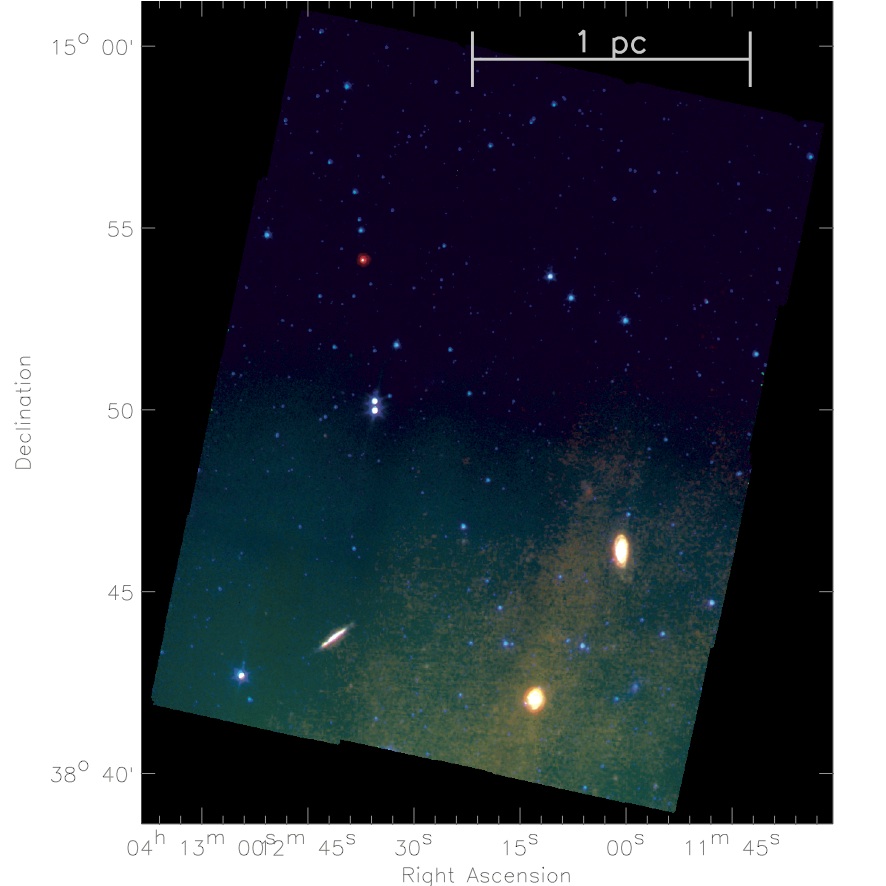}
\end{centering}
\caption{
False colour image with 4.5 \micron\ (blue), 8 \micron\ (green), and 24 \micron\ (red) of the IRAC fields 1a, 1b, 3b, and 4b (left to right, top to bottom) with YSO positions are overlaid. These regions do not contain YSOs.
(Similar figures for other IRAC regions are shown in Figures \ref{fig:rgb1} -- \ref{fig:rgb3}.)}
\label{fig:rgb4}
\end{figure*}

Figures \ref{fig:rgb1} -- \ref{fig:rgb4} show RGB mosaics for the IRAC covered regions using 4.5 \micron\ (blue), 8.0 \micron\ (green) and 24 \micron\ (red) data with the positions of YSOs overlaid. The diffuse 8.0 \micron\ emission is strongly concentrated at the eastern edge of the cloud, near the well-known object \lkha 101. The \lkha 101 data are taken from and have been discussed by \cite{Gutermuthetal2009}. 

%%%%%%%%%%%%%%%%%%%%%%%%%%%%%%%%%%%%%%%%%%%%%%%%
\subsection{YSO Selection}
\label{sec:ysoselect}

The majority of objects in our fields are not YSOs. The maps are contaminated by background/foreground stars and background galaxies. We have selected our YSO candidates (YSOcs) by various methods, augmenting the list where possible based on data outside the \Spitzer\ IRAC/MIPS wavelength bands. The fundamental criteria use IRAC, MIPS and 2MASS data \citep{Cutri2003} and are based on identification of infrared excess and brightness limits below which the probability of detection of external galaxies becomes high. The total number of sources is 704,045. In regions observed by both IRAC and MIPS,
the YSOc selection follows that of \cite{Harvey2008}. We refer to these as IRAC+MIPS YSOcs. For objects with upper limits on the MIPS 24 \micron\ flux, we follow the method outlined by \cite{Harvey2006}. We refer to these as IRAC-only YSOcs. In regions observed only by MIPS and not IRAC, we have used the formalism of \cite{Rebull2007}, except we use a tighter 2MASS K$_{\rm{S}}$ cut of [K$_{\rm{S}}$] $< 13.5$. This tighter magnitude cut removed objects that were similar in color and magnitude to others that had already been eliminated. We further remove galaxies from the MIPS-only source list by including photometry from the Wide-field Infrared Survey Explorer (WISE; \citealt{Wright2010}) and applying color cuts suggested by \cite{Koenig2012} (see their Figure 7) and requiring the WISE Band 2 magnitude criterion of [4.6] $<$ 12. We refer to these as MIPS-only YSOcs. Note that the MIPS-only YSOcs were not observed with IRAC, as opposed to the IRAC-only YSOcs which were observed, but not detected, with MIPS.

\begin{figure*}%[h]
\includegraphics[width=7 in]{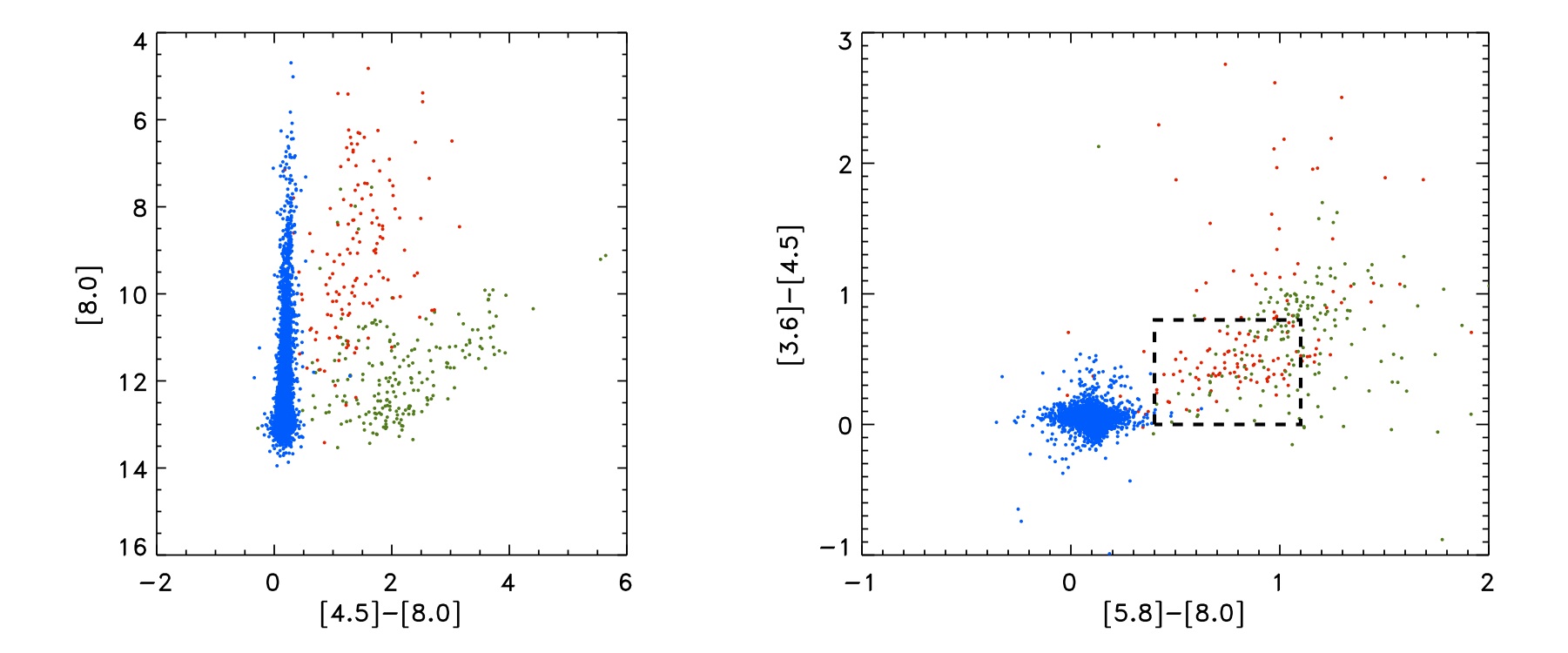}
\caption{IRAC colors of the sources in the the regions observed with IRAC. Stars are in blue; YSOs are in red; and ``other sources'' (e.g., galaxies) are in green. 
 The boxed region on the right panel marks the approximate domain of Class II sources identified by \cite{Allen2004}.}\label{fig:cc_dia}
\end{figure*}

\begin{figure*}%[h]
\includegraphics[width=5.5  in,clip=True, trim=0.1cm 0cm 2.5cm 0.25cm]{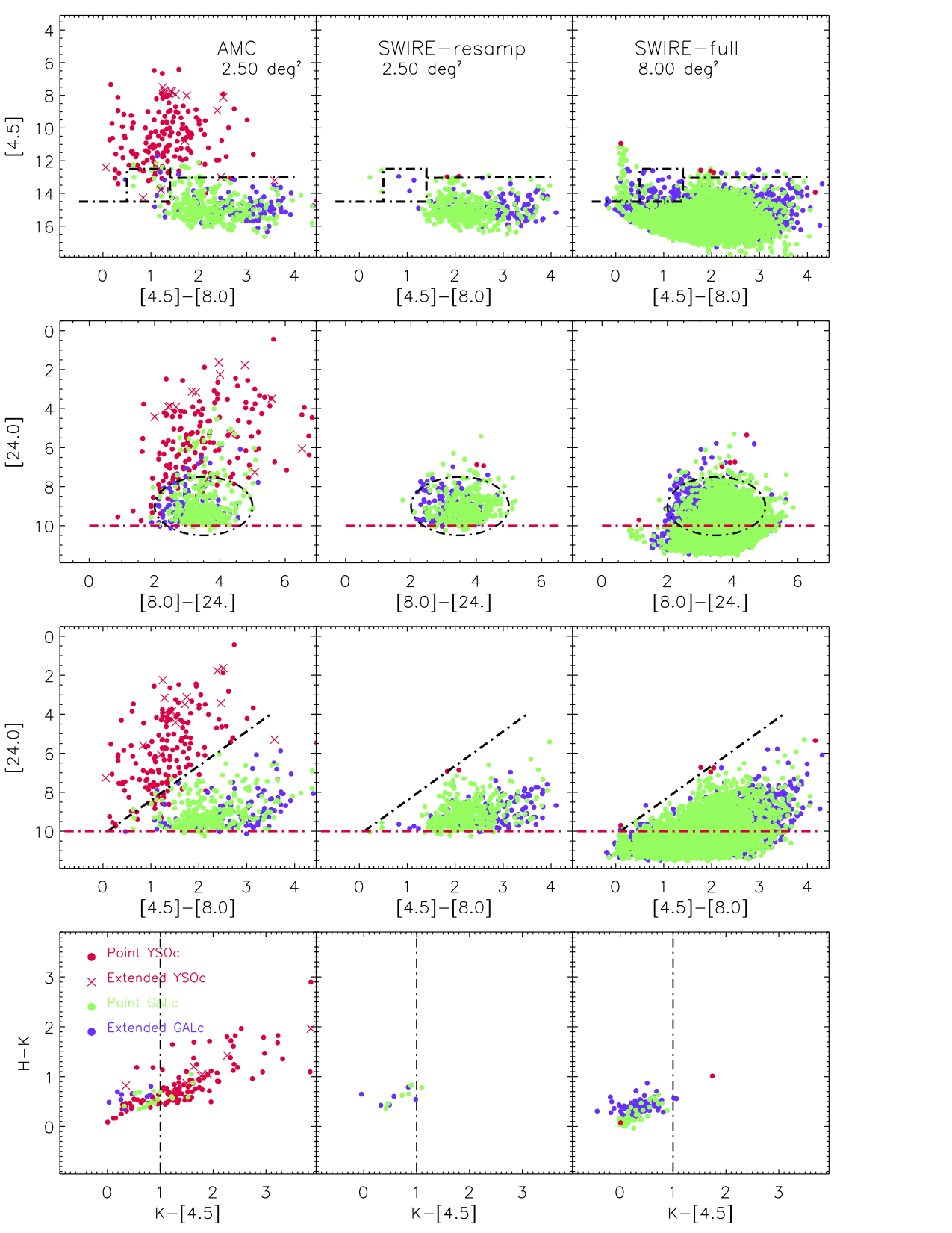}\centering
\caption{Color-magnitude and color-color diagrams for the AMC (left), the SWIRE dataset resampled to match our sensitivities and measured extinction (middle), and the full SWIRE dataset (right). The black dash-dot lines show soft boundaries for YSO candidates whereas the red dash-dot lines show hard limits, fainter than which objects are not included as YSO candidates.}\label{fig:colmag}
\end{figure*}

Figure~\ref{fig:cc_dia} shows the IRAC color-magnitude and color-color diagrams relevant for classifying IRAC-only sources. 
The different domains occupied by stars, YSOcs, and other (e.g., extragalactic) sources are are shown.

For sources in regions observed by both IRAC and MIPS, Figure~\ref{fig:colmag} shows the color and magnitude boundaries used to remove sources that are likely extragalactic. This identification is done by comparing the observed fluxes and colors to results from the SWIRE extragalactic survey \citep{Surace2004}. The sources in the AMC field are compared to a control catalogue from the SWIRE dataset that is resampled to match our sensitivity limits and the extinction level derived for the AMC. (See \citealt{Evans2007} for a complete description.)

Finally, we vetted the YSOcs through individual inspection of the \Spitzer\ maps (and optical images where available), and determined that 24 of the original 159 IRAC+MIPS YSOcs, 14 of the original 17 IRAC-only YSOcs, and 56 (26 based on WISE and other photometric criteria) of the original 84 MIPS-only YSOcs were unlikely to be YSOs. Henceforth we refer to the list of vetted YSOcs, totalling 166, as YSOs to distinguish them from the raw unvetted list. While we have undergone an extensive process to construct a list of sources that are very likely to be YSOs, we stress that these YSOs have not been confirmed spectroscopically. 
Table~\ref{tbl:SourceSummary} lists the final source counts for objects in the observed fields.
The IRAC and MIPS fluxes of the IRAC+MIPS and IRAC-only YSOs are listed in Table~\ref{tbl:irac}.
The 70 \micron\ fluxes have been listed where available. (There are fewer YSOs with fluxes at 70 \micron\ because of the lower sensitivity and, in some cases, the bright background.) The fluxes of MIPS-only vetted YSOs are listed in Table~\ref{tbl:wisemips} with their WISE and MIPS fluxes (and IRAC fluxes where available).
In Tables~\ref{tbl:irac} and~\ref{tbl:wisemips}, we have noted which YSOs are in regions of low column density ($N_{\rm{H2}} < 5 \times 10^{21}$ cm$^{-2}$) according to the column density maps by \cite{Harveyetal2013}, as these are more likely to be contaminants than YSOs in regions of high column density.

%            IRAC+MIPS  IRAC-only   MIPS-only
%AMC         14 of 147  0 of 3      52 of 79
%AMC-North   10 of 12   14 of 14    4 of 5
%Vetted      135        3           28

We compare our final YSO source list to those found for \lkha 101 in \cite{Gutermuthetal2009}. All 103 YSOs in \cite{Gutermuthetal2009} are identified as sources in our catalogue with positions that are within a couple tenths of an arcsecond agreement. Where this work and \cite{Gutermuthetal2009} provide fluxes, they agree at the  shorter IRAC bands (IRAC1-3) typically within 0.05 -- 0.1 mag.  At IRAC4 and MIPS1, the agreement is typically within 0.2 mag. These differences are what one might expect for PSF-fitting (used here) versus aperture fluxes (used by \citealt{Gutermuthetal2009}) at wavelengths where there is substantial diffuse emission. (Recall that we have incorporated their dataset into our own.) Therefore no previously identified sources have been missed in this study, and our measurements agree well with those of \cite{Gutermuthetal2009}. 
Note, however, that the different classification methods used in this work and by \cite{Gutermuthetal2009} each yield a different total number of YSOs in this region;
we have identified 42 YSOs whereas \cite{Gutermuthetal2009} identified 103. Our total breaks down into 7 YSOs identified here that were not identified by \cite{Gutermuthetal2009} and 35 YSOs shared between the two lists. (The c2d pipeline identified 47 YSOcs that were listed as YSOs by \cite{Gutermuthetal2009}, but 12 were removed during the vetting process.)
The major source of this discrepancy is that we require 
4 (or 5) band photometry with S/N $\geq 3$ in IRAC (and MIPS 24 \micron\ bands) to identify YSO candidates. 
Such criteria are especially difficult to satisfy in the region of bright and diffuse emission around \lkha 101. 
Therefore, our results do not contradict those in \cite{Gutermuthetal2009}, rather we believe that the stringent criteria used here have excluded some YSOs. We keep these criteria for consistency with other c2d and \Spitzer\ GBS observations and analyses, but note the limitations in such a bright region.

\begin{figure*}%[h]
\includegraphics[width=6.5 in, clip=True,trim=0.5cm 10.3cm 0.5cm 10.5cm]{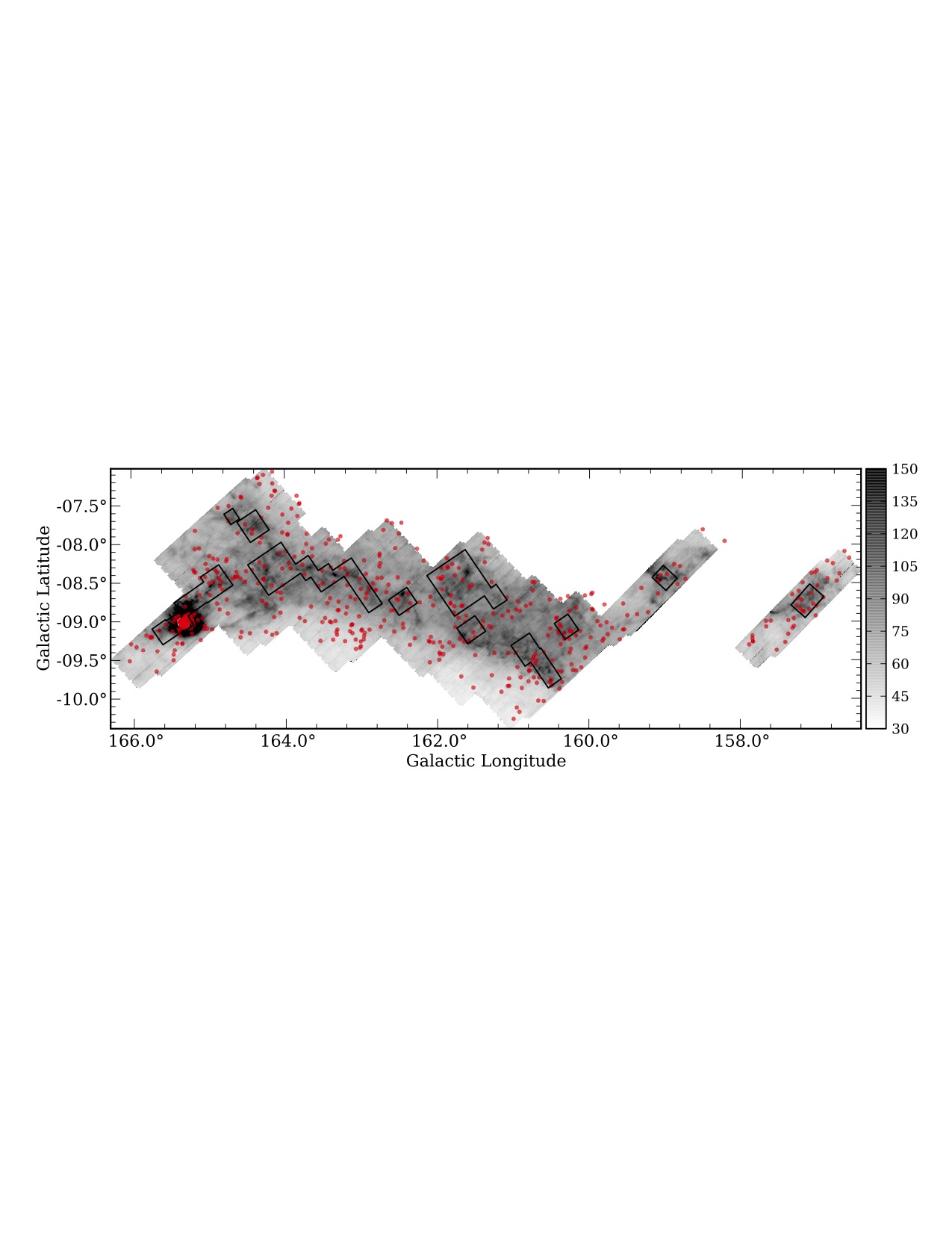}
\caption{Sources with SEDs consistent with a reddened stellar photosphere and a dust component (IR excess) but for which detections with S/N $\geq 3$ across all 4 IRAC bands, 
required to considered a YSOc, did not exist (see text). 
The positions of these sources are plotted against the 160 \micron\ greyscale (colorbar units are MJy\,sr$^{-1}$). The striking over-density at the center of \lkha 101 compared to other IRAC+MIPS regions (marked by black lines) suggests that we are missing veritable YSOs in this region. The robust set of measurements required to identify whether a source is a likely YSO or background galaxy is difficult to attain in this region of very bright emission.}
\label{fig:stardust}
\end{figure*}

The diffuse emission problem is isolated to the immediate vicinity of \lkha 101. To demonstrate this point, in Figure~\ref{fig:stardust} we have plotted the location of all the sources having an SED consistent with being a reddened stellar photosphere and an associated dust component, which do not have S/N $\geq 3$ at all IRAC bands.
The SEDs of these sources are classified as `star+dust' in our catalogue. 
Of the 56 YSOs listed by \cite{Gutermuthetal2009} that were not identified as YSOs in this work, the majority of them (34 of 56) have a `star+dust' SED.
There is a total of 465 `star+dust' sources without robust 4-band IRAC fluxes in the AMC field. 
These sources are relatively evenly distributed throughout the field, with the exception of a striking over-density at the center of \lkha 101 compared to other IRAC regions. Therefore, we believe this over-density is an effect of the difficulty in getting detections with S/N $\geq 3$ across 4 bands in the bright \lkha 101 region and not that there are significantly fewer YSOs than suggested by \cite{Gutermuthetal2009}.

\cite{Harveyetal2013} identified 60 YSOs in the AMC with \Herschel/PACS, 49 of which are also identified in this work. Four of these \Spitzer-identified YSOs are members of pairs of YSOs that are blended in the \Herschel\ images. \Herschel\ is more sensitive to the rising- and flat-spectrum sources, i.e., of the other 45 \Spitzer-identified YSOs that are also detected in the \Herschel\ maps, most (76\%) are Class I/F objects, and the remaining 24\% are Class IIs.
%Class I+F (26+8) with only (11)

%%%%%%%%%%%%%%%%%%%%%%%%%%%%%%%%%%%%%%%%%%%%%%%%
\subsection{YSO classification}\label{sec:classification}

\begin{figure*}%[h]
\includegraphics[width=4.5 in, clip=True,trim=1cm 0.5cm 0cm 0cm]{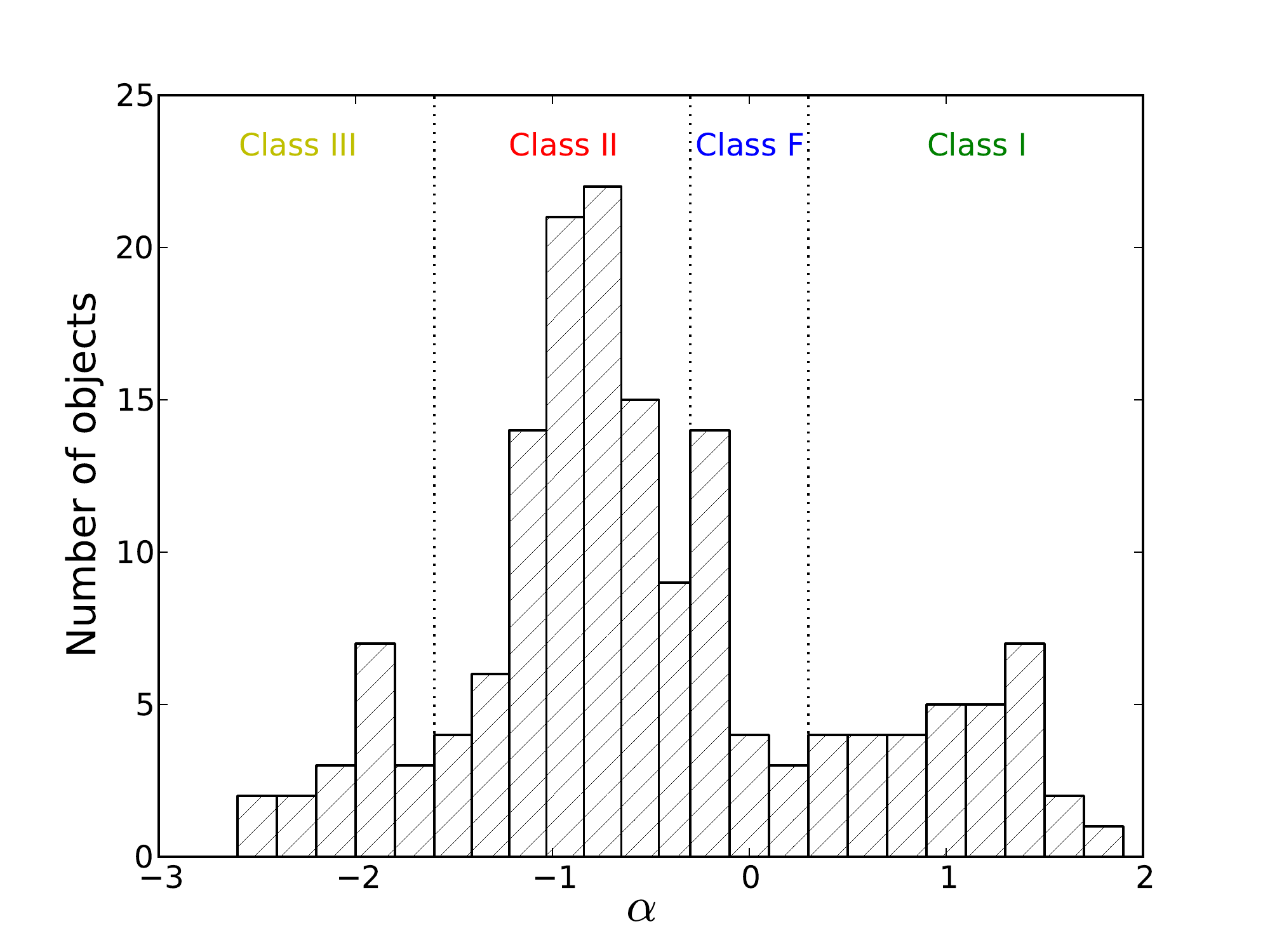}
\includegraphics[width=2 in, clip=True,trim=3cm 3.5cm 3.5cm 5cm]{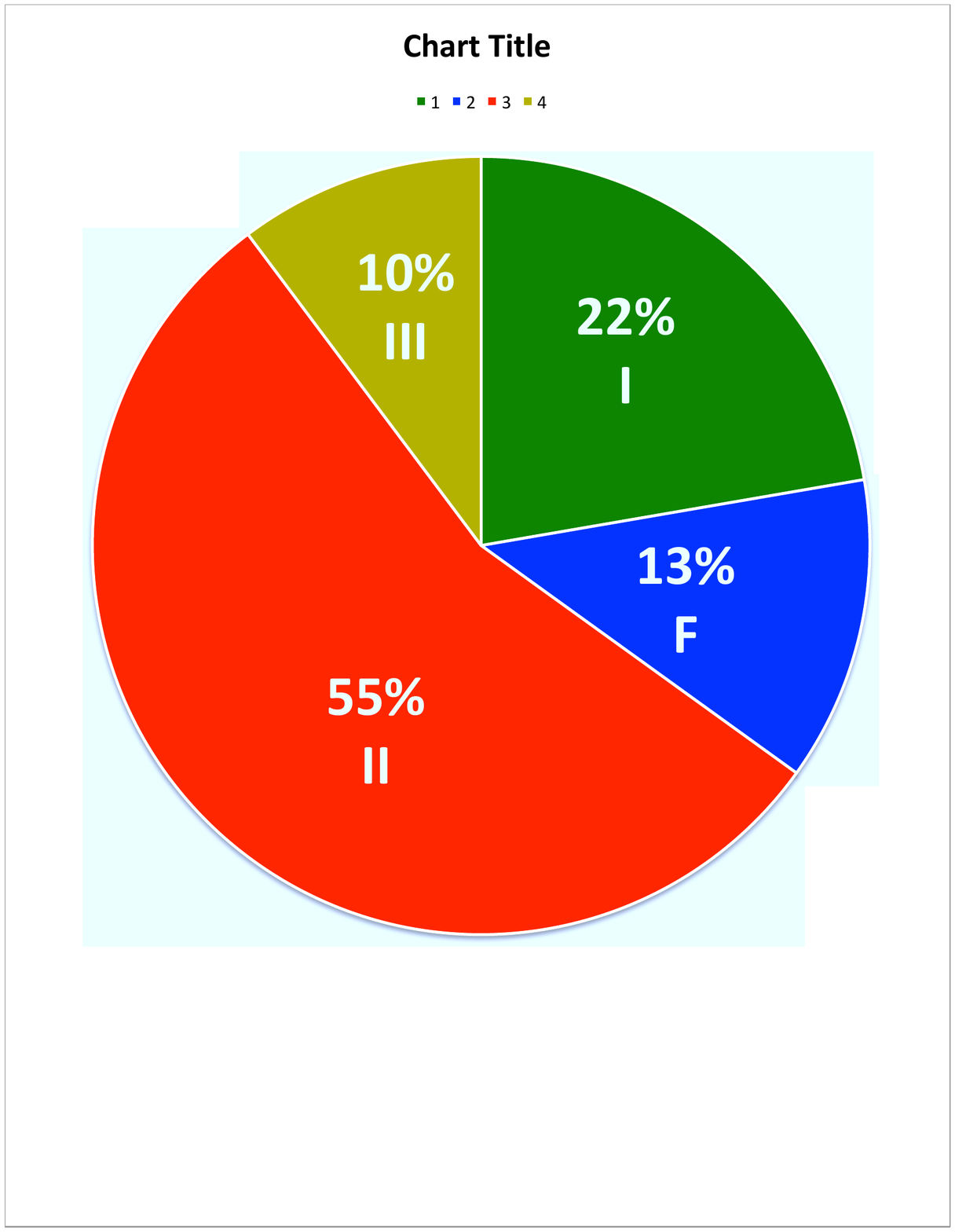}
\caption{Left: Distribution of $\alpha$ values (the slope of the SED in the IR) used to determine the `class' of the YSOs in the AMC. 
The vertical dotted lines mark the boundaries between the different classes as defined by \cite{Greene1994}. Right: Pie chart for the AMC showing the percentage of sources in each SED class. Green is Class I; blue is Flat; red is Class II; and yellow is Class III (colors are the same as in Figure~\ref{fig:ysodistribution}).}
\label{fig:alpha_hist}
\end{figure*}

The YSOs are classified according to the slope of their SED in the infrared (see \citealt{Evans2009} for a description). The spectral index, $\alpha$, is given by
\begin{equation}
\label{eq:alpha}
\alpha \equiv {{d \log(\lambda S(\lambda))} \over {d \log(\lambda)}}
\end{equation}
and determined by fitting the photometry between 2 \micron\ and 24 \micron. The distribution of $\alpha$ values is shown in Figure~\ref{fig:alpha_hist} along with the relative number of YSOs in each SED class. The majority of YSOs identified in the cloud are Class II objects (55\%). The percentage of sources in each SED class for the AMC is strikingly similar to that of Perseus (23\%, 11\%, 58\%, and 8\% for Class Is, Fs, IIs and IIIs, respectively; \citealt{Evans2009}).

%%%%%%%%%%%%%%%%%%%%%%%%%%%%%%%%%%%%%%%%%%%%%%%%
%\subsection{Comparison to other star-forming clouds}\label{sec:compare}

Table~\ref{tbl:ages} lists the breakdown of Class Is, Fs, and IIs for the AMC and other clouds in the GB and c2d surveys to estimate their relative ages. We did not include Class IIIs in this analysis since this population is typically incomplete in \Spitzer\ surveys (e.g., see discussions in \citealt{Harvey2008,Evans2009,Gutermuthetal2009}) due to their weak IR excess. This simplifies the comparison to other clouds where the completeness limits may vary. We compared the ratio of Class Is and Fs to Class IIs, \NIF/\NII, for the different cloud populations in other GB and c2d surveys which use the same classification scheme. We also include YSOs in the OMC identified with \Spitzer\ by \cite{Megeathetal2012}; since they use a different classification scheme however, we have re-calculated the $\alpha$ values for their sample. The Class I/F lifetime is relatively short compared to the Class II lifetime, and therefore a higher ratio indicates a younger population (see discussion in \citealt{Evans2009}).  The high number of Class Is and Fs suggests that the AMC is relatively young compared to other clouds.

Finally, we also compared the number of YSOs per square degree in the AMC (11.5~deg$^2$)\footnote{Here we use the total coverage of IRAC + MIPS1, the five bands used to identify YSOs. This differs from the overlapping MIPS1, MIPS2 and MIPS3 coverage of 10.47~deg$^2$ described in Section~\ref{sec:intro}.} to that in the OMC (14~deg$^2$). The OMC is forming vastly larger amounts of stars. It has 237 YSOs per deg$^2$ whereas the AMC only has 13 YSOs per deg$^2$, a factor of about 20 fewer. 
Even if we only compare the number of YSOs in the OMC with 4 band photometry (as this was the source of the discrepancy between the total number of YSOs around \lkha 101 identified in this work and by \citealt{Gutermuthetal2009}, who use a similar identification method to \citealt{Megeathetal2012}), this  still suggests that there is at least a factor 15 more YSOs in the OMC than in the AMC.
Despite the differences in identification methods used for the OMC and for the AMC, it is clear that the OMC is forming far more stars than the AMC is. The YSOs in the OMC are also concentrated much more strongly than the AMC, despite both clouds having comparable sizes and masses.
We note that \cite{Ladaetal2009} attribute the difference between the amount of star formation to the different amounts of material at high \Av/column density.

%%%%%%%%%%%%%%%%%%%%%%%%%%%%%%%%%%%%%%%%%%%%%%%%
%%%%%%%%%%%%%%%%%%%%%%%%%%%%%%%%%%%%%%%%%%%%%%%%

\section{Spectral Energy Distribution Modeling}\label{sec:sed}

Optical data of the YSOs were downloaded from the USNO NOMAD catalogue (Zacharias et al. 2004). SEDs of the YSOs are shown in Figures~\ref{fig:classIF1} and \ref{fig:classIF2} (Class Is and Class Fs), \ref{fig:classII1} -- \ref{fig:classII3} (Class IIs) and \ref{fig:classIII} (Class IIIs). We were able to perform relatively detailed modelling of the stellar and dust components of the Class II and Class III sources (YSOs which are not heavily obscured by dust). The luminosities of sources in the earlier classes are presented in \cite{Dunham2013}. The majority of the Class II and Class III sources are likely in the physical stage where the stellar source and circumstellar disk are no longer enshrouded by a circumstellar envelope. We note that the observed ``class" does not always correspond to the associated physical stage of the YSO (see discussion in \citealt{Evans2009}) and that some Class IIs may be sources, viewed pole-on, with circumstellar envelopes that are only beginning to dissipate. Conversely, an edge-on disk without an envelope could look like a Class I object.

%Cloud                    [log Ldisk/Lstar] Ldisk/Lstar    [log Lstar/Lsun]
% AMC                      -0.8 and -1.0    0.1 and 0.16    
% IC. 5146                 -0.4 and -0.6    0.25 and 0.40     
% Serpens (Harvey2007)     -0.7 and -0.9    0.2 and 0.16 
% Lupus (Merin2008)        -0.3 and -0.5                     -0.5 and -1.0
% Cepheus Flare (Kirk2009) -0.2 and -0.5                     -0.5 and -1.0
% Chameleon II (Alcala2008)-0.3 and -0.6

Our SED modelling methods follow those used by \cite{Harvey2007} (and similar works since, e.g., \citealt{Merin2008}, \citealt{Kirk2009}) to model the SEDs. The stellar spectrum of a K7 star was fit to the SEDs by normalizing it to the de-reddened fluxes in the shortest available IR band of J, K or IRAC1. We use the extinction law of \cite{WeingartnerDraine2001} with \Rv$ = 5.5$ to calculate the de-reddened fluxes. The \Av~value was estimated by matching the de-reddened fluxes with the stellar spectrum. In eight cases, we used an A0 spectrum when the K7 spectrum was unable to produce a reasonable fit. The use of only two stellar spectra is of course over-simplified; how-

%\placefigure{fig:classIF1} 

%\begin{minipage}{\linewidth}
\begin{figure*}%[H]
\includegraphics[width=6 in,clip=True, trim=1cm 0cm 1cm 1cm]{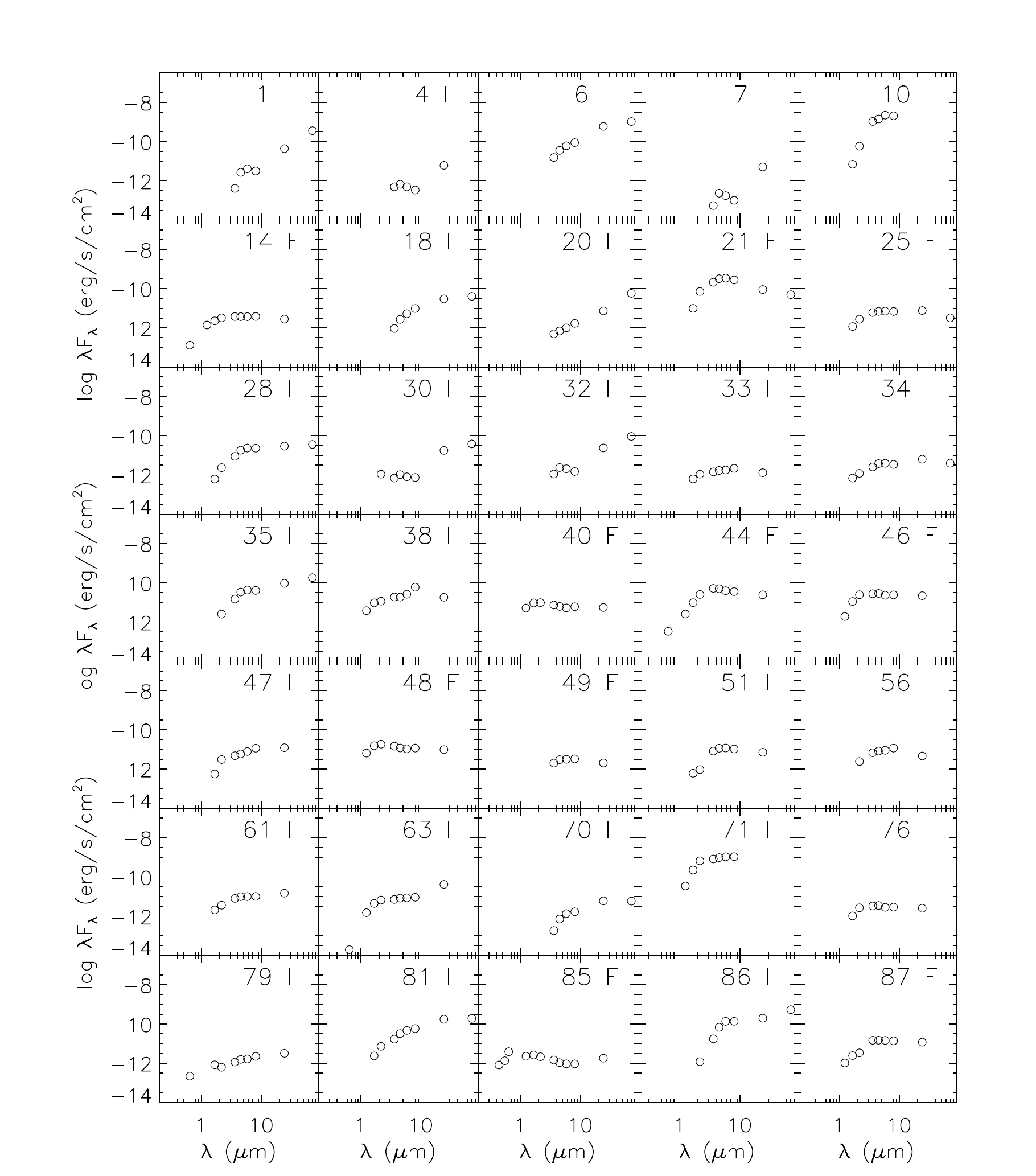}\centering
\caption{SEDs of Class I and Flat sources. The YSO ID, from Tables~\ref{tbl:irac} and~\ref{tbl:wisemips}, is shown in the upper right of each panel along with the Class (I or F) of the YSO.}
\label{fig:classIF1}
\end{figure*}
%\end{minipage}

\begin{figure*}%[H]
%\ContinuedFloat
\includegraphics[width=6 in,clip=True, trim=1cm 5.5cm 1cm 1cm]{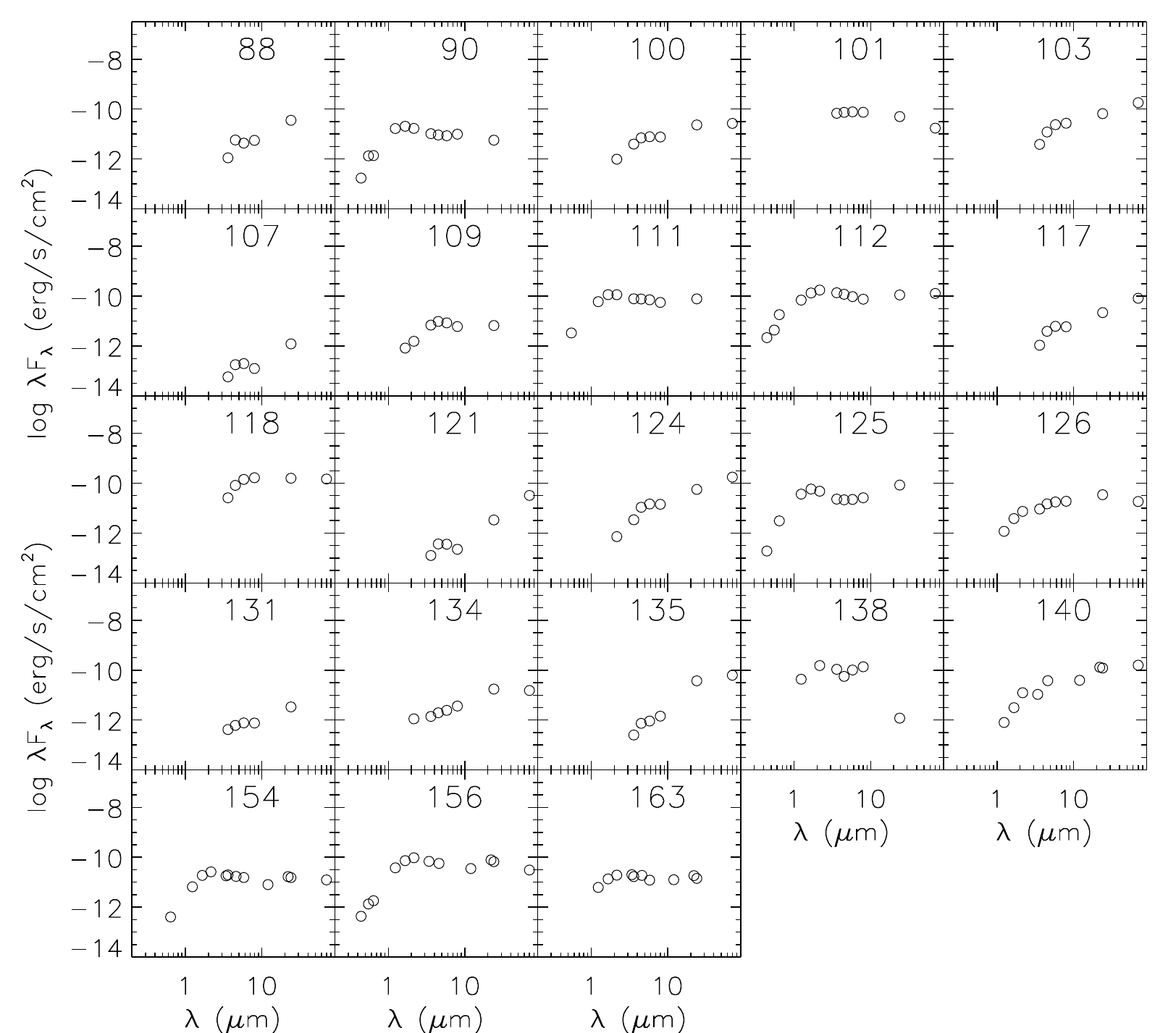}\centering
\caption{\it continued from Figure~\ref{fig:classIF1}.}
%SEDs of Class I and Flat sources (continued). The YSO ID, from Tables~\ref{tbl:irac} and~\ref{tbl:wisemips}, is shown in the upper right of each panel along with the Class (I or F) of the YSO.}
\label{fig:classIF2}
\end{figure*}

\begin{figure*}%[H]
\includegraphics[width=6 in,clip=True, trim=1cm 0cm 1cm 1cm]{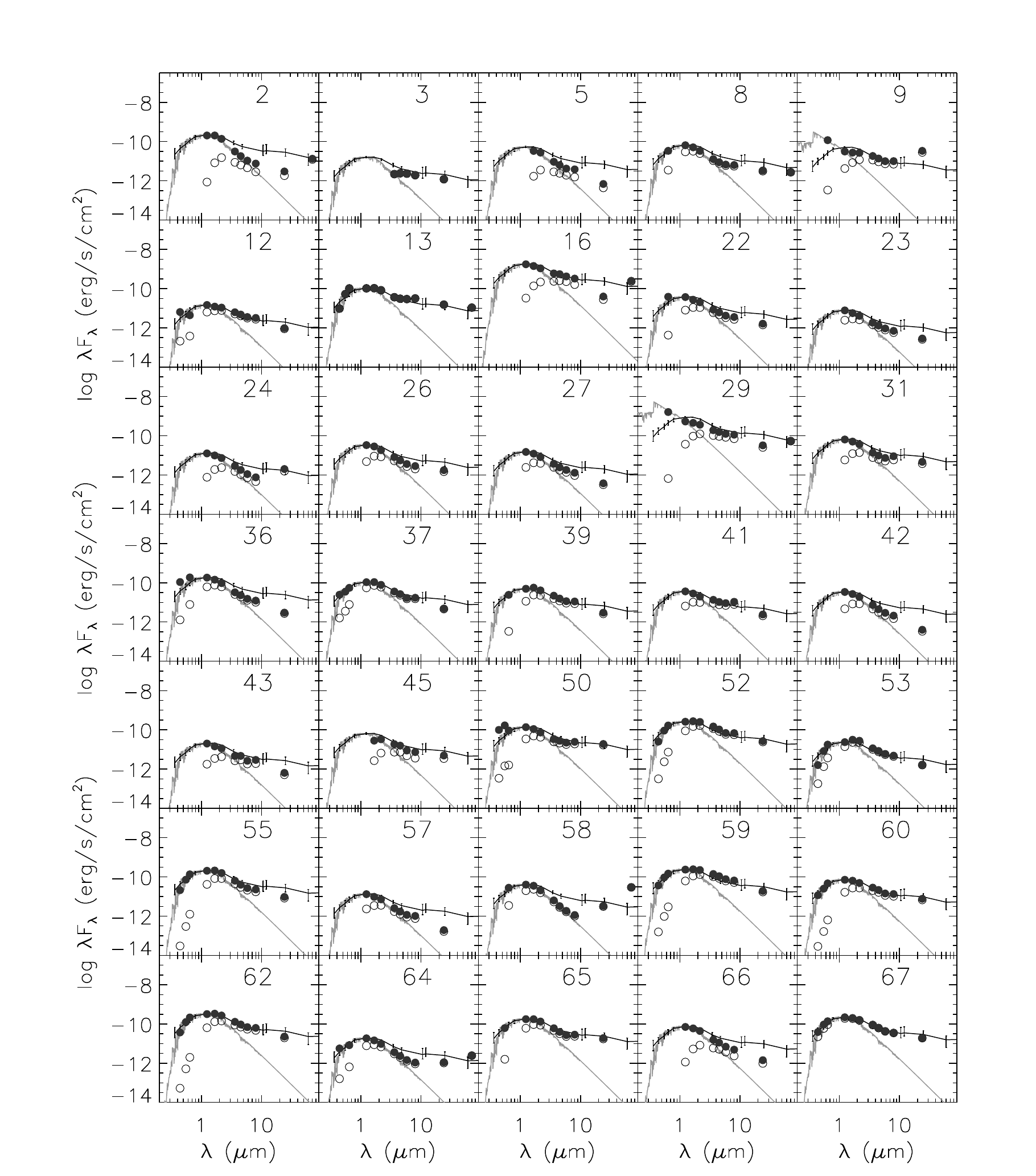}\centering
\caption{SEDs of Class II sources. The YSO ID, from Tables~\ref{tbl:irac} and~\ref{tbl:wisemips}, is shown in the upper right of each panel. The observed fluxes are plotted with unfilled circles. The de-reddened fluxes are plotted with filled circles. The grey line plots the model stellar spectrum fit to the shorter wavelengths. The black line shows the median SED of T Tauri stars in Taurus (with error bars denoting quartiles of the distribution, \citealt{DAlessioetal1999}) normalized to the B band flux and J band flux of the K7 and A0 stellar spectrum models, respectively.}\label{fig:classII1}
\end{figure*}

\begin{figure*}%[H]
%\ContinuedFloat
\includegraphics[width=6 in,clip=True, trim=1cm 0cm 1cm 1cm]{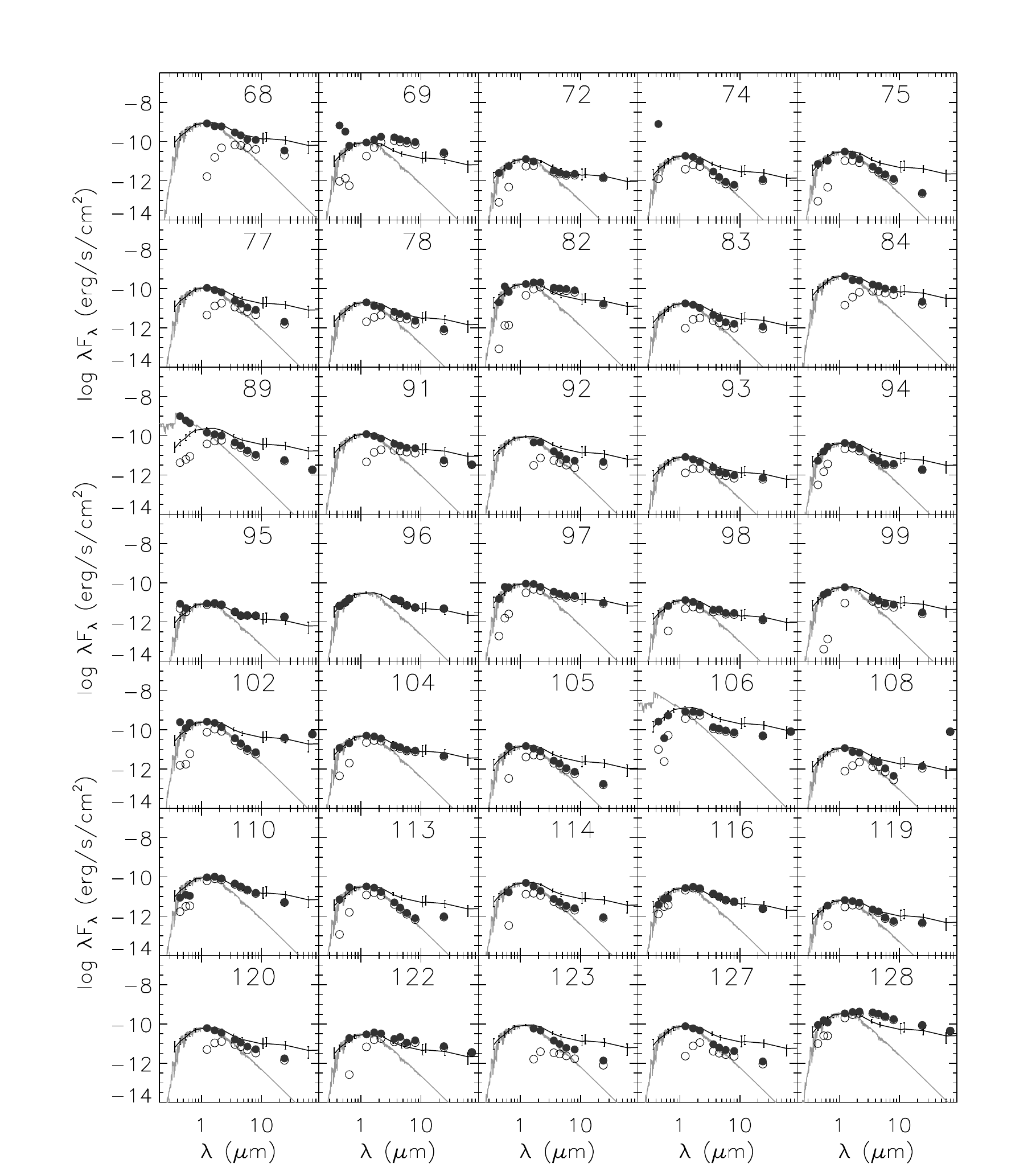}\centering
\caption{\it continued from Figure~\ref{fig:classII1}.}
%{SEDs of Class II sources (continued). The YSO ID, from Tables~\ref{tbl:irac} and~\ref{tbl:wisemips}, is shown in the upper right of each panel. The observed fluxes are plotted with unfilled circles. The de-reddened fluxes are plotted with filled circles. The grey line plots the model stellar spectrum fit to the shorter wavelengths. The black line shows the median SED of T Tauri stars in Taurus (with error bars denoting quartiles of the distribution, \citealt{DAlessioetal1999}) normalized to the B band flux and J band flux of the K7 and A0 stellar spectrum models, respectively.}
\label{fig:classII2}
\end{figure*}

\begin{figure*}%[H]
%\ContinuedFloat
\includegraphics[width=6 in,clip=True, trim=1cm 5.5cm 1cm 1cm]{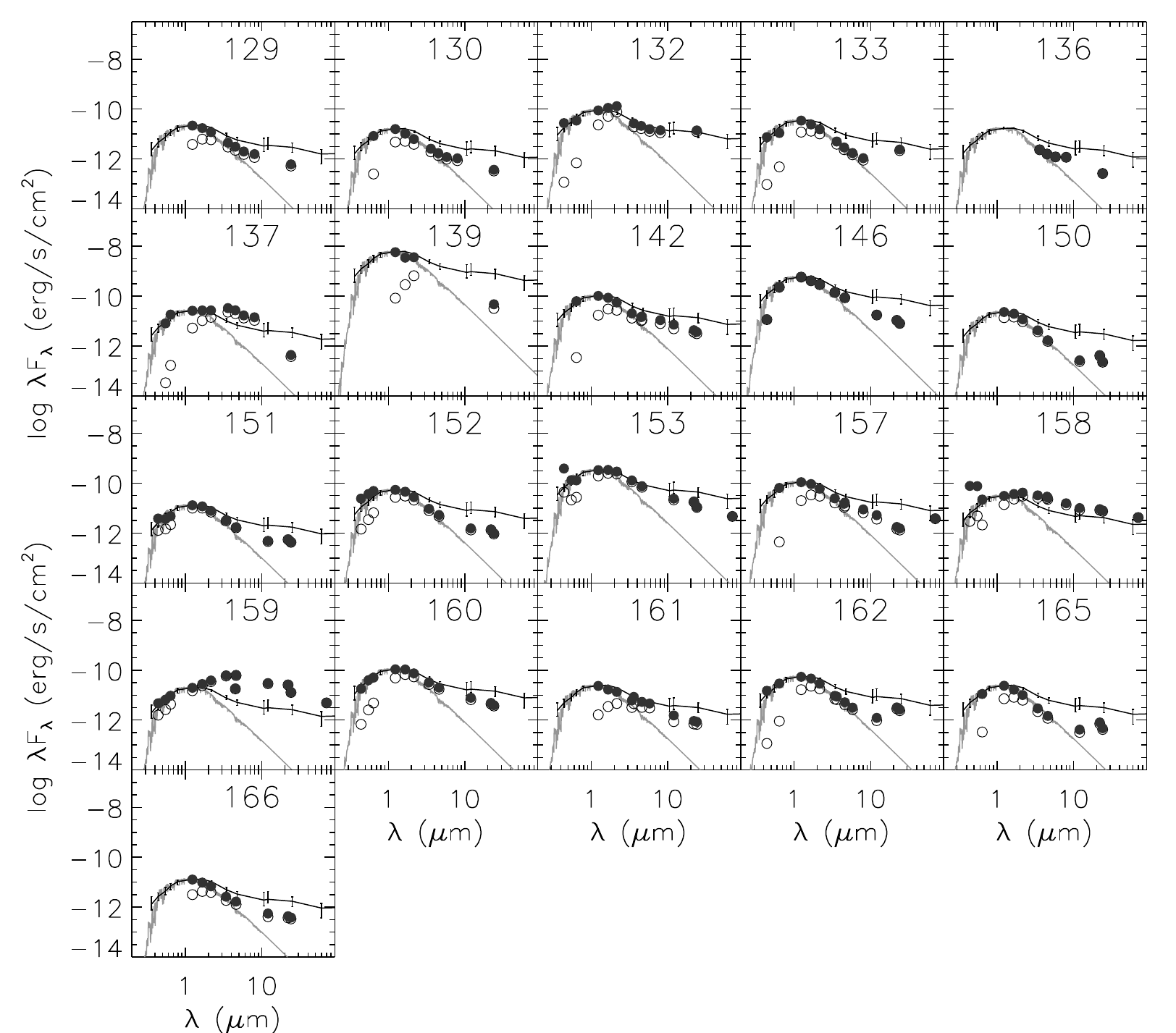}\centering
\caption{\it continued from Figure~\ref{fig:classII1}.}
%{SEDs of Class II sources (continued). The YSO ID, from Tables~\ref{tbl:irac} and~\ref{tbl:wisemips}, is shown in the upper right of each panel. The observed fluxes are plotted with unfilled circles. The de-reddened fluxes are plotted with filled circles. The grey line plots the model stellar spectrum fit to the shorter wavelengths. The black line shows the median SED of T Tauri stars in Taurus (with error bars denoting quartiles of the distribution, \citealt{DAlessioetal1999}) normalized to the B band flux and J band flux of the K7 and A0 stellar spectrum models, respectively.}
\label{fig:classII3}
\end{figure*}

\begin{figure*}%[H]
\includegraphics[width=6 in,clip=True, trim=1cm 8cm 1cm 1cm]{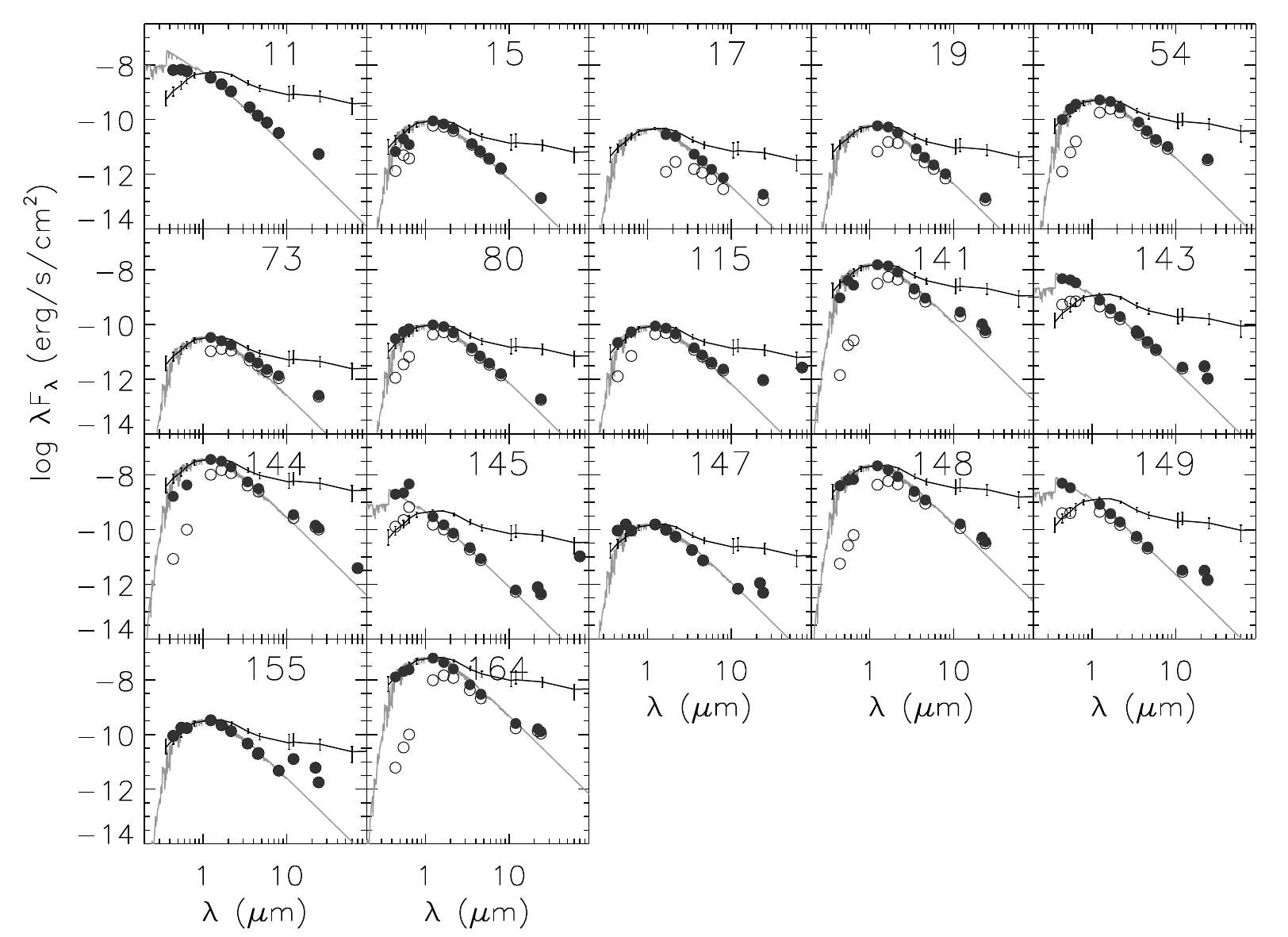}\centering
\caption{SEDs of Class III sources. The YSO ID, from Tables~\ref{tbl:irac} and~\ref{tbl:wisemips}, is shown in the upper right of each panel. The observed fluxes are plotted with unfilled circles. The de-reddened fluxes are plotted with filled circles. The grey line plots the model stellar spectrum fit to the shorter wavelengths. The black line shows the median SED of T Tauri stars in Taurus (with error bars denoting quartiles of the distribution, \citealt{DAlessioetal1999}) normalized to the B band flux and J band flux of the K7 and A0 stellar spectrum models, respectively.}\label{fig:classIII}
\end{figure*}

\noindent ever, it produces adequate results for the purposes of this study. More exact spectral typing is difficult with only the photometric data presented here and the uncertainties in \Av. We nevertheless obtain a broad overview of the disk population with the applied assumptions. Tables~\ref{tbl:diskpropII} and~\ref{tbl:diskpropIII} list the stellar spectrum, the \Av~value, and stellar luminosity (\Lstar) used for the stellar models of each source's SED for the Class II and Class III YSOs, respectively.

%%%%%%%%%%%%%%%%%%%%%%%%%%%%%%%%%%%%%%%%%%%%%%%%
\subsection{Second order SED parameters \alphaexcess~and \lambdaturnoff}
\label{sec:transitiondisks}
The first order SED parameter $\alpha$ is used as a primary diagnostic of the excess and circumstellar environment and to separate the YSOs into different ``classes'' (\S~\ref{sec:classification}). Once we have a model of the stellar source, however, we are able to characterize the circumstellar dust better. For each source we determined the values of \alphaexcess\ and \lambdaturnoff\ defined by \cite{Cieza2007} and \cite{Harvey2007} and used in many works since. \lambdaturnoff\ is the longest measured wavelength before an excess greater than 80\% of the stellar model is observed. If no excess $>$ 80\% is observed, than \lambdaturnoff\ is set to 24 \micron. \alphaexcess\ is the slope of the SED at wavelengths longward of \lambdaturnoff. \alphaexcess\ is not calculated for YSOs with \lambdaturnoff~$= 24$ \micron\, as there are not enough data points to determine the slope of the excess. These parameters provide a better characterization of the excess since $\alpha$ can include varying contributions from the stellar and dust components. 

\begin{figure}[h]
\includegraphics[width=3.5 in,clip=True, trim=1cm 0cm 0cm 0cm]{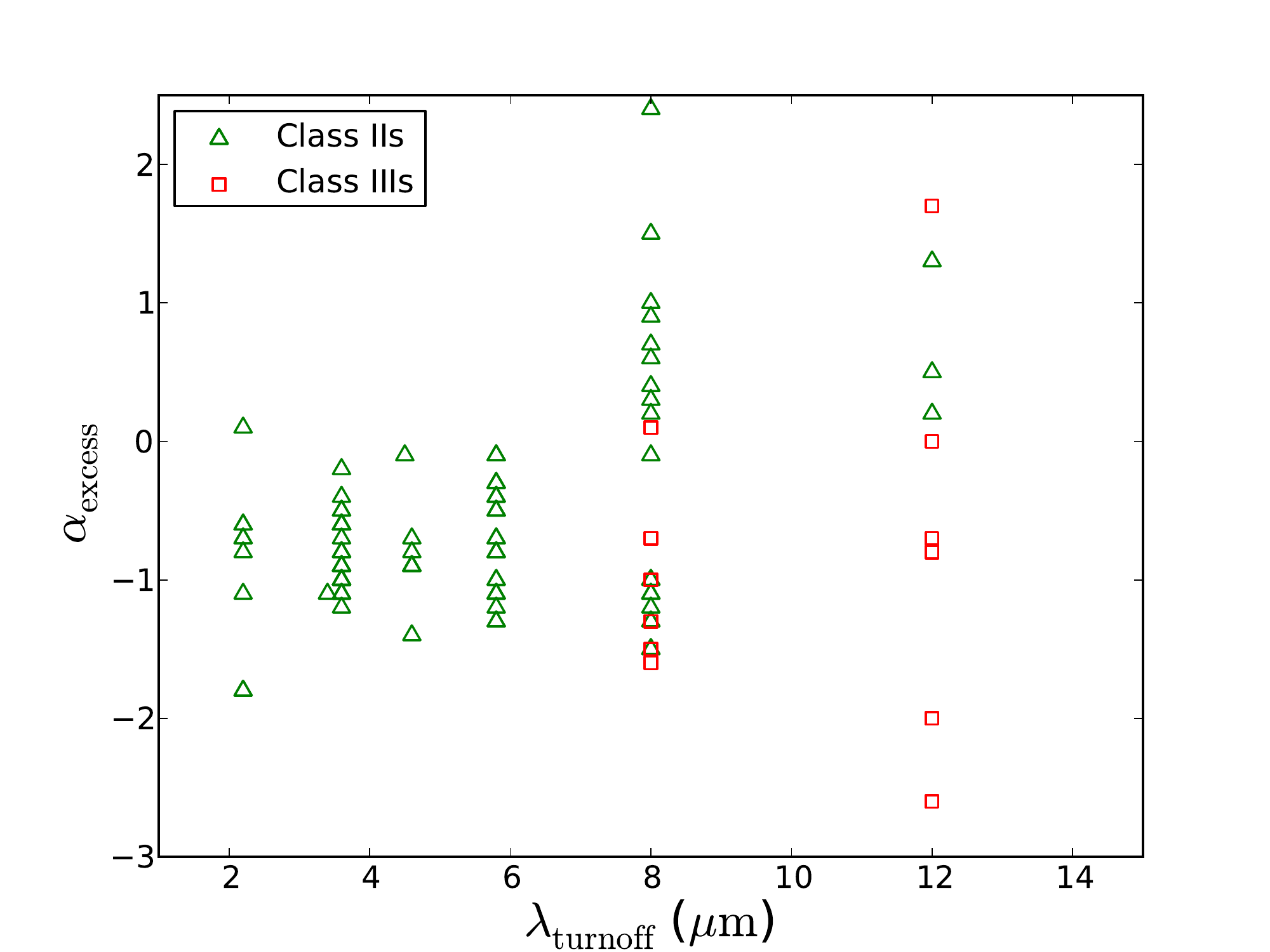}\centering
\caption{Distribution of \alphaexcess\ and \lambdaturnoff\ for Class II and Class III sources. The Class IIIs with \lambdaturnoff$ = 24$~\micron\ (IDs 15, 19, 80, and 148) are not shown as those sources typically do not have excess measured across a wide enough range to calculate reliable values of \alphaexcess.}
\label{fig:l_exc_vs_a_exc}
\end{figure}

Figure~\ref{fig:l_exc_vs_a_exc} shows the distribution of \alphaexcess~and \lambdaturnoff~for the Class IIs and Class IIIs. Class II and Class III YSOs with long \lambdaturnoff\ and positive \alphaexcess\, (YSOs 2, 24, 58, 64, 74, 102, 108, 113, 115, and 133 in the 8 \micron\ bin and YSOs 145, 150, 162, and 165 in the 12 \micron\ bin of Figure~\ref{fig:l_exc_vs_a_exc}) are good classical transition disk candidates;
the lack of near-IR excess but large mid-IR excess is a sign of a deficit of material close to the star within a substantial disk. \cite{Cieza2012} have recently done a study on the transition disks in the AMC, Perseus and Taurus and identify six transition disk candidates in the AMC, three of which are also in our list of candidates (YSOs 58, 102 and 115). Of their remaining candidates, two were debris-like disks (YSOs 11 and 54) and the other was not identified in our YSO list.
The larger distribution of \alphaexcess\ for sources with longer \lambdaturnoff\ is consistent with distributions found for other disk populations (e.g., \citealt{Cieza2007,Alcala2008,Harvey2008,Merin2008}).

%%%%%%%%%%%%%%%%%%%%%%%%%%%%%%%%%%%%%%%%%%%%%%%%
\subsection{Disk luminosities}\label{sec:diskprop}

\begin{figure}[h]
\includegraphics[width=3.5 in,clip=True, trim=1cm 0cm 0cm 0cm]{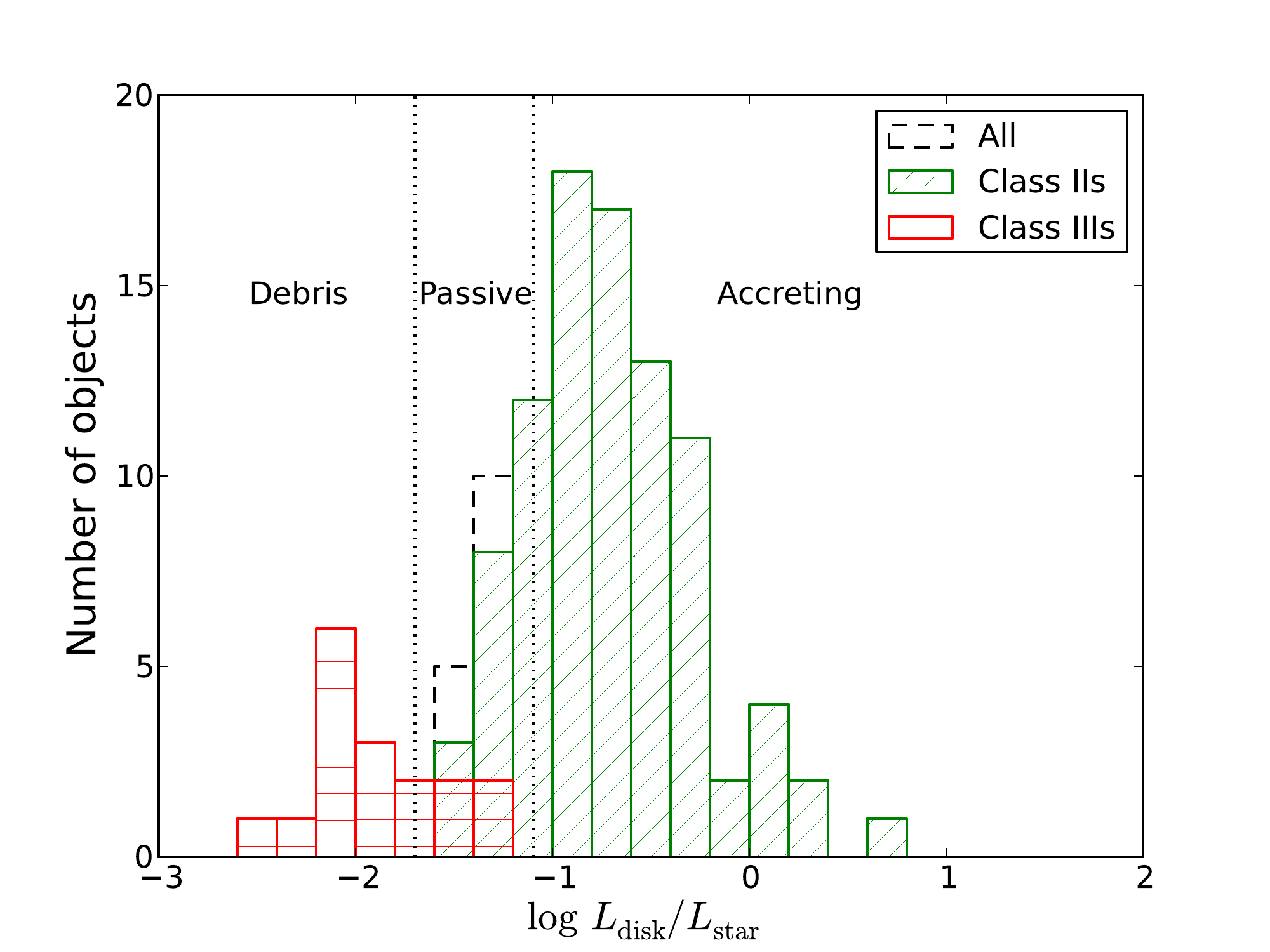} % UPDATE
\caption{The ratio of the disk luminosity to the stellar luminosity for Class II and Class III sources. Also shown are the typical boundaries found for accreting disks, passive disks and debris disks \citep{KenyonHartmann1987}.}
%(A color version of this figure is available in the online journal.)
\label{fig:Ldisk}
\end{figure}

Figure~\ref{fig:Ldisk} shows the ratio of the disk luminosities to stellar luminosities for the Class II and Class III sources. The disk luminosity is the integral of the observed excesses. (The excess at a given wavelength is calculated by subtracting the flux of the stellar model at that wavelength from the observed flux). The distribution of \Ldisk$/$\Lstar\ for Class II and III sources in the AMC is similar to that found for other \textit{c2d} and GB surveys with \Spitzer\ (Serpens: \citealt{Harvey2007}, IC 5146: \citealt{Harvey2008}, Chameleon II: \citealt{Alcala2008}, Lupus: \citealt{Merin2008}, and the Cepheus Flare: \citealt{Kirk2009}). We find the Class III sources in the regions typically occupied by sources with passive disks and debris disks (e.g., 0.02 $<$ \Ldisk$/$\Lstar $<$ 0.08 for passive disks; \citealt{KenyonHartmann1987}). The low disk luminosity may be attributable to the lack of mid-IR excess at IRAC wavelengths in these sources' SEDs.

%Cloud                    [log Ldisk/Lstar] Ldisk/Lstar    [log Lstar/Lsun]
% AMC                      -0.8 and -1.0    0.1 and 0.16    
% IC. 5146                 -0.4 and -0.6    0.25 and 0.40     
% Serpens (Harvey2007)     -0.7 and -0.9    0.2 and 0.16 
% Lupus (Merin2008)        -0.3 and -0.5                     -0.5 and -1.0
% Cepheus Flare (Kirk2009) -0.2 and -0.5                     -0.5 and -1.0
% Chameleon II (Alcala2008)-0.3 and -0.6

%%%%%%%%%%%%%%%%%%%%%%%%%%%%%%%%%%%%%%%%%%%%%%%%
\subsection{Questionable Class III sources}
\label{sec:classIIIs}

It is possible that some of the Class III sources identified here are field giants. 
\cite{Oliveira2009} followed up on 150 \Spitzer\ identified YSOs in Serpens and obtained 78 optical spectra with sufficient signal-to-noise. They showed that there were at least 20 giant contaminants in this list, 18 of which were identified as Class III sources. The more scattered spatial distribution of Class IIIs throughout the AMC is consistent with this idea that they are contaminants. Additionally, five of our Class III objects (YSOs 11, 141, 144, 148, 164) have very high luminosities ($> 100$ \Lsun). Four of these objects (YSOs 141, 144, 148, 164), as well as YSO 149 which is not of particularly high luminosity, are quite removed from the areas of high extinction towards the AMC (see Figure~\ref{fig:ysodistribution} in the following section) and regions of low column density ($N_{\rm{H2}}< 5 \times 10^{21}$ cm$^{-2}$, see \S~\ref{sec:ysoselect}). 
%Far from filament: 7, ~19, 23, 29, 164
%Far from filament: 141, ~144, 148, 149, 164
% high luminosity: 7, 15, 19, 23, 164
% high luminosity: 141, 11, 144, 148, 164
%maybe do a search of wise)

%%%%%%%%%%%%%%%%%%%%%%%%%%%%%%%%%%%%%%%%%%%%%%%%
%%%%%%%%%%%%%%%%%%%%%%%%%%%%%%%%%%%%%%%%%%%%%%%%

\section{Spatial Distribution of Star Formation}
\label{sec:spatial}

\begin{figure*}%[h]
\includegraphics[width=6.5 in, clip=True, trim=1cm 9.9cm 0.75cm 4.5cm]{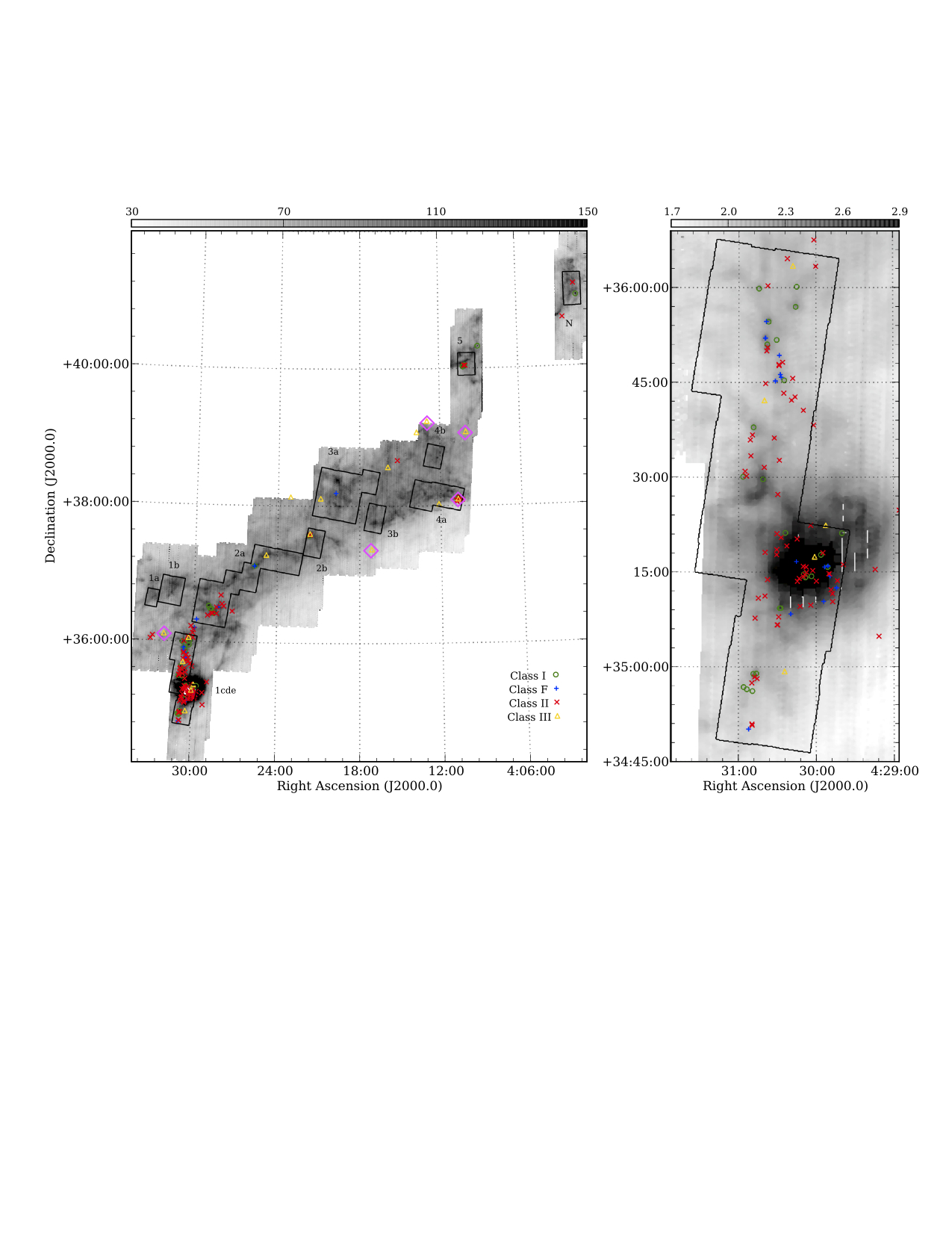}\centering
\caption{Left: The positions of YSOs and IRAC fields in Auriga. The greyscale is the MIPS 160\,$\mu$m map (colorbar units are MJy\,sr$^{-1}$) and the YSOs are marked according to their classification: green circles denote Class~Is; blue $+$s denote Class~Fs; red $\times$s denote Class~IIs; yellow triangles denote Class~IIIs. The magenta diamonds mark the Class III sources of high luminosities that are likely contaminants (see \S~\ref{sec:classIIIs}). IRAC fields are outlined in black and labelled. (Note that some YSOs fall beyond the 160 \micron\ coverage because it is slightly offset from the 24 \micron\ coverage that is used for YSO identification.)
Right: Close-up of the region around LkH$\alpha$\,101. The greyscale is the log (base 10) of the flux (colorbar units are $\log$(MJy\,sr$^{-1}$)). The centre of the field is entirely saturated. As is evident, there are some YSOs outside the IRAC coverage area. This list of MIPS-only YSOs has been trimmed by using WISE data to remove more objects that are likely background galaxies. }
\label{fig:ysodistribution}
\end{figure*}

The spatial distribution of IRAC/MIPS-identified YSOs by class is shown in Figure \ref{fig:ysodistribution}. A close-up of the region surrounding the \lkha 101 cluster and the cluster extension along the filament is also included so the relatively densely clustered YSOs can be better distinguished. Figure \ref{fig:ysodistribution} shows that the bulk of star formation in the AMC has been concentrated 
in  this southern region of the cloud;
the majority of the identified YSOs (79\%) are in this area. (Note that the number of YSOs in that region is a lower limit as it is likely that a significant number of YSOs in the \lkha 101 region are not identified, see discussion at the end of $\S$ \ref{sec:ysoselect}.)

\subsection{Identification of YSO groups}

\begin{figure}[h]
\includegraphics[width=3.0 in,clip=True,trim=0.8cm 0.25cm 0.2cm 0.65cm]{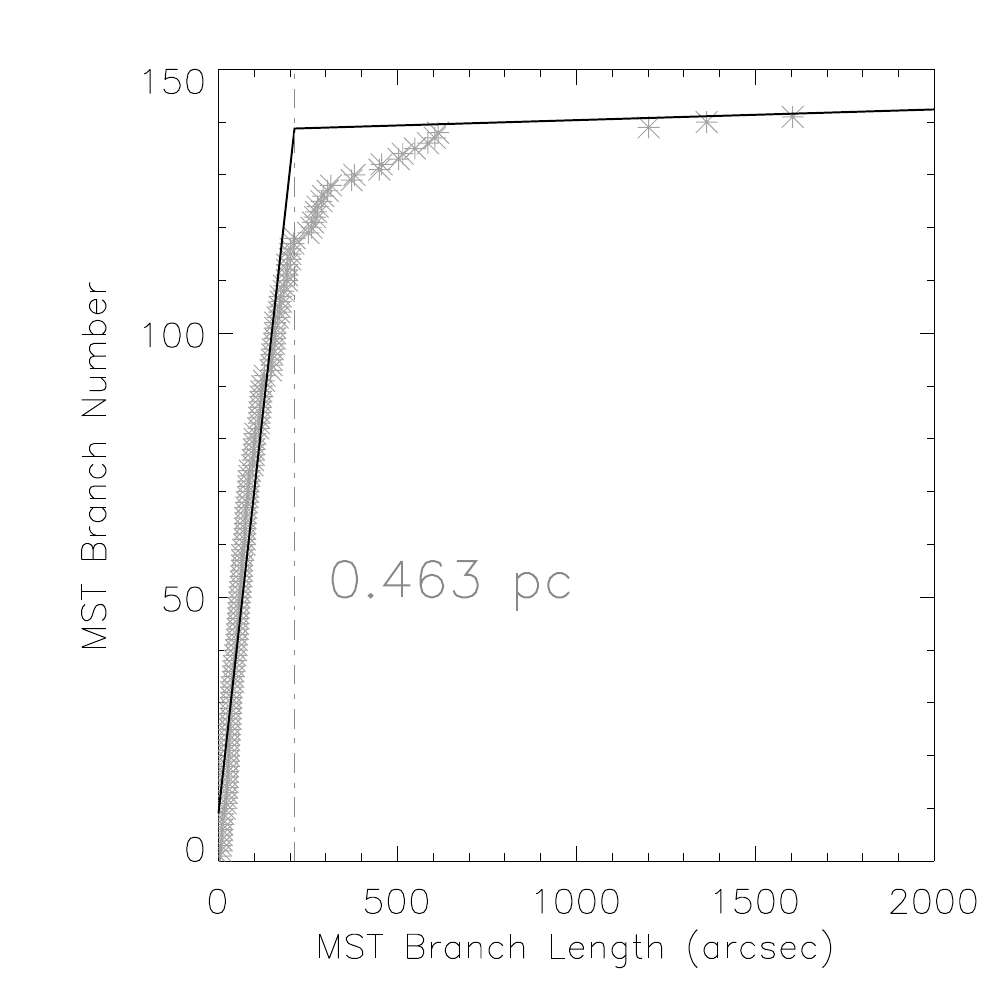} %3.7
\caption{Cumulative distribution function (CDF) of MST branch lengths (asterisks). The solid lines represent linear fits to each end of the CDF. The dot-dash line marks \Lcrit\ where the solid lines meet. The solid lines follow the CDF in the dense regions (steep line) and the sparser regions (shallow line). }
\label{fig:brkpt}
\end{figure}

\begin{figure*}[h]
\includegraphics[width=6.5 in,clip=True,trim=0.3cm 9.4cm 0.3cm 2.8cm]{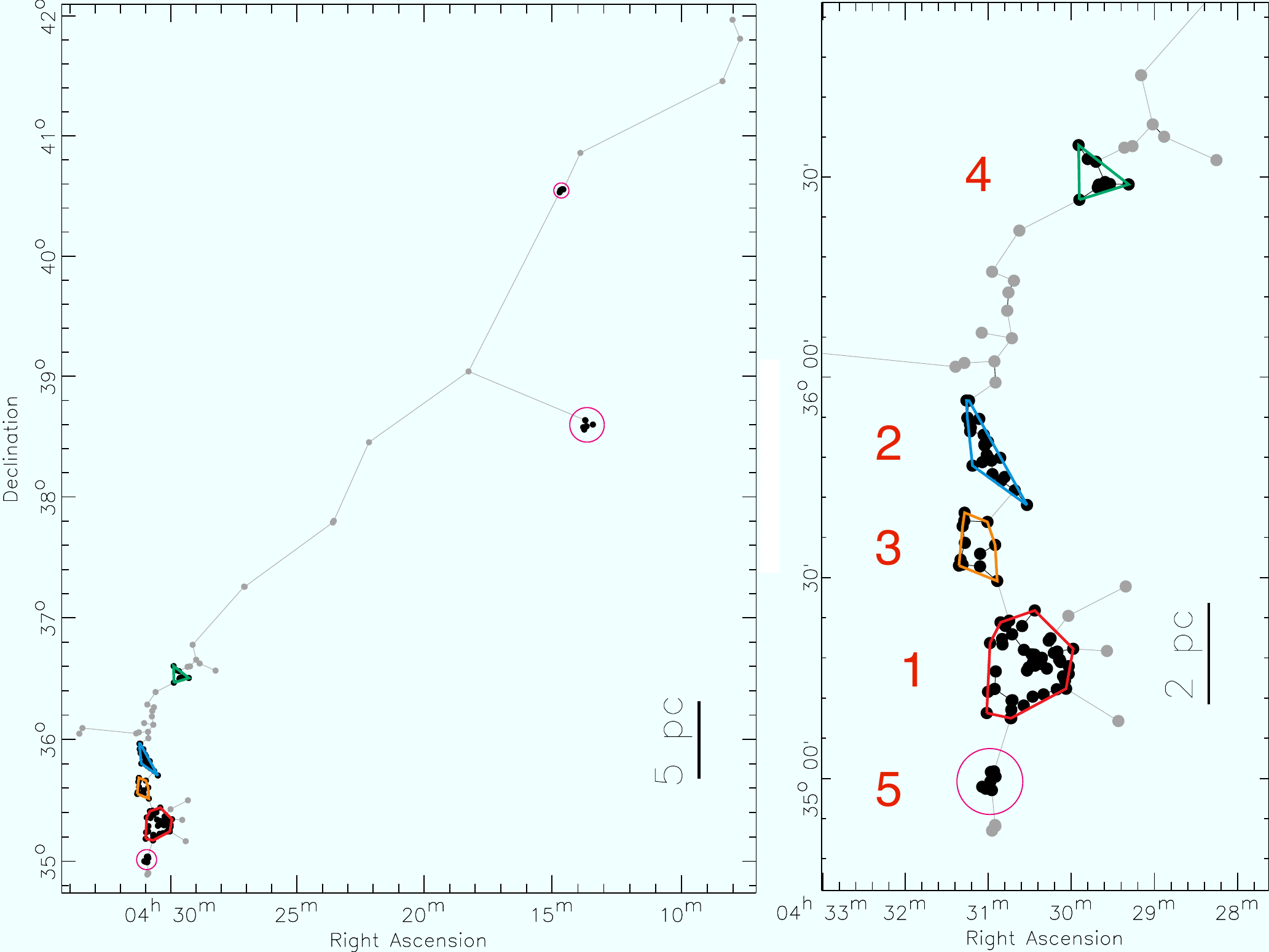}\centering %3
\caption{We extract four groups with 10 or more members (colored convex hulls) and three groups with 5--9 members (magenta circles) using an MST analysis. The right hand panel shows the enlarged southern region of the cloud where most of the groups are located. The red numbers adjacent to the groups correspond to the group number listed in Table~\ref{tbl:grps}.}
\label{grps}
\end{figure*}

We performed a clustering analysis on the identified Class I, F, and II sources in the AMC to identify the densest regions of YSOs and the largest groups. The details of the analysis are described in \cite{Masiunas2012}. We omit the Class III sources from the analysis to avoid the risk of including field giants (see for example \S~\ref{sec:classIIIs}). 
We performed a minimum spanning tree (MST) analysis to identify groups of YSOs within the region. 
This analysis connects YSOs by the minimum distance to the next YSO to form a ``branch'' \citep{CartwrightWhitworth2004}.
Figure~\ref{fig:brkpt} shows the cumulative distribution function (CDF) of the branch lengths between YSOs. This is used to determine the MST critical branch length, \Lcrit, that defines the transition between the branch lengths in the denser regions to the branch lengths in the sparser regions \citep{Gutermuthetal2009}. Therefore \Lcrit\ is based on relative over densities of objects. We measure an \Lcrit\ of 210\arcsec\ for the AMC.
Group memberships are defined by members which are all connected by branches of lengths less than \Lcrit. The boundary of a group is defined where the branch length between adjacent sources exceeds \Lcrit.
Figure~\ref{grps} shows that we have extracted four groups with 10 or more members (marked by colored convex hulls) and three groups with 5--9 members (marked with magenta circles). Table~\ref{tbl:grps} lists the properties of these groups. 
The position of the group is given by its geometric center. The group's effective radius, \Reff, defines the radius of a circle with the same area as the convex hull containing the group members. 
The maximum radial distance to a member from the median position gives \Rcirc, therefore a circle with this radius would contain all group members.  Finally, the elongation of the group is determined by comparing \Rcirc\ to \Reff\ and represented by the aspect ratio, \Rcirc$^2/$\Reff$^2$.
The MST analysis on the full cloud recovers the clustering %NGC 1579} 
surrounding \lkha 101. The cluster subtends a larger area than that measured in \cite{Gutermuthetal2009} confirming their claim that there was star formation extended beyond their field of view. The star formation is mostly extended along the North-South direction of the cluster and therefore we measure a more elongated group than measured by \cite{Gutermuthetal2009}.
This is still the largest group in the AMC in terms of area and the number of members.

As discussed in $\S$~\ref{sec:ysoselect}, our analysis is likely to have underestimated the number of YSOs in the region around \lkha 101. To check the consistency of our analysis with \cite{Gutermuthetal2009}, we ran the MST analysis on both YSO lists within the \cite{Gutermuthetal2009} area of  4-channel IRAC coverage. This leaves us with 41 of the YSOs presented here and 102 of those presented in \cite{Gutermuthetal2009}. (There is one bright YSO in \citealt{Gutermuthetal2009} that lies just outside their 4-channel IRAC coverage to the south. It was only observed at IRAC1 and IRAC3.) 
We get an \Lcrit~of 120\arcsec~for our cropped list of YSOs and an \Lcrit~of 73\arcsec~for the cropped \cite{Gutermuthetal2009} YSO list. (Note that running the analysis on the cropped field, which is dense compared to the rest of the cloud, yields a smaller \Lcrit\ than when the analysis is run on the whole cloud. This is expected as \Lcrit\ is based on over densities, as discussed above.) The ratio of the \Lcrit~values for the two YSO lists ($73/120 = 0.61$) agrees with our expectation that it should scale with the square-root of the density, and hence the cropped YSO count ($\sqrt{102/41} = 0.63$). Therefore we report that the derived properties are consistent with those measured by \cite{Gutermuthetal2009}. (Differences are expected as shown by \cite{Gutermuthetal2009} with their comparisons among several shared regions.) However, the missing YSOs at the centre of the cluster complicate any further comparison their results.  

\subsection{Comparison of grouped and non-grouped YSOs}

We find 76\% (113 of the 149) of the Class Is, Class Fs, and Class IIs are found in groups. Rather than compare the class fractions, given by $N_{\rm{I+F}}$/$N_{\rm{II}}$ in Table~\ref{tbl:grps}, we directly compare the underlying distribution of $\alpha$ to determine whether the distribution of YSOs within groups is consistent with the whole cloud. We get the same result for each group:
a KS test on the $\alpha$ distribution of the group and the $\alpha$ distribution of the whole cloud shows that we cannot reject the hypothesis that they are drawn from the same sample (p-values $>$ 0.13). (We also did a KS test for each group with the extended population and found the same result.)

Similarly, we compared the properties of disks within groups and those not in groups by performing a KS test on the distributions of disk luminosities (p-value of 0.08), \alphaexcess\ (p-value of 0.9), and \lambdaturnoff\ (p-value of 0.9) and find no evidence that the two populations are drawn from different parent populations.

%%%%%%%%%%%%%%%%%%%%%%%%%%%%%%%%%%%%%%%%%%%%%%%%
%%%%%%%%%%%%%%%%%%%%%%%%%%%%%%%%%%%%%%%%%%%%%%%%

\section{Summary}
\label{sec:summary}

We observed the AMC with IRAC and MIPS aboard the \textit{Spitzer Space Telescope} and identify 138 YSOs in the cloud. As our IRAC coverage is segmented, we complemented our more contiguous MIPS coverage with WISE data to further eliminate galaxies from the sample, leaving 28 MIPS-only YSOs remaining, bringing the total number of YSOs in the AMC to 166. We classified the YSOs based on the spectral slope of their SEDs between 2 \micron\ and 24 \micron\ and find 37 Class I objects, 21 Class F objects (flat spectrum sources), 91 Class II objects, and 17 Class III objects. The high fraction of Class Is and Class Fs suggests that the AMC is relatively unevolved compared to other star-forming clouds. Despite the similarity in cloud properties between the AMC and the OMC, there is a distinct difference in the star formation properties. The star formation in the AMC is also concentrated along its filament, however, it is also forming a factor of about 20 fewer stars than the OMC.  \cite{Ladaetal2009} find that there is much less material at high density in the AMC than in the OMC and attribute the difference in star formation to this. Further studies of the star formation and YSO population in the AMC are needed to highlight the differences of the two clouds given their similar age.

We modelled the SEDs of the Class II and Class III sources and their excesses by first fitting a K7 stellar spectrum to the optical and near-IR fluxes. The spectrum is normalized to the 2MASS flux (or the IRAC1 flux when 2MASS is unavailable) and we use an \Av~value to match the spectrum of the stellar model to the de-reddened observed optical fluxes. An A0 stellar spectrum is used in the eight cases where a K7 spectrum is unable to provide a reasonable fit. Fitting a stellar spectrum allows us to measure the disk luminosities and characterize the excess. The excesses of the Class II and Class III sources were further parameterized by \lambdaturnoff, the longest wavelength before an excess greater that 80\% is measured, and \alphaexcess, the slope of the SED at wavelengths longward of \lambdaturnoff. \lambdaturnoff\ is a useful tracer for the proximity of dust to the star and consequently we identify fourteen classical transition disk candidates.

The bulk of the star formation in the AMC is in the southern region of the cloud.
We included a clustering analysis to quantify the densest areas of star formation and to identify groups within the cloud. 
%groups separated by ~<1pc
%groups in Perseus separated by ~3pc
We find four groups with 10 or more members all in the region around \lkha 101 and its adjoining filament. We find three smaller groups with 5 -- 9 members scattered throughout the cloud. The largest group is that around \lkha 101 and contains 49 members. We note that there are likely even more YSOs in this group since our YSO identification criteria of S/N $\geq 3$ in IRAC1-4 and MIPS1 are difficult to attain in this bright region.

\acknowledgements

We thank the referee whose comments and suggestions greatly helped improve the paper and its clarity.
H.B.F gratefully acknowledges research support from an NSERC Discovery Grant.
This research made use of APLpy, an open-source plotting package for Python hosted at http://aplpy.github.com
This research made use of Montage, funded by the National Aeronautics and Space Administration's Earth Science Technology Office, Computation Technologies Project, under Cooperative Agreement Number NCC5-626 between NASA and the California Institute of Technology. Montage is maintained by the NASA/IPAC Infrared Science Archive. 

%%%%%%%%%%%%%%%%%%%%%%%%%%%%%%%%%%%%%%%%%%%%%%%%%%%%%%%%%%%%%%%%%%%%%%%%
%%%%%%%%%%%%%%%%%%%%%%%%%%%%%%%%%%%%%%%%%%%%%%%%%%%%%%%%%%%%%%%%%%%%%%%%

\bibliographystyle{apj}
\bibliography{library}

\newpage

%Table 1
\begin{deluxetable}{lclc}
\tablecolumns{4} 
\tablewidth{0pc} 
\tablecaption{Summary of IRAC Observations}
\tablehead{\colhead{IRAC Sub-region} & \colhead{Size} & \colhead{AOR Sub-region ID} & \colhead{AOR Key (1st epoch, 2nd
    epoch)} \\
& \colhead{(sq. deg.)} & & }
\startdata
AUR\_1a & 0.3 $\times$ 0.2 & auri\_irac6b &  19972096, 19971584 \\
AUR\_1b & 0.4 $\times$ 0.3 & auri\_irac6  &  20014336, 20014080 \\
AUR\_1c & 0.9 $\times$ 0.3 & auri\_irac7  &  19980544, 19980288 \\ 
& & auri\_irac7b &  19984384, 19984128 \\
AUR\_1d & 0.3 $\times$ 0.2 &  non-GB data & 03654144 \\
AUR\_1e & 0.3 $\times$ 0.3 & auri\_irac8  &  20013312, 20013056 \\
% 1cde_ALL size is 1.3 x 0.5 sq. deg
AUR\_2a & 1.3 $\times$ 1.4 & auri\_irac3  &  19983360, 19983104 \\
& & auri\_irac4  &  20016640, 20016384 \\
& & auri\_irac5  &  19981824, 19981568 \\
& & auri\_irac5b &  19956480, 19956224 \\
AUR\_2b & 0.4 $\times$ 0.3 & auri\_irac2  &  20018432, 20017920 \\
AUR\_3a & 0.8 $\times$ 0.9 & auri\_irac1  &  19984640, 19967744 \\
& & auri\_irac9  &  19978240, 19977984 \\
AUR\_3b & 0.4 $\times$ 0.3 & auri\_irac9b &  20012288, 20011776 \\
& & auri\_irac9c &  19976960, 19976192 \\
AUR\_4a & 0.4 $\times$ 0.7 & auri\_irac10 &  19993344, 19993088 \\
& & auri\_irac10b &  19988992, 19988736 \\
AUR\_4b & 0.3 $\times$ 0.3 & auri\_irac11 &  19961088, 19960832 \\
AUR\_5 & 0.3 $\times$ 0.3 & auri\_irac12 &  19992576, 19992064 \\
AUR\_NORTH & 0.5 $\times$ 0.3 & auri\_irac13 &  19960320, 19959808 \\
\enddata 
\label{iracobssum}
\end{deluxetable}

%Table 2
\begin{deluxetable}{lcl}
\tablecolumns{3} 
\tablewidth{0pc} 
\tablecaption{Summary of MIPS Observations}
\tablehead{\colhead{MIPS Sub-region} & \colhead{Size} & \colhead{AOR
    Key} \\
& \colhead{(sq. deg)} & }
\startdata
AUR\_1 & 1.2 $\times$ 3.2 & 20019712,19983872,20019456,19983616 \\
AUR\_2 & 1.6 $\times$ 2.6 & 20017152,19982336,20016896,19982080 \\
AUR\_3 & 1.0 $\times$ 2.0 & 20015360,20014848 \\
AUR\_4 & 1.4 $\times$ 2.2 & 19981312,19979520,19981056,19979008 \\
AUR\_5 & 0.5 $\times$ 1.9 & 20013824,20013568 \\
% AUR_ALL is 5.0 x 6.9 sq. deg
AUR\_NORTH & 0.5 $\times$ 1.9 & 20011520 \\
\enddata 
\label{mipsobssum}
\end{deluxetable}

%\begin{float}[h]
\begin{deluxetable}{lr}
%\tablecolumns{2} 
\tablewidth{0pc} 
\tablecaption{Sources in the AMC Field}
\tablehead{\colhead{Sources} & \colhead{Number}	}
\startdata 
Total			& 704045	 	\\
YSO				& 166		\\ 
Galc				& 322		\\
Stellar			& 32579		\\
2MASS			& 87745		\\
Zero	\tnm{a}		& 247257		\\
Something else	& 335976		\\
\enddata
\tnt{a}{Sources that do not have detections in the combined epochs data in any of the 2MASS, IRAC or MIPS bands. (It may have been detected in one or both of the epochs at different bands.)}
\label{tbl:SourceSummary}
\end{deluxetable}
%\end{float}

\begin{deluxetable}{lccccccccc}
\tablecolumns{8} 
\tabletypesize{\footnotesize}
\tablewidth{0pc} 
\tablecaption{YSOs in the AMC Based on IRAC and MIPS}
\tablehead{
\colhead{}   & \colhead{}     & \colhead{}      & \colhead{}         & \colhead{3.6 \micron} & \colhead{4.5 \micron} & \colhead{5.8 \micron} & \colhead{8.0 \micron} & \colhead{24.0 \micron} & \colhead{70.0 \micron} \\
\colhead{ID} & \colhead{Name} & \colhead{Class} & \colhead{$\alpha$} & \colhead{(mJy)}       & \colhead{(mJy)}       & \colhead{(mJy)}       & \colhead{(mJy)}       & \colhead{(mJy)}        & \colhead{(mJy)}} 
\startdata 
  1\tnm{N} & 04012455$+$4101490 &    I &  2.04 & 0.50$\pm$0.03 & 4.06$\pm$0.20 & 7.94$\pm$0.38 & 8.49$\pm$0.41 &  352$\pm$  32 & 8410$\pm$  965\\
  2\tnm{N} & 04013436$+$4111430 &   II & -1.00 & 10.5$\pm$ 0.5 & 9.66$\pm$0.46 & 8.99$\pm$0.44 & 7.65$\pm$0.36 & 14.7$\pm$ 1.4 &  288$\pm$  30\\
  3 & 04100064$+$4002361 &   II & -0.31 & 2.64$\pm$0.13 & 3.63$\pm$0.18 & 4.50$\pm$0.22 & 5.16$\pm$0.25 & 9.66$\pm$0.93 & \nodata\\
%\mc{8}{c}{...}\\
  4 & 04100263$+$4002482 &    I &  0.98 & 0.60$\pm$0.04 & 0.99$\pm$0.05 & 0.96$\pm$0.06 & 0.91$\pm$0.06 & 49.3$\pm$ 4.6 & \nodata\\
  5 & 04100562$+$4002386 &   II & -0.79 & 3.46$\pm$0.20 & 3.65$\pm$0.20 & 3.74$\pm$0.20 & 4.18$\pm$0.20 & 3.51$\pm$0.68 & \nodata\\
  6 & 04100841$+$4002244 &    I &  1.70 & 18.6$\pm$ 2.8 & 53.3$\pm$ 3.0 &  119$\pm$   6 &  237$\pm$  11 & 4770$\pm$  470 & 24600$\pm$  3550\\
  7 & 04101116$+$4001262 &    I &  1.99 &0.066$\pm$0.006 & 0.35$\pm$0.02 & 0.34$\pm$0.03 & 0.27$\pm$0.04 & 41.1$\pm$ 3.8 & \nodata\\
  8 & 04104051$+$3805004 &   II & -0.78 & 12.9$\pm$ 0.6 & 11.5$\pm$ 0.5 & 11.6$\pm$ 0.6 & 14.9$\pm$ 0.7 & 25.1$\pm$ 2.3 & 64.0$\pm$13.1\\
  9 & 04104163$+$3808058 &   II & -0.32 & 13.9$\pm$ 0.7 & 14.1$\pm$ 0.9 & 14.3$\pm$ 1.0 & 18.9$\pm$ 1.6 &  226$\pm$  75 & \nodata\\
 10 & 04104109$+$3807545 &    I &  2.03 & 1280$\pm$   78 & 2150$\pm$  132 & 4330$\pm$  244 & 5530$\pm$  304 & 11000$\pm$  2200 & \nodata\\
 11 & 04104211$+$3805599 &  III & -2.26 &  341$\pm$  25 &  210$\pm$  12 &  151$\pm$   9 & 87.1$\pm$ 4.4 & 43.9$\pm$ 4.2 & \nodata\\
 12 & 04104761$+$3803338 &   II & -0.87 & 6.21$\pm$0.32 & 6.14$\pm$0.33 & 6.11$\pm$0.30 & 7.42$\pm$0.37 & 7.18$\pm$0.70 & \nodata\\
 13 & 04104916$+$3804458 &   II & -0.49 & 44.1$\pm$ 2.2 & 46.3$\pm$ 2.2 & 57.0$\pm$ 2.7 & 85.0$\pm$ 4.1 &  123$\pm$  11 &  253$\pm$  26\\
 14 & 04194467$+$3811219 &    F & -0.07 & 4.54$\pm$0.23 & 5.62$\pm$0.27 & 7.26$\pm$0.36 & 10.2$\pm$ 0.5 & 22.5$\pm$ 2.1 & \nodata\\
 15 & 04205246$+$3806358 &  III & -2.42 & 14.5$\pm$ 0.7 & 10.1$\pm$ 0.5 & 7.13$\pm$0.35 & 4.33$\pm$0.23 & 1.08$\pm$0.17 & \nodata\\
 16 & 04213795$+$3734418 &   II & -0.85 &  284$\pm$  14 &  387$\pm$  20 &  443$\pm$  21 &  432$\pm$  20 &  223$\pm$  20 & 5530$\pm$ 1230\\
 17 & 04213808$+$3735409 &  III & -1.64 & 1.88$\pm$0.09 & 1.71$\pm$0.08 & 1.29$\pm$0.07 & 0.77$\pm$0.06 & 0.92$\pm$0.20 & \nodata\\
 18 & 04214080$+$3733590 &    I &  1.99 & 1.12$\pm$0.06 & 4.07$\pm$0.20 & 10.2$\pm$ 0.5 & 26.1$\pm$ 1.2 &  241$\pm$  22 &  945$\pm$ 112\\
 19 & 04244934$+$3716464 &  III & -2.27 & 6.19$\pm$0.30 & 4.14$\pm$0.20 & 3.05$\pm$0.16 & 1.89$\pm$0.11 & 0.91$\pm$0.20 & \nodata\\
 20 & 04253848$+$3707012 &    I &  1.43 & 0.60$\pm$0.03 & 1.04$\pm$0.08 & 1.95$\pm$0.13 & 4.50$\pm$0.25 & 59.1$\pm$ 5.5 & 1350$\pm$  152\\
 21 & 04253979$+$3707082 &    F & -0.01 &  254$\pm$  12 &  485$\pm$  25 &  671$\pm$  33 &  744$\pm$  39 &  727$\pm$  68 & 1160$\pm$  157\\
 22 & 04275080$+$3631264 &   II & -0.86 & 7.15$\pm$0.35 & 6.93$\pm$0.34 & 6.45$\pm$0.32 & 7.28$\pm$0.35 & 11.4$\pm$ 1.1 & \nodata\\
 23 & 04275826$+$3633265 &   II & -1.03 & 1.72$\pm$0.08 & 1.59$\pm$0.08 & 1.52$\pm$0.08 & 1.54$\pm$0.09 & 2.08$\pm$0.30 & \nodata\\
 24 & 04280289$+$3640586 &   II & -0.37 & 1.81$\pm$0.09 & 1.59$\pm$0.08 & 1.39$\pm$0.08 & 1.23$\pm$0.08 & 12.5$\pm$ 1.2 & \nodata\\
 25 & 04281515$+$3630286 &    F &  0.25 & 7.28$\pm$0.36 & 10.4$\pm$ 0.5 & 13.8$\pm$ 0.7 & 18.3$\pm$ 0.9 & 61.3$\pm$ 5.7 & 75.4$\pm$13.8\\
 26 & 04282116$+$3624478 &   II & -0.83 & 6.30$\pm$0.31 & 5.79$\pm$0.28 & 5.20$\pm$0.26 & 5.40$\pm$0.26 & 11.8$\pm$ 1.1 & \nodata\\
 27 & 04282136$+$3630215 &   II & -1.09 & 2.93$\pm$0.14 & 2.60$\pm$0.13 & 2.63$\pm$0.14 & 2.50$\pm$0.13 & 2.55$\pm$0.30 & \nodata\\
 28 & 04283509$+$3625064 &    I &  0.88 & 10.7$\pm$ 0.5 & 27.5$\pm$ 1.3 & 45.8$\pm$ 2.2 & 62.1$\pm$ 2.9 &  237$\pm$  21 &  840$\pm$  92\\
 29 & 04283789$+$3624553 &   II & -0.63 &  124$\pm$   6 &  140$\pm$   6 &  161$\pm$   8 &  188$\pm$   9 &  204$\pm$  18 & 1240$\pm$  123\\
 30 & 04283856$+$3625289 &    I &  1.14 & 0.83$\pm$0.04 & 1.56$\pm$0.08 & 1.56$\pm$0.09 & 1.99$\pm$0.11 &  143$\pm$  13 &  896$\pm$  96\\
 31 & 04284335$+$3625117 &   II & -0.44 & 9.32$\pm$0.45 & 9.36$\pm$0.45 & 9.92$\pm$0.48 & 15.7$\pm$ 0.7 & 30.1$\pm$ 2.8 & \nodata\\
 32 & 04284367$+$3628393 &    I &  1.16 & 1.34$\pm$0.07 & 3.58$\pm$0.18 & 4.01$\pm$0.19 & 4.01$\pm$0.20 &  192$\pm$  17 & 2170$\pm$  244\\
 33 & 04284443$+$3624456 &    F &  0.12 & 1.73$\pm$0.08 & 2.55$\pm$0.12 & 3.46$\pm$0.17 & 5.78$\pm$0.28 & 10.4$\pm$ 1.0 & \nodata\\
 34 & 04284958$+$3629107 &    I &  0.47 & 3.08$\pm$0.15 & 5.70$\pm$0.27 & 7.60$\pm$0.37 & 9.14$\pm$0.44 & 51.0$\pm$ 4.7 & 93.2$\pm$16.9\\
 35 & 04285530$+$3631225 &    I &  1.18 & 17.9$\pm$ 0.9 & 51.0$\pm$ 2.4 & 83.1$\pm$ 4.0 &  109$\pm$   5 &  752$\pm$  70 & 4290$\pm$  476\\
 36\tnm{*} & 04285911$+$3623112 &   II & -1.23 & 30.7$\pm$ 1.5 & 28.5$\pm$ 1.4 & 24.9$\pm$ 1.2 & 27.2$\pm$ 1.3 & 21.9$\pm$ 2.0 & \nodata\\
 37 & 04293901$+$3516105 &   II & -1.00 & 37.0$\pm$ 2.1 & 35.0$\pm$ 1.8 & 29.5$\pm$ 1.6 & 41.0$\pm$ 2.1 & 36.8$\pm$ 3.5 & \nodata\\
 38 & 04294001$+$3521089 &    I &  0.51 & 22.7$\pm$ 1.1 & 28.1$\pm$ 1.4 & 50.6$\pm$ 2.5 &  160$\pm$   8 &  147$\pm$  14 & \nodata\\
 39 & 04294358$+$3513386 &   II & -0.86 & 18.8$\pm$ 0.9 & 17.9$\pm$ 0.9 & 17.8$\pm$ 0.9 & 23.0$\pm$ 1.1 & 21.0$\pm$ 2.1 & \nodata\\
 40 & 04294421$+$3512300 &    F & -0.21 & 8.89$\pm$0.43 & 9.39$\pm$0.46 & 10.2$\pm$ 0.5 & 16.2$\pm$ 0.8 & 44.2$\pm$ 4.1 & \nodata\\
 41 & 04294728$+$3510192 &   II & -0.54 & 10.9$\pm$ 0.5 & 11.4$\pm$ 0.5 & 13.6$\pm$ 0.7 & 21.1$\pm$ 1.0 & 16.6$\pm$ 1.6 & \nodata\\
 42 & 04294742$+$3511335 &   II & -1.37 & 5.59$\pm$0.27 & 4.66$\pm$0.22 & 4.19$\pm$0.22 & 3.99$\pm$0.20 & 2.70$\pm$0.36 & \nodata\\
 43 & 04294854$+$3512125 &   II & -0.75 & 3.21$\pm$0.15 & 4.47$\pm$0.22 & 3.54$\pm$0.18 & 5.09$\pm$0.25 & 4.13$\pm$0.50 & \nodata\\
 44 & 04294921$+$3514227 &    F & -0.24 & 62.6$\pm$ 3.2 & 74.9$\pm$ 3.7 & 76.9$\pm$ 3.7 & 94.9$\pm$ 4.7 &  196$\pm$  18 & \nodata\\
 45 & 04294961$+$3514438 &   II & -0.51 & 8.80$\pm$0.44 & 11.0$\pm$ 0.5 & 9.26$\pm$0.46 & 9.66$\pm$0.48 & 27.6$\pm$ 2.8 & \nodata\\
 46 & 04295084$+$3515579 &    F & -0.11 & 33.3$\pm$ 1.7 & 43.1$\pm$ 2.2 & 45.1$\pm$ 2.2 & 65.0$\pm$ 3.1 &  177$\pm$  17 & \nodata\\
 47 & 04295101$+$3515475 &    I &  0.74 & 5.80$\pm$0.30 & 8.90$\pm$0.44 & 15.3$\pm$ 0.8 & 31.1$\pm$ 1.5 & 98.6$\pm$11.0 & \nodata\\
 48 & 04295346$+$3515485 &    F & -0.26 & 17.7$\pm$ 0.9 & 17.9$\pm$ 0.9 & 21.0$\pm$ 1.1 & 31.5$\pm$ 1.9 & 79.0$\pm$ 8.4 & \nodata\\
 49 & 04295415$+$3510216 &    F &  0.08 & 2.44$\pm$0.12 & 4.56$\pm$0.22 & 6.10$\pm$0.32 & 8.85$\pm$0.44 & 16.5$\pm$ 1.6 & \nodata\\
 50 & 04295479$+$3518025 &   II & -0.32 & 29.6$\pm$ 1.5 & 32.0$\pm$ 1.6 & 35.0$\pm$ 1.8 & 51.2$\pm$ 3.4 &  135$\pm$  12 & \nodata\\
 51 & 04295627$+$3517429 &    I &  0.61 & 10.0$\pm$ 0.5 & 17.4$\pm$ 0.9 & 23.1$\pm$ 1.3 & 28.3$\pm$ 2.6 & 57.7$\pm$11.2 & \nodata\\
 52 & 04295976$+$3513342 &   II & -0.81 &  139$\pm$   8 &  128$\pm$   6 &  114$\pm$   5 &  151$\pm$   8 &  195$\pm$  18 & \nodata\\
 53\tnm{*} & 04300016$+$3603227 &   II & -1.05 & 12.2$\pm$ 0.6 & 11.4$\pm$ 0.5 & 10.6$\pm$ 0.5 & 11.8$\pm$ 0.6 & 12.9$\pm$ 1.2 & \nodata\\
 54 & 04300114$+$3517246 &  III & -1.93 & 75.7$\pm$ 4.0 & 48.0$\pm$ 2.5 & 30.8$\pm$ 1.6 & 22.9$\pm$ 1.5 & 26.5$\pm$ 3.4 & \nodata\\
 55 & 04300263$+$3515143 &   II & -1.00 & 53.3$\pm$ 2.7 & 44.5$\pm$ 2.2 & 41.0$\pm$ 2.1 & 46.4$\pm$ 2.7 & 66.4$\pm$ 7.3 & \nodata\\
 56 & 04300363$+$3514201 &    I &  0.75 & 8.19$\pm$0.57 & 12.5$\pm$ 0.6 & 17.6$\pm$ 1.3 & 31.4$\pm$ 4.1 & 37.4$\pm$11.2 & \nodata\\
 57 & 04300423$+$3509459 &   II & -1.17 & 2.06$\pm$0.10 & 1.85$\pm$0.09 & 1.73$\pm$0.10 & 2.02$\pm$0.12 & 1.33$\pm$0.26 & \nodata\\
 58\tnm{*} & 04300425$+$3522238 &   II & -0.62 & 6.47$\pm$0.37 & 4.25$\pm$0.21 & 3.23$\pm$0.25 & 2.69$\pm$0.60 & 24.8$\pm$ 2.5 &  688$\pm$ 118\\
 59 & 04300743$+$3514579 &   II & -0.88 &  124$\pm$   6 &  127$\pm$   6 &  121$\pm$   5 &  139$\pm$   7 &  137$\pm$  13 & \nodata\\
 60 & 04300773$+$3515484 &   II & -0.77 & 24.3$\pm$ 1.2 & 24.6$\pm$ 1.2 & 22.2$\pm$ 1.3 & 27.6$\pm$ 2.0 & 53.8$\pm$12.8 & \nodata\\
 61 & 04300825$+$3514100 &    I &  0.46 & 9.62$\pm$0.48 & 14.9$\pm$ 0.7 & 19.3$\pm$ 1.1 & 27.1$\pm$ 2.0 &  119$\pm$  13 & \nodata\\
 62 & 04300874$+$3514375 &   II & -0.84 &  107$\pm$   5 &  105$\pm$   5 &  105$\pm$   5 &  127$\pm$   8 &  158$\pm$  38 & \nodata\\
 63 & 04300951$+$3514403 &    I &  0.81 & 8.53$\pm$0.51 & 12.5$\pm$ 0.6 & 16.6$\pm$ 1.6 & 24.6$\pm$ 5.2 &  334$\pm$  33 & \nodata\\
 64\tnm{*} & 04300980$+$3540355 &   II & -0.89 & 3.61$\pm$0.18 & 2.96$\pm$0.14 & 2.46$\pm$0.13 & 2.47$\pm$0.13 & 8.30$\pm$0.79 & 57.0$\pm$ 9.0\\
 65 & 04300991$+$3515539 &   II & -0.94 & 56.8$\pm$ 3.2 & 48.4$\pm$ 2.9 & 44.8$\pm$ 2.8 & 63.2$\pm$ 4.0 &  139$\pm$  42 & \nodata\\
 66 & 04301234$+$3509346 &   II & -0.99 & 7.18$\pm$0.35 & 7.73$\pm$0.37 & 7.17$\pm$0.35 & 6.30$\pm$0.30 & 7.86$\pm$0.76 & \nodata\\
 67 & 04301309$+$3513586 &   II & -0.90 &  107$\pm$   7 & 93.5$\pm$ 4.8 & 81.6$\pm$ 4.6 & 93.2$\pm$ 7.1 &  153$\pm$  15 & \nodata\\
 68 & 04301453$+$3513326 &   II & -0.39 & 81.4$\pm$ 7.7 & 96.1$\pm$ 4.9 & 96.3$\pm$ 5.0 &  108$\pm$   7 &  160$\pm$  27 & \nodata\\
 69 & 04301474$+$3520143 &   II & -0.60 &  136$\pm$  10 &  146$\pm$   8 &  166$\pm$   8 &  191$\pm$  11 &  197$\pm$  19 & \nodata\\
 70 & 04301495$+$3600085 &    I &  1.77 & 0.22$\pm$0.01 & 1.07$\pm$0.05 & 2.61$\pm$0.14 & 4.45$\pm$0.22 & 48.2$\pm$ 4.5 &  137$\pm$  16\\
 71 & 04301576$+$3556578 &    I &  0.40 &  996$\pm$  51 & 1450$\pm$   79 & 2080$\pm$  103 & 2910$\pm$  163 & 5500$\pm$ 1100 & \nodata\\
 72\tnm{*} & 04301627$+$3542429 &   II & -0.35 & 3.34$\pm$0.16 & 3.31$\pm$0.16 & 3.66$\pm$0.19 & 5.18$\pm$0.25 & 11.2$\pm$ 1.1 & \nodata\\
 73 & 04301784$+$3603266 &  III & -1.72 & 5.90$\pm$0.29 & 4.77$\pm$0.23 & 3.72$\pm$0.19 & 2.94$\pm$0.15 & 1.87$\pm$0.25 & \nodata\\
 74 & 04301808$+$3545389 &   II & -0.82 & 2.46$\pm$0.12 & 1.76$\pm$0.09 & 1.34$\pm$0.08 & 1.27$\pm$0.08 & 8.13$\pm$0.78 & \nodata\\
 75 & 04301899$+$3542120 &   II & -1.60 & 4.95$\pm$0.24 & 4.09$\pm$0.20 & 3.53$\pm$0.18 & 2.79$\pm$0.14 & 1.78$\pm$0.28 & \nodata\\
 76 & 04301959$+$3508216 &    F & -0.11 & 3.96$\pm$0.19 & 5.25$\pm$0.25 & 5.47$\pm$0.28 & 7.74$\pm$0.38 & 20.3$\pm$ 1.9 & \nodata\\
 77 & 04302219$+$3604359 &   II & -1.07 & 13.9$\pm$ 0.7 & 13.9$\pm$ 0.7 & 13.0$\pm$ 0.6 & 12.5$\pm$ 0.6 & 12.3$\pm$ 1.1 & \nodata\\
 78 & 04302268$+$3519081 &   II & -0.72 & 4.64$\pm$0.22 & 5.02$\pm$0.24 & 5.45$\pm$0.28 & 4.07$\pm$0.56 & 5.66$\pm$1.81 & \nodata\\
 79 & 04302382$+$3521123 &    I &  0.61 & 1.37$\pm$0.07 & 2.35$\pm$0.11 & 3.20$\pm$0.16 & 5.95$\pm$0.29 & 25.7$\pm$ 2.4 & \nodata\\
 80\tnm{*} & 04302433$+$3459165 &  III & -2.43 & 13.9$\pm$ 0.7 & 9.19$\pm$0.44 & 6.62$\pm$0.32 & 3.73$\pm$0.18 & 1.43$\pm$0.24 & \nodata\\
 81 & 04302468$+$3545206 &    I &  1.32 & 20.3$\pm$ 1.0 & 48.6$\pm$ 2.4 & 90.9$\pm$ 4.3 &  156$\pm$   7 & 1400$\pm$  131 & 4530$\pm$  520\\
 82 & 04302503$+$3543179 &   II & -0.73 & 97.7$\pm$ 4.9 &  120$\pm$   5 &  152$\pm$   7 &  173$\pm$   8 &  122$\pm$  11 & \nodata\\
 83 & 04302589$+$3548113 &   II & -0.68 & 2.82$\pm$0.14 & 2.71$\pm$0.13 & 2.40$\pm$0.13 & 2.56$\pm$0.13 & 7.15$\pm$0.68 & \nodata\\
 84 & 04302702$+$3520284 &   II & -0.63 & 88.4$\pm$ 4.2 &  109$\pm$   5 &  116$\pm$   5 &  132$\pm$   6 &  128$\pm$  11 & \nodata\\
 85 & 04302704$+$3545505 &    F & -0.11 & 1.75$\pm$0.09 & 1.61$\pm$0.08 & 1.80$\pm$0.10 & 2.47$\pm$0.13 & 14.4$\pm$ 1.4 & \nodata\\
 86 & 04302741$+$3509178 &    I &  1.60 & 21.1$\pm$ 1.5 &  102$\pm$   6 &  265$\pm$  13 &  367$\pm$  17 & 1580$\pm$  155 & 12600$\pm$  1490\\
 87 & 04302775$+$3546150 &    F &  0.16 & 17.8$\pm$ 0.9 & 22.4$\pm$ 1.1 & 28.2$\pm$ 1.4 & 37.0$\pm$ 1.7 & 95.0$\pm$ 8.8 & \nodata\\
 88 & 04302809$+$3509164 &    I &  1.43 & 1.34$\pm$0.09 & 8.70$\pm$0.47 & 8.40$\pm$0.45 & 15.0$\pm$ 0.8 &  287$\pm$  33 & \nodata\\
 89\tnm{*} & 04302842$+$3532419 &   II & -1.21 & 40.0$\pm$ 2.0 & 36.8$\pm$ 1.8 & 27.7$\pm$ 1.4 & 22.6$\pm$ 1.1 & 41.0$\pm$ 3.8 & 43.1$\pm$12.2\\
 90 & 04302844$+$3549176 &    F & -0.30 & 12.5$\pm$ 0.6 & 13.8$\pm$ 0.7 & 16.6$\pm$ 1.0 & 26.5$\pm$ 1.3 & 45.7$\pm$ 4.2 & \nodata\\
 91 & 04302861$+$3547407 &   II & -0.62 & 21.7$\pm$ 1.1 & 25.3$\pm$ 1.2 & 28.4$\pm$ 1.4 & 34.5$\pm$ 1.6 & 33.4$\pm$ 3.1 & 76.8$\pm$ 9.9\\
 92 & 04302871$+$3547498 &   II & -0.54 & 6.66$\pm$0.32 & 6.42$\pm$0.31 & 6.11$\pm$0.30 & 6.24$\pm$0.30 & 24.4$\pm$ 2.3 & \nodata\\
 93 & 04302898$+$3507540 &   II & -0.66 & 1.92$\pm$0.10 & 1.46$\pm$0.07 & 1.76$\pm$0.10 & 1.81$\pm$0.10 & 4.78$\pm$0.54 & \nodata\\
 94 & 04302961$+$3527172 &   II & -0.82 & 7.65$\pm$0.38 & 6.83$\pm$0.32 & 6.33$\pm$0.31 & 8.76$\pm$0.41 & 14.7$\pm$ 1.4 & \nodata\\
 95 & 04302966$+$3506390 &   II & -0.49 & 3.97$\pm$0.19 & 3.16$\pm$0.16 & 4.17$\pm$0.20 & 5.58$\pm$0.27 & 14.7$\pm$ 1.5 & \nodata\\
 96 & 04303014$+$3506392 &   II & -0.79 & 18.1$\pm$ 0.9 & 18.1$\pm$ 0.9 & 13.7$\pm$ 0.6 & 14.6$\pm$ 0.7 & 38.5$\pm$ 3.7 & \nodata\\
 97 & 04303028$+$3521040 &   II & -0.60 & 32.8$\pm$ 1.6 & 33.8$\pm$ 1.6 & 34.9$\pm$ 1.6 & 47.1$\pm$ 2.2 & 70.0$\pm$ 6.5 & \nodata\\
 98\tnm{*} & 04303043$+$3518337 &   II & -0.52 & 3.92$\pm$0.19 & 5.44$\pm$0.27 & 4.82$\pm$0.25 & 6.46$\pm$0.56 & 10.3$\pm$ 1.2 & \nodata\\
 99\tnm{*} & 04303051$+$3517447 &   II & -0.80 & 13.5$\pm$ 0.6 & 12.9$\pm$ 0.6 & 12.7$\pm$ 0.6 & 14.9$\pm$ 0.8 & 20.8$\pm$ 2.4 & \nodata\\
100 & 04303056$+$3551440 &    I &  0.93 & 4.77$\pm$0.24 & 10.6$\pm$ 0.5 & 15.2$\pm$ 0.7 & 20.4$\pm$ 1.0 &  187$\pm$  17 &  628$\pm$  64\\
101 & 04303158$+$3545137 &    F & -0.11 & 82.0$\pm$ 4.0 &  112$\pm$   5 &  151$\pm$   7 &  200$\pm$   9 &  403$\pm$  37 &  405$\pm$  42\\
102 & 04303235$+$3536134 &   II & -0.64 & 33.7$\pm$ 1.6 & 24.2$\pm$ 1.2 & 17.2$\pm$ 0.8 & 15.5$\pm$ 0.7 &  292$\pm$  27 & 1400$\pm$  146\\
103 & 04303680$+$3554362 &    I &  1.49 & 4.65$\pm$0.26 & 18.2$\pm$ 0.9 & 46.2$\pm$ 2.2 & 72.6$\pm$ 3.5 &  529$\pm$  49 & 4260$\pm$  476\\
104 & 04303740$+$3600180 &   II & -0.69 & 16.0$\pm$ 0.8 & 16.8$\pm$ 0.8 & 16.3$\pm$ 0.8 & 20.3$\pm$ 1.0 & 35.8$\pm$ 3.3 & \nodata\\
105 & 04303751$+$3513486 &   II & -1.54 & 2.38$\pm$0.11 & 2.24$\pm$0.11 & 1.77$\pm$0.09 & 1.55$\pm$0.09 & 1.27$\pm$0.31 & \nodata\\
106 & 04303751$+$3550317 &   II & -0.81 &  138$\pm$   7 &  149$\pm$   7 &  160$\pm$   8 &  173$\pm$   8 &  390$\pm$  36 & 1910$\pm$  314\\
107 & 04303789$+$3551014 &    I &  1.28 &0.070$\pm$0.009 & 0.27$\pm$0.02 & 0.38$\pm$0.04 & 0.34$\pm$0.05 & 9.82$\pm$0.92 & \nodata\\
108 & 04303826$+$3549593 &   II & -1.08 & 1.60$\pm$0.13 & 1.98$\pm$0.15 & 1.39$\pm$0.09 & 0.74$\pm$0.06 & 8.84$\pm$2.08 & 1880$\pm$  214\\
109 & 04303865$+$3554391 &    F &  0.10 & 8.32$\pm$0.40 & 14.6$\pm$ 0.7 & 16.5$\pm$ 0.8 & 16.2$\pm$ 0.8 & 53.4$\pm$ 4.9 & \nodata\\
110 & 04303912$+$3544498 &   II & -1.17 & 50.4$\pm$ 2.5 & 45.4$\pm$ 2.2 & 40.9$\pm$ 2.0 & 38.4$\pm$ 1.8 & 39.8$\pm$ 3.7 & \nodata\\
111 & 04303916$+$3552038 &    F & -0.14 & 94.6$\pm$ 4.8 &  116$\pm$   5 &  139$\pm$   6 &  150$\pm$  17 &  625$\pm$  63 & \nodata\\
112 & 04303931$+$3552007 &    F & -0.27 &  165$\pm$   8 &  179$\pm$   8 &  186$\pm$   8 &  202$\pm$  13 &  899$\pm$  84 & 3010$\pm$  327\\
113\tnm{*} & 04303956$+$3518069 &   II & -1.35 & 4.77$\pm$0.23 & 3.38$\pm$0.16 & 2.43$\pm$0.12 & 1.77$\pm$0.14 & 7.17$\pm$0.87 & \nodata\\
114\tnm{*} & 04303958$+$3511128 &   II & -1.14 & 6.71$\pm$0.34 & 6.11$\pm$0.29 & 5.24$\pm$0.26 & 5.37$\pm$0.26 & 6.32$\pm$0.63 & \nodata\\
115\tnm{*} & 04304005$+$3542103 &  III & -1.64 & 14.4$\pm$ 0.7 & 10.2$\pm$ 0.5 & 7.32$\pm$0.37 & 5.57$\pm$0.28 & 7.36$\pm$0.70 & 62.9$\pm$10.5\\
116 & 04304014$+$3531341 &   II & -1.00 & 16.2$\pm$ 0.8 & 14.7$\pm$ 0.7 & 12.7$\pm$ 0.6 & 14.4$\pm$ 0.7 & 18.9$\pm$ 1.8 & \nodata\\
117 & 04304116$+$3529410 &    I &  1.49 & 1.30$\pm$0.07 & 5.87$\pm$0.28 & 12.0$\pm$ 0.6 & 15.9$\pm$ 0.8 &  176$\pm$  16 & 1930$\pm$  204\\
118 & 04304423$+$3559511 &    I &  1.08 & 31.2$\pm$ 1.6 &  123$\pm$   6 &  276$\pm$  13 &  443$\pm$  21 & 1270$\pm$  119 & 3440$\pm$  371\\
119\tnm{*} & 04304469$+$3510521 &   II & -1.06 & 2.17$\pm$0.11 & 2.23$\pm$0.11 & 1.46$\pm$0.09 & 1.31$\pm$0.08 & 3.54$\pm$0.38 & \nodata\\
120 & 04304558$+$3458080 &   II & -1.03 & 10.9$\pm$ 0.5 & 10.5$\pm$ 0.5 & 9.33$\pm$0.44 & 8.82$\pm$0.42 & 11.3$\pm$ 1.1 & \nodata\\
121 & 04304625$+$3458562 &    I &  1.41 & 0.15$\pm$0.01 & 0.55$\pm$0.03 & 0.69$\pm$0.05 & 0.60$\pm$0.05 & 26.9$\pm$ 2.5 &  756$\pm$ 101\\
122 & 04304723$+$3507432 &   II & -0.39 & 14.9$\pm$ 0.7 & 24.6$\pm$ 1.2 & 17.5$\pm$ 0.8 & 30.1$\pm$ 1.4 & 51.8$\pm$ 4.8 & 83.4$\pm$11.1\\
123 & 04304757$+$3458242 &   II & -0.76 & 4.11$\pm$0.20 & 4.47$\pm$0.22 & 4.58$\pm$0.23 & 4.53$\pm$0.22 & 6.31$\pm$0.63 & \nodata\\
124 & 04304852$+$3537537 &    I &  1.46 & 4.12$\pm$0.22 & 16.3$\pm$ 0.9 & 28.4$\pm$ 1.3 & 38.1$\pm$ 1.8 &  452$\pm$  42 & 4120$\pm$  451\\
125 & 04304861$+$3458535 &    I &  0.34 & 27.7$\pm$ 1.4 & 32.5$\pm$ 1.6 & 43.2$\pm$ 2.1 & 69.7$\pm$ 3.3 &  677$\pm$  63 & \nodata\\
126 & 04304922$+$3456103 &    I &  0.69 & 11.1$\pm$ 0.5 & 22.3$\pm$ 1.1 & 34.3$\pm$ 1.6 & 50.5$\pm$ 2.4 &  277$\pm$  25 &  432$\pm$  46\\
127 & 04304934$+$3536419 &   II & -0.90 & 4.90$\pm$0.24 & 4.86$\pm$0.24 & 5.34$\pm$0.26 & 6.02$\pm$0.29 & 7.21$\pm$0.70 & \nodata\\
128 & 04304968$+$3457277 &   II & -0.72 &  416$\pm$  21 &  458$\pm$  30 &  438$\pm$  21 &  437$\pm$  21 &  677$\pm$  62 & 1030$\pm$  108\\
129 & 04305057$+$3533235 &   II & -1.05 & 3.62$\pm$0.18 & 3.36$\pm$0.17 & 2.96$\pm$0.15 & 3.09$\pm$0.16 & 4.19$\pm$0.43 & \nodata\\
130 & 04305098$+$3535548 &   II & -1.01 & 2.34$\pm$0.11 & 2.07$\pm$0.10 & 1.93$\pm$0.10 & 2.31$\pm$0.12 & 2.66$\pm$0.31 & \nodata\\
131 & 04305350$+$3456274 &    I &  0.98 & 0.50$\pm$0.03 & 0.92$\pm$0.05 & 1.50$\pm$0.09 & 2.04$\pm$0.11 & 27.2$\pm$ 2.5 & \nodata\\
132 & 04305390$+$3530110 &   II & -0.62 & 23.7$\pm$ 1.2 & 24.9$\pm$ 1.2 & 24.9$\pm$ 1.2 & 30.0$\pm$ 1.4 & 99.7$\pm$ 9.3 & \nodata\\
133 & 04305501$+$3530562 &   II & -0.85 & 4.71$\pm$0.23 & 3.56$\pm$0.17 & 2.86$\pm$0.15 & 2.39$\pm$0.12 & 17.5$\pm$ 1.6 & \nodata\\
134 & 04305599$+$3456478 &    I &  1.23 & 1.69$\pm$0.08 & 2.96$\pm$0.14 & 4.77$\pm$0.24 & 9.80$\pm$0.47 &  141$\pm$  13 &  360$\pm$  40\\
135 & 04305661$+$3530045 &    I &  2.35 & 0.30$\pm$0.02 & 1.12$\pm$0.06 & 1.78$\pm$0.10 & 3.85$\pm$0.19 &  302$\pm$  28 & 1470$\pm$  153\\
%\mc{8}{c}{...}\\
136 & 04295017$+$3514445 &   II & -0.90 & 2.82$\pm$0.14 & 2.38$\pm$0.12 & 2.35$\pm$0.15 & 3.14$\pm$0.20 & $<$  7.75 & \nodata\\
137 & 04300986$+$3514163 &   II & -0.47 & 27.6$\pm$ 1.4 & 30.2$\pm$ 1.5 & 25.7$\pm$ 1.5 & 28.6$\pm$ 2.5 & $<$  40.4 & \nodata\\
138 & 04301521$+$3516398 &    F & -0.22 &  131$\pm$  12 & 85.4$\pm$13.2 &  198$\pm$  28 &  368$\pm$  52 & $<$   196 & \nodata\\
\enddata
\tnt{*}{The YSO is in a region of low column density ($N_{\rm{H2}}< 5 \times 10^{21}$ cm$^{-2}$) and so is a possible contaminant.}
\tnt{N}{The YSO lies beyond the $N_{\rm{H2}}$ column density map from \cite{Harveyetal2013} and so $N_{\rm{H2}}$ at its position is unknown.}
\tablecomments{The names of the YSOs give their J2000 positions. Note that YSOs with 24 \micron\ upper limits are identified according to the IRAC-only criteria.}
\label{tbl:irac}
\end{deluxetable}

\begin{deluxetable}{lccccccccccccc}
\tablecolumns{12} 
\rotate
\tabletypesize{\scriptsize}
\setlength{\tabcolsep}{0.04in} 
\tablewidth{0pc} 
\tablecaption{YSO Candidates in the AMC Based on WISE and MIPS}
\tablehead{
\colhead{}   & \colhead{}     & \colhead{}      & \colhead{}         & \colhead{IRAC}        & \colhead{IRAC}        & \colhead{IRAC}        & \colhead{IRAC}        & \colhead{WISE}        & \colhead{WISE}        & \colhead{WISE}       & \colhead{WISE}       & \colhead{MIPS}         & \colhead{MIPS} \\
\colhead{}   & \colhead{}     & \colhead{}      & \colhead{}         & \colhead{3.6 \micron} & \colhead{4.5 \micron} & \colhead{5.8 \micron} & \colhead{8.0 \micron} & \colhead{3.4 \micron} & \colhead{4.6 \micron} & \colhead{12 \micron} & \colhead{22 \micron} & \colhead{24.0 \micron} & \colhead{70.0 \micron} \\
\colhead{ID} & \colhead{Name} & \colhead{Class} & \colhead{$\alpha$} & \colhead{mJy}         & \colhead{mJy}         & \colhead{mJy}         & \colhead{mJy}         & \colhead{(mJy)}       & \colhead{(mJy)}       & \colhead{(mJy)}      & \colhead{(mJy)}      & \colhead{(mJy)}        & \colhead{(mJy)}} 
\startdata 
139\tnm{N} & 04022975$+$4042419 &   II & -1.25 & \nodata & \nodata & \nodata & \nodata & 1631$\pm$   84 & 2646$\pm$   99 &  964$\pm$  13 &  350$\pm$   8 &  259$\pm$  24 & \nodata\\
140 & 04090200$+$4019131 &    I &  0.95 & \nodata & \nodata & \nodata & \nodata & 12.3$\pm$ 0.3 & 58.8$\pm$ 1.1 &  165$\pm$   2 &  977$\pm$  18 &  980$\pm$  91 & 3730$\pm$  434\\
141\tnm{*} & 04100343$+$3904495 &  III & -1.85 & \nodata & \nodata & \nodata & \nodata & 1576$\pm$   81 & 1072$\pm$   24 &  865$\pm$  12 &  667$\pm$  13 &  418$\pm$  43 & \nodata\\
142 & 04102441$+$3805227 &   II & -0.81 & \nodata & 15.5$\pm$ 0.8 & \nodata & 21.3$\pm$ 1.1 & 15.0$\pm$ 0.3 & 17.0$\pm$ 0.3 & 20.7$\pm$ 0.4 & 26.5$\pm$ 1.4 & 25.0$\pm$ 2.3 & \nodata\\
143\tnm{*} & 04120847$+$3801466 &  III & -2.08 & 50.7$\pm$ 2.5 & \nodata & 22.6$\pm$ 1.1 & \nodata & 59.7$\pm$ 1.2 & 33.4$\pm$ 0.6 & 10.5$\pm$ 0.3 & 21.9$\pm$ 1.6 & 8.49$\pm$0.82 & \nodata\\
144\tnm{*} & 04125764$+$3914183 &  III & -1.97 & \nodata & \nodata & \nodata & \nodata & 4653$\pm$  342 & 3817$\pm$  162 & 1118$\pm$   17 &  930$\pm$  17 &  809$\pm$  76 & 90.7$\pm$10.2\\
145\tnm{*} & 04134457$+$3904357 &  III & -2.02 & \nodata & \nodata & \nodata & \nodata & 21.3$\pm$ 0.4 & 11.8$\pm$ 0.2 & 2.26$\pm$0.22 & 5.64$\pm$1.71 & 3.48$\pm$0.46 &  246$\pm$  28\\
146\tnm{*} & 04151120$+$3839571 &   II & -1.49 & \nodata & \nodata & \nodata & \nodata &  160$\pm$   3 &  133$\pm$   2 & 73.9$\pm$ 1.0 & 79.1$\pm$ 2.3 & 64.9$\pm$ 6.0 & \nodata\\
147\tnm{*} & 04155405$+$3834131 &  III & -1.96 & \nodata & \nodata & \nodata & \nodata & 20.3$\pm$ 0.4 & 11.5$\pm$ 0.2 & 2.89$\pm$0.18 & 8.14$\pm$1.00 & 3.93$\pm$0.41 & \nodata\\
148\tnm{*} & 04170593$+$3722187 &  III & -2.07 & \nodata & \nodata & \nodata & \nodata & 1927$\pm$  107 & 1381$\pm$   40 &  479$\pm$   6 &  329$\pm$   8 &  250$\pm$  23 & \nodata\\
149\tnm{*} & 04230546$+$3807369 &  III & -1.93 & \nodata & \nodata & \nodata & \nodata & 55.5$\pm$ 1.2 & 31.6$\pm$ 0.6 & 12.1$\pm$ 0.3 & 22.7$\pm$ 1.3 & 11.5$\pm$ 1.2 & \nodata\\
150\tnm{*} & 04271374$+$3627107 &   II & -1.56 & \nodata & \nodata & \nodata & \nodata & 4.28$\pm$0.20 & 2.39$\pm$0.11 & 1.01$\pm$0.16 & $<$  12.6 & 1.81$\pm$0.29 & \nodata\\
151\tnm{*} & 04285556$+$3524460 &   II & -1.16 & \nodata & \nodata & \nodata & \nodata & 3.51$\pm$0.08 & 2.52$\pm$0.06 & 1.94$\pm$0.21 & 4.02$\pm$1.22 & 3.42$\pm$0.37 & \nodata\\
152\tnm{*} & 04291153$+$3504495 &   II & -1.30 & \nodata & \nodata & \nodata & \nodata & 9.03$\pm$0.20 & 7.20$\pm$0.14 & 5.59$\pm$0.20 & 9.73$\pm$1.08 & 7.39$\pm$0.73 & \nodata\\
153\tnm{*} & 04291438$+$3515245 &   II & -1.28 & \nodata & \nodata & \nodata & \nodata &  129$\pm$   2 &  106$\pm$   1 & 90.3$\pm$ 1.5 &  129$\pm$   4 & 85.9$\pm$ 8.0 &  109$\pm$  21\\
154\tnm{*} & 04294628$+$3619235 &    F & -0.21 & 23.7$\pm$ 1.2 & \nodata & 29.9$\pm$ 1.4 & \nodata & 20.5$\pm$ 0.4 & 26.3$\pm$ 0.4 & 33.8$\pm$ 0.5 &  123$\pm$   3 &  123$\pm$  11 &  290$\pm$  30\\
155\tnm{*} & 04295254$+$3522236 &  III & -1.89 & \nodata & 29.4$\pm$ 1.4 & \nodata & 12.8$\pm$ 0.7 & 53.3$\pm$ 1.2 & 32.5$\pm$ 0.7 & 53.5$\pm$ 1.3 & 45.2$\pm$12.2 & 14.3$\pm$ 1.6 & \nodata\\
156 & 04295418$+$3611573 &    F & -0.15 & \nodata & \nodata & \nodata & \nodata & 78.4$\pm$ 1.6 & 87.1$\pm$ 1.4 &  149$\pm$   1 &  582$\pm$   7 &  534$\pm$  50 &  721$\pm$  75\\
157 & 04295919$+$3610161 &   II & -1.22 & \nodata & 16.3$\pm$ 0.8 & \nodata & 17.6$\pm$ 0.8 & 18.5$\pm$ 0.4 & 18.8$\pm$ 0.4 & 15.3$\pm$ 0.3 & 10.8$\pm$ 1.6 & 10.6$\pm$ 1.0 & 88.1$\pm$11.3\\
158\tnm{*} & 04300152$+$3607333 &   II & -0.67 & \nodata & 37.2$\pm$ 1.8 & \nodata & 36.4$\pm$ 1.7 & 30.6$\pm$ 0.6 & 34.7$\pm$ 0.7 & 35.7$\pm$ 0.6 & 61.7$\pm$ 1.9 & 60.0$\pm$ 5.6 & 98.6$\pm$14.7\\
159\tnm{*} & 04300188$+$3538147 &   II & -0.39 & \nodata & 26.7$\pm$ 3.2 & \nodata & \nodata & 66.5$\pm$ 1.4 & 95.2$\pm$ 1.7 &  121$\pm$   1 &  191$\pm$   4 &  103$\pm$   9 &  115$\pm$  15\\
160\tnm{*} & 04300980$+$3613354 &   II & -1.12 & \nodata & \nodata & \nodata & \nodata & 30.4$\pm$ 0.6 & 26.5$\pm$ 0.5 & 27.8$\pm$ 0.5 & 33.1$\pm$ 1.4 & 29.5$\pm$ 2.7 & \nodata\\
161 & 04304933$+$3450460 &   II & -0.87 & 5.52$\pm$0.27 & \nodata & 5.95$\pm$0.29 & \nodata & 3.67$\pm$0.08 & 5.02$\pm$0.10 & 3.65$\pm$0.18 & 5.16$\pm$1.12 & 5.40$\pm$0.58 & \nodata\\
162 & 04304948$+$3450562 &   II & -0.81 & 8.02$\pm$0.38 & \nodata & 5.12$\pm$0.25 & \nodata & 8.03$\pm$0.18 & 6.40$\pm$0.14 & 4.02$\pm$0.20 & 21.1$\pm$ 1.4 & 19.2$\pm$ 1.8 & \nodata\\
163 & 04305208$+$3450089 &    F & -0.15 & 20.2$\pm$ 1.0 & \nodata & 23.5$\pm$ 1.1 & \nodata & 22.9$\pm$ 0.7 & 28.9$\pm$ 0.8 & 53.1$\pm$ 1.0 &  136$\pm$   5 &  114$\pm$  10 & \nodata\\
164\tnm{*} & 04320577$+$3606375 &  III & -1.95 & \nodata & \nodata & \nodata & \nodata & 4827$\pm$  402 & 3258$\pm$  129 &  715$\pm$   9 &  992$\pm$  16 &  874$\pm$  82 & \nodata\\
165\tnm{*} & 04325431$+$3604440 &   II & -1.12 & \nodata & \nodata & \nodata & \nodata & 2.51$\pm$0.06 & 1.83$\pm$0.05 & 1.36$\pm$0.13 & 5.26$\pm$0.89 & 3.39$\pm$0.40 & \nodata\\
166\tnm{*} & 04330315$+$3602045 &   II & -1.02 & \nodata & \nodata & \nodata & \nodata & 2.14$\pm$0.06 & 1.98$\pm$0.05 & 1.74$\pm$0.14 & 2.74$\pm$0.98 & 2.71$\pm$0.32 & \nodata\\
\enddata
\tnt{*}{The YSO is in a region of low column density ($N_{\rm{H2}}< 5 \times 10^{21}$ cm$^{-2}$) and so is a possible contaminant.}
\tnt{N}{The YSO lies beyond the $N_{\rm{H2}}$ column density map from \cite{Harveyetal2013} and so $N_{\rm{H2}}$ at its position is unknown.}
\tablecomments{The names of the YSOs give their J2000 positions. These YSOs are outside the 4 band IRAC coverage area and so are identified based on their WISE and MIPS fluxes. The coverage of individual IRAC bands are slightly offset from each other. Therefore some YSOs at the edges of the IRAC coverage have fluxes at some IRAC wavelengths. }\label{tbl:wisemips}
\end{deluxetable}

\begin{deluxetable}{cccccccccccc}
\tabletypesize{\scriptsize}
\tablecaption{Relative ages}
\tablewidth{0pt}
\tablehead{\ch{Region} & \ch{$N_{\rm{YSO}}$} & \ch{$N_{\rm{I}}$} & \ch{$N_{\rm{F}}$} & \ch{$N_{\rm{II}}$} &  \ch{$N_{\rm{I+F}}$/$N_{\rm{II}}$} }
\startdata
AMC				& 149 & 37 & 21 & 91 & 0.64 \\
OMC				& 3330 & 668 & 467 & 2195 & 0.52 \\
Perseus			&  368 & 54 & 71 & 243 & 0.51 \\ %c2d III Jorgensen IRAC
Serpens			& 196 & 39 & 25 & 132 & 0.49 \\
Ophiuchus		& 258 & 35 & 47 & 176 & 0.47 \\
IC 5146			& 128 & 29 & 12 & 87 & 0.47 \\ % GBS I Paul
Cepheus Flare	& 122 & 21 & 14 & 87 & 0.40 \\ % GBS II J Kirk
Corona Australis	& 37 & 7 & 2 & 28 & 0.32 \\ % GBS III Peterson
Lupus			& 95 & 8 & 12 & 75 & 0.27 \\
Chameleon II		& 22 & 2 & 1 & 19 & 0.16 \\
%Lupus			& 0 & 0 & 0 & 0 & 0.0 \\ %c2d XI Merin IRAC and MIPS
%Chameleon II		& 0 & 0 & 0 & 0 & 0.0 \\ %c2d X Alcala IRAC and MIPS
%Serpens			& 0 & 0 & 0 & 0 & 0.0 \\ %c2d XI Harvey IRAC and MIPS
%Serpens			& 0 & 0 & 0 & 0 & 0.0 \\ %c2d VIII Harvey MIPS
%Ophiuchus		& 0 & 0 & 0 & 0 & 0.0 \\ %c2d VII Padgett MIPS
%Perseus			& 0 & 0 & 0 & 0 & 0.0 \\ %c2d VI Rebull MIPS
%Chameleon II	& 0 & 0 & 0 & 0 & 0.0 \\ %c2d V Porras IRAC
%Lupus			& 0 & 0 & 0 & 0 & 0.0 \\ %c2d IV Chapman MIPS
%Serpens			& 0 & 0 & 0 & 0 & 0.0 \\ %c2d II Harvey IRAC
%Chameleon II	& 0 & 0 & 0 & 0 & 0.0 \\ %c2d I Young
%Lupus V and VI	& 258 & 35 & 47 & 176 & 0.47 \\ % GBS IV
%Ophiuchus North	& 258 & 35 & 47 & 176 & 0.47 \\ % GBS V
\enddata
\tablerefs{AMC: this work, OMC: \cite{Megeathetal2012}, Perseus: \cite{Jorgensen2006}, Serpens: \cite{Harvey2007}, Ophiuchus: L. Allen, in preparation (see \citealt{Evans2009}), IC 5146: \cite{Harvey2008}, Cepheus Flare: \cite{Kirk2009}, Corona Australis: \cite{Peterson2011}, Lupus: \cite{Merin2008}, Chameleon II: \cite{Alcala2008}}\label{tbl:ages}
\end{deluxetable}

\begin{deluxetable}{lcccccccc}
\tablecolumns{7} 
\tabletypesize{\footnotesize}
\tablewidth{0pc} 
\tablecaption{SED modelling results for Class II sources}
\tablehead{\ch{ID} & \ch{Fitted stellar } & \ch{\Av}	& \ch{\Lstar} & \ch{\lambdaturnoff} & \ch{\alphaexcess} & \ch{\Ldisk/\Lstar} \\
\ch{} 	& \ch{spectrum} 				   & \ch{(mag)}	& \ch{(\Lsun)} & \ch{(\micron)} & \ch{} & \ch{}} 
\startdata 
2  & K7 &  20.5  &  1.89  &  8.0  &  0.3  &  0.086  \\
3  & K7 &  0.0  &  0.14  &  5.8  &  -0.5  &  0.150  \\
5  & K7 &  19.0  &  0.46  &  5.8  &  -1.3  &  0.083  \\
8  & K7 &  2.9  &  0.57  &  5.8  &  -0.4  &  0.169  \\
9  & A0 &  7.5  &  1.46  &  2.2  &  0.1  &  0.205  \\
12  & K7 &  3.1  &  0.13  &  3.6  &  -1.0  &  0.402  \\
13  & K7 &  0.0  &  0.91  &  3.6  &  -0.4  &  0.634  \\
16  & K7 &  14.9  &  15.91  &  3.6  &  -0.5  &  0.338  \\
22  & K7 &  5.8  &  0.33  &  5.8  &  -0.7  &  0.133  \\
23  & K7 &  4.3  &  0.07  &  5.8  &  -0.8  &  0.133  \\
24  & K7 &  10.5  &  0.11  &  8.0  &  0.9  &  0.128  \\
26  & K7 &  7.2  &  0.30  &  5.8  &  -0.5  &  0.130  \\
27  & K7 &  6.8  &  0.13  &  5.8  &  -1.1  &  0.127  \\
29  & K7 &  10.0  &  25.06  &  2.2  &  -0.6  &  0.124  \\
31  & K7 &  9.0  &  0.57  &  5.8  &  -0.4  &  0.172  \\
36  & K7 &  4.1  &  1.62  &  8.0  &  -1.3  &  0.080  \\
37  & K7 &  2.5  &  0.95  &  3.6  &  -1.0  &  0.296  \\
39  & K7 &  5.5  &  0.44  &  3.6  &  -1.0  &  0.454  \\
41  & K7 &  6.5  &  0.32  &  3.6  &  -0.9  &  0.383  \\
42  & K7 &  7.4  &  0.30  &  8.0  &  -1.5  &  0.084  \\
43  & K7 &  9.1  &  0.18  &  5.8  &  -1.1  &  0.191  \\
45  & K7 &  15.1  &  0.56  &  3.6  &  -0.7  &  0.195  \\
50  & K7 &  5.2  &  1.22  &  5.8  &  -0.1  &  0.244  \\
52  & K7 &  4.0  &  2.29  &  3.6  &  -0.8  &  0.582  \\
53  & K7 &  2.0  &  0.21  &  3.6  &  -1.0  &  0.615  \\
55  & K7 &  6.0  &  1.83  &  5.8  &  -0.7  &  0.244  \\
57  & K7 &  6.5  &  0.12  &  5.8  &  -1.3  &  0.092  \\
58  & K7 &  2.7  &  0.36  &  8.0  &  1.5  &  0.271  \\
59  & K7 &  5.0  &  2.10  &  3.6  &  -1.0  &  0.600  \\
60  & K7 &  5.5  &  0.63  &  3.6  &  -0.6  &  0.348  \\
62  & K7 &  6.0  &  2.83  &  3.6  &  -0.9  &  0.408  \\
64  & K7 &  3.3  &  0.16  &  8.0  &  0.4  &  0.158  \\
65  & K7 &  4.0  &  1.56  &  5.8  &  -0.3  &  0.316  \\
66  & K7 &  15.5  &  0.64  &  8.0  &  -1.1  &  0.115  \\
67  & K7 &  0.5  &  1.93  &  3.6  &  -0.8  &  0.346  \\
68  & K7 &  23.4  &  7.82  &  5.8  &  -1.0  &  0.249  \\
69  & K7 &  6.0  &  0.80  &  2.2  &  -0.8  &  2.331  \\
72  & K7 &  3.2  &  0.11  &  5.8  &  -0.3  &  0.248  \\
74  & K7 &  5.9  &  0.17  &  8.0  &  0.6  &  0.043  \\
75  & K7 &  4.0  &  0.27  &  8.0  &  -1.5  &  0.048  \\
77  & K7 &  12.0  &  0.99  &  5.8  &  -1.2  &  0.111  \\
78  & K7 &  8.5  &  0.18  &  3.6  &  -1.1  &  0.206  \\
82  & K7 &  5.0  &  1.53  &  3.6  &  -1.0  &  0.960  \\
83  & K7 &  10.9  &  0.16  &  5.8  &  -0.3  &  0.149  \\
84  & K7 &  12.8  &  3.90  &  3.6  &  -1.1  &  0.319  \\
89  & K7 &  5.0  &  6.74  &  2.2  &  -1.1  &  0.087  \\
91  & K7 &  12.2  &  1.09  &  3.6  &  -0.9  &  0.310  \\
92  & K7 &  17.2  &  0.78  &  8.0  &  -0.1  &  0.079  \\
93  & K7 &  7.0  &  0.07  &  5.8  &  -0.4  &  0.182  \\
94  & K7 &  2.6  &  0.38  &  5.8  &  -0.5  &  0.102  \\
95  & K7 &  0.5  &  0.08  &  3.6  &  -0.2  &  0.499  \\
96  & K7 &  0.0  &  0.26  &  3.6  &  -0.6  &  0.434  \\
97  & K7 &  4.0  &  0.80  &  3.6  &  -0.6  &  0.394  \\
98  & K7 &  3.8  &  0.12  &  3.6  &  -0.6  &  0.321  \\
99  & K7 &  7.0  &  0.53  &  5.8  &  -0.8  &  0.204  \\
102  & K7 &  4.6  &  2.31  &  8.0  &  1.0  &  0.215  \\
104  & K7 &  3.0  &  0.45  &  3.6  &  -0.6  &  0.319  \\
105  & K7 &  4.8  &  0.13  &  8.0  &  -1.3  &  0.062  \\
106  & K7 &  3.0  &  38.62  &  4.5  &  -0.1  &  0.110  \\
108  & K7 &  10.2  &  0.11  &  8.0  &  2.4  &  1.830  \\
110  & K7 &  1.5  &  0.83  &  3.6  &  -1.1  &  0.428  \\
113  & K7 &  3.8  &  0.29  &  8.0  &  0.2  &  0.027  \\
114  & K7 &  5.0  &  0.44  &  8.0  &  -1.0  &  0.053  \\
116  & K7 &  1.0  &  0.24  &  3.6  &  -0.9  &  0.541  \\
119  & K7 &  3.0  &  0.06  &  3.6  &  -0.8  &  0.241  \\
120  & K7 &  9.4  &  0.55  &  5.8  &  -1.0  &  0.134  \\
122  & K7 &  5.5  &  0.26  &  3.6  &  -0.5  &  1.164  \\
123  & K7 &  23.0  &  0.76  &  8.0  &  -1.2  &  0.064  \\
127  & K7 &  13.2  &  0.70  &  8.0  &  -1.1  &  0.059  \\
128  & K7 &  2.0  &  3.04  &  2.2  &  -0.7  &  1.477  \\
129  & K7 &  6.5  &  0.19  &  5.8  &  -0.8  &  0.096  \\
130  & K7 &  4.5  &  0.14  &  8.0  &  -1.0  &  0.072  \\
132  & K7 &  5.0  &  0.79  &  5.8  &  -0.1  &  0.599  \\
133  & K7 &  4.0  &  0.31  &  8.0  &  0.7  &  0.040  \\
136  & K7 &  0.0  &  0.15  &  5.8  &  -1.1  &  0.066  \\
137  & K7 &  6.0  &  0.23  &  2.2  &  -1.8  &  1.156  \\
139  & K7 &  15.9  &  52.81  &  2.2  &  -1.8  &  0.228  \\
142  & K7 &  6.7  &  0.91  &  4.6  &  -0.9  &  0.112  \\
146  & K7 &  0.0  &  5.13  &  4.6  &  -1.4  &  0.109  \\
150  & K7 &  2.0  &  0.20  &  12.0  &  0.2  &  0.033  \\
151  & K7 &  1.0  &  0.11  &  4.6  &  -0.8  &  0.123  \\
152  & K7 &  2.6  &  0.48  &  4.6  &  -0.9  &  0.057  \\
153  & K7 &  2.0  &  2.94  &  3.4  &  -1.1  &  0.349  \\
157  & K7 &  6.4  &  0.98  &  4.6  &  -0.7  &  0.112  \\
158  & K7 &  3.0  &  0.27  &  2.2  &  -0.7  &  1.353  \\
159  & K7 &  1.0  &  0.17  &  2.2  &  -0.6  &  4.827  \\
160  & K7 &  3.0  &  0.96  &  4.6  &  -0.9  &  0.215  \\
161  & K7 &  10.0  &  0.21  &  3.6  &  -1.2  &  0.185  \\
162  & K7 &  4.4  &  0.48  &  12.0  &  1.3  &  0.044  \\
165  & K7 &  4.5  &  0.21  &  12.0  &  0.5  &  0.026  \\
166  & K7 &  5.2  &  0.11  &  4.6  &  -0.9  &  0.095  \\
\enddata
\label{tbl:diskpropII}
%\tablecomments{Table~\ref{tbl:diskpropII} is published in its entirety in the electronic edition of the {\it Astrophysical Journal}. A portion is shown here for guidance regarding its form and content.}
\end{deluxetable}

\begin{deluxetable}{lcccccccc}
\tablecolumns{7}
\tabletypesize{\footnotesize}
\tablewidth{0pc} 
\tablecaption{SED modelling results for Class III sources}
\tablehead{\ch{ID} & \ch{Fitted stellar } & \ch{\Av}	& \ch{\Lstar} & \ch{\lambdaturnoff} & \ch{\alphaexcess} & \ch{\Ldisk/\Lstar} \\
\ch{} 	& \ch{spectrum} 				   & \ch{(mag)}	& \ch{(\Lsun)} & \ch{(\micron)} & \ch{} & \ch{}} 
\startdata 
11  & A0 &  0.0  &  156.74  &  8.0  &  -1.6  &  0.019  \\
15  & K7 &  1.5  &  0.79  &  24.0  &  \nodata  &  0.015  \\
17  & K7 &  20.0  &  0.40  &  8.0  &  -1.3  &  0.006  \\
19  & K7 &  8.1  &  0.53  &  24.0  &  \nodata  &  0.009  \\
54  & K7 &  4.0  &  4.66  &  8.0  &  -1.0  &  0.014  \\
73  & K7 &  4.3  &  0.30  &  8.0  &  -1.5  &  0.053  \\
80  & K7 &  3.0  &  0.86  &  24.0  &  -99.0  &  0.008  \\
115  & K7 &  2.6  &  0.78  &  8.0  &  0.1  &  0.041  \\
141  & K7 &  6.0  &  137.96  &  12.0  &  -2.0  &  0.019  \\
143  & K7 &  2.0  &  36.17  &  12.0  &  -0.8  &  0.009  \\
144  & K7 &  4.8  &  326.26  &  12.0  &  -2.6  &  0.032  \\
145  & K7 &  2.5  &  13.61  &  12.0  &  1.7  &  0.012  \\
147  & K7 &  0.0  &  1.35  &  12.0  &  -0.0  &  0.007  \\
148  & K7 &  6.0  &  191.61  &  24.0  &  -99.0  &  0.010  \\
149  & K7 &  2.3  &  38.39  &  12.0  &  -0.7  &  0.007  \\
155  & K7 &  0.0  &  2.92  &  8.0  &  -0.7  &  0.026  \\
164  & K7 &  7.0  &  558.63  &  24.0  &  -99.0  &  0.003  \\
\enddata
\label{tbl:diskpropIII}
%\tablecomments{Table~\ref{tbl:diskpropIII} is published in its entirety in the electronic edition of the {\it Astrophysical Journal}. A portion is shown here for guidance regarding its form and content.}
\end{deluxetable}

\begin{deluxetable}{cccccccccccc}
\tabletypesize{\scriptsize}
\tablecaption{AMC Groups Summary}
\tablewidth{0pt}
\tablehead{
\ch{Group} & \ch{Position}		& \ch{\NYSO} & \ch{\NII}	& \ch{\NF}	& \ch{\NI}	& \ch{\NIF/\NII}	& \ch{\Reff}	& \ch{\Rcirc}	& \ch{Aspect Ratio}	& \ch{Mean Surf. Dens.} \\
\ch{}	& \ch{(RA, Dec)}			& \ch{}		& \ch{}		& \ch{}		& \ch{}		& \ch{}			& \ch{(pc)}	& \ch{(pc)}		& \ch{}				& \ch{(pc$^{-2}$)} }
\startdata
1\tnm{a}	& 67.562286, 35.239391	& 49			& 34			& 7			& 8			& 0.44			& 0.99		& 1.22			& 1.52				& 15.8 \\ 
2		& 67.610970, 35.770126	& 23			& 12			& 7			& 4			& 0.92			& 0.55		& 1.23			& 5.01				& 24.1 \\ 
3		& 67.671758, 35.541806	& 12			& 9			& 0			& 3			& 0.33			& 0.66		& 0.74			& 1.26				& 8.55 \\ 
4		& 67.188288, 36.440921	& 10			& 4			& 1			& 5			& 1.5			& 0.48		& 0.69			& 2.03				& 13.3 \\ 
5		& 67.708443, 34.958037	& 8			& 3			& 0			& 5			& \nodata		& \nodata	& \nodata		& \nodata			& \nodata \\ 
6		& 62.662345, 38.094258	& 6			& 5			& 0			& 1			& \nodata		& \nodata	& \nodata		& \nodata			& \nodata \\ 
7		& 62.525460, 40.037669	& 5			& 2			& 0			& 3			& \nodata		& \nodata	& \nodata		& \nodata			& \nodata \\ 
\enddata
\tnt{a}{Several known members near \lkha 101 are missing in our YSO list, affecting the values reported for this group.}
\label{tbl:grps}
\end{deluxetable}

\end{document}